\documentclass[]{aastex631}
\usepackage{xcolor}
\usepackage{tabularx}
\usepackage{multirow}
\usepackage{amsmath}
\usepackage{nicefrac}
\usepackage[h]{esvect}
\usepackage[normalem]{ulem}
\usepackage[all]{hypcap} 
\usepackage{tablefootnote}

\newcommand{\htwoplus}{\mathrm{H}_2^+}
\newcommand{\hthreeplus}{\mathrm{H}_3^+}
\newcommand{\heplus}{\mathrm{He}^+}
\newcommand{\hehplus}{\mathrm{HeH}^+}
\newcommand{\htwo}{\mathrm{H}_2}
\newcommand{\he}{\mathrm{He}}
\newcommand{\hplus}{\mathrm{H}^+}
\newcommand{\h}{\mathrm{H}}
\newcommand{\e}{\mathrm{e}^{-}}
\newcommand{\kb}{k_{\mathrm{B}}}

\newcommand{\xt}{\tilde{X}}

\shorttitle{Efficiency of Hydrodynamic Atmospheric Escape}
\shortauthors{Frelikh and Murray-Clay}

\begin{document}

\title{Efficiency of Hydrodynamic Atmospheric Escape in Hot Jupiters and Super Earths}
\author[0000-0003-3290-818X]{Renata Frelikh}
\affiliation{Laboratory for Atmospheric and Space Physics, University of Colorado, Boulder, 3665 Discovery Dr, Boulder, CO 80303}
\affiliation{Astrophysics Group, Department of Physics, Imperial College London, Prince Consort Rd, London SW7 2AZ, UK}
\email{renata.frelikh@lasp.colorado.edu}
\author[0000-0001-5061-0462]{Ruth A. Murray-Clay}
\affiliation{Department of Astronomy and Astrophysics, University of California, Santa Cruz, CA 95064, USA}
\email{rmc@ucsc.edu}
\begin{abstract}

We develop a flexible one-dimensional code to model the escape of hydrogen and helium from a hot Jupiter as a result of photoionization from extreme-ultraviolet (EUV) radiation. We include stellar spectrum heating and ionization, radiative cooling by Lyman-$\alpha$ and $\hthreeplus$, heat conduction, tidal gravity, a $\h$-$\he$ reaction network, and account for the secondary ionization of species by photoelectrons. For a fiducial hot Jupiter, we uncover a three-layer structure: an $\hthreeplus$-cooled layer of molecular hydrogen at the base, enveloped by a Lyman-$\alpha$-cooled layer of neutral hydrogen, which transitions into an ionized wind layer that is cooled by adiabatic expansion. The highest spectral energy photons are deposited in the molecular layer, where, after accounting for energy loss via photoelectrons and ionization, $\hthreeplus$ is a substantial radiative coolant. We run a grid of models, varying the distance of our fiducial planet from the star. We find that heat conduction at the base starts to have an effect at distances $\gtrsim 0.2$ au, increasing the $\hthreeplus$ cooling relative to the EUV input flux. At increasing stellocentric distances, the outflow becomes increasingly more neutral. The neutral hydrogen starts to decouple from the ionized outflow, free-streaming out. In pure H-He mini-Neptune/super-Earth simulations, the outflows are significantly cooler, allowing molecules to survive throughout the outflow, pointing toward the likely importance of molecular cooling in determining whether these planets can maintain massive atmospheres. The analysis in this paper provides a framework for understanding the impact of molecular radiative cooling on atmospheric outflows.

\end{abstract}



\section{Introduction} \label{sec:intro}
In this paper we build a framework for evaluating the effect of radiative cooling on planets experiencing stellar-irradiation-driven atmospheric mass loss. Hot and warm Jupiters are gas giants on orbits $\lesssim$ 1 au from their host stars. We demonstrate that for such planets, radiative cooling can act as a substantial energy sink, diverting energy away from powering the outflow. We adopt published estimates for heating efficiency and secondary ionization due to photoelectrons, whose interactions with the gas act as an important energy sink, especially for photon energies approaching the X-ray range of the stellar spectrum. We set the stage for understanding the role of radiative cooling due to molecules in sub-Neptune and super-Earth atmospheres.
Our goal in this paper is to guide the use of the energy-limited escape approximation that is widely used in the literature. For Jupiter-mass planets, we provide improved estimates of the efficiency factor for escape and the spectral energy range that is available to drive the wind. We illustrate how our simulation results diverge from this approximation, as radiative cooling becomes more dominant with increasing semimajor axis.

Atmospheric escape is known to affect the atmospheres of highly-irradiated planets. XUV (extreme ultraviolet and X-ray) radiation drives hydrodynamic escape from planets with hydrogen-dominated thermospheres by photoionizing hydrogen and heating the upper layer of the atmosphere \citep{watson1981dynamics}. Atmospheric escape has been theoretically predicted to shape the bulk properties of super-Earth-sized planets (\citealt{owen2013kepler} and \citealt{lopez2014understanding}) and has been later inferred from population analyses (e.g. \citealt{fulton2017california}). It has produced observable signatures in the hot Jupiters and warm Neptunes. Excess Lyman-$\alpha$ absorption in transit that is interpreted as escaping hydrogen has been detected for gas giants (e.g., \citealt{vidal2003extended}, 
\citealt{vidal2004detection}, \citealt{des2010evaporation}, \citealt{bourrier2013atmospheric}, \citealt{dos2023hydrodynamic}). Metals have been detected (\citealt{sing2019hubble}) in the upper atmosphere extending beyond the Roche lobe (i.e. the region surrounding a planet within which material is gravitationally bound to it), with several studies additionally providing tentative evidence for the presence of escaping metals (e.g. \citealt{vidal2004detection}, \citealt{linsky2010observations}, \citealt{ben2013hubble}, \citealt{salz2019swift}, \citealt{cubillos2020near}, \citealt{dos2023hydrodynamic}). 
Broad-band X-ray observations of hot Jupiter HD 209458b suggest the presence of metals in its extended upper atmosphere \citep{poppenhaeger2013transit}. Observations of the stellar Balmer line series (e.g. \citealt{jensen2018hydrogen}, \citealt{cabot2020detection}, \citealt{wyttenbach2020mass}, \citealt{yan2021detection}) provide an interesting probe of the upper atmospheres of planets potentially undergoing atmospheric escape. Escape has also been observed for warm Neptunes, with the striking Lyman-$\alpha$ transit of \citealt{ehrenreich2015giant}, detections of Lyman-$\alpha$ and metal excess absorption (\citealt{ben2022signatures}), and detections of the He 10833 \AA $\,$ line (\citealt{spake2018helium}, \citealt{allart2018spectrally}, \citealt{ninan2020evidence}, and \citealt{ben2022signatures}).

In the case of the hot Jupiters, winds reaching velocities of order 10 km s$^{-1}$ are driven from these strongly irradiated, heated planets, resulting in atmospheric mass loss as their hydrogen-dominated atmospheres are ionized as they are exposed to stellar irradiation. Theoretical work on atmospheric escape includes two important categories of hydrodynamic models: core-powered mass loss (e.g., \citealt{ginzburg2016super}, \citealt{ginzburg2018core}) and photoevaporative mass loss driven by stellar EUV and/or X-ray radiation (e.g. \citealt{lammer2003atmospheric}, \citealt{baraffe2004effect}, \citealt{yelle2004aeronomy}, \citealt{munoz2007physical}, \citealt{koskinen2007stability}, \citealt{rmc2009}, \citealt{adams2011magnetically}, \citealt{owen2014magnetically}, \citealt{trammell2014magnetohydrodynamic},  \citealt{tripathi2015simulated}, \citealt{mccann2019morphology}, \citealt{oklopvcic2018new}), as well as Balmer-driven escape for planets around high-mass stars (\citealt{munoz2019rapid}). Our focus is on EUV-driven photoevaporative mass loss. \citet{yelle2004aeronomy} and \citet{munoz2007physical} noted the presence of the $\hthreeplus$ ion, which is an effective radiative coolant. They found that the escape approaches the energy-limited (\citealt{watson1981dynamics}, \citealt{lammer2003atmospheric}) regime for strongly irradiated hot Jupiter planets. \citet{koskinen2007stability} additionally found that $\hthreeplus$ radiative cooling can significantly limit outflows. \citet{rmc2009} found that Lyman-$\alpha$ cooling can be significant, particularly for higher stellar ionizing fluxes.

The goal of our work is to understand what fraction of a planet's incoming stellar flux is radiatively cooled, and what fraction is available to power the hydrodynamic outflow, after accounting for primary and photoelectron-incuded secondary ionization. Thus, we aim to investigate in depth how radiative cooling from $\hthreeplus$ and Lyman-$\alpha$ affects the photoevaporative outflow process, and how the effect varies with distance from the star. In a hot Jupiter, it is necessary to understand the interface layer from molecular hydrogen gas (that is efficiently radiatively cooled and largely shielded from EUV radiation) to the photodissociated and/or thermally dissociated atomic layer that comprises that outflow. To this end, we couple photochemistry to an outflow model. Going to smaller planet masses, the bulk properties of sub-Neptunes and super-Earths will be significantly shaped by atmospheric escape. These planets are likely enriched in metals, which may be a source of additional radiative cooling, particularly if the outflows remain cool enough to host molecules. Though we do not include the full molecular chemistry necessary to model super-Earths, we present a series of preliminary models in this paper. The lower temperatures and the presence of $\hthreeplus$ throughout the outflow demonstrate the necessity of including molecular cooling for these smaller planets in follow-up work. We emphasize that we only include species of $\h$ and $\he$ in our models, which is more likely to be a significant limitation for these smaller planets than for hot Jupiters.

We analyze the results of our model in terms of where in the atmosphere the EUV photons of different energies are absorbed, as well as which cooling terms dominate. $\hthreeplus$ is formed and contributes to the radiative cooling of the highest-energy photons near the base of our atmosphere. For Jupiter-mass planets, Lyman-$\alpha$ cooling is important in the atomic-hydrogen-dominated layer just above the $\hthreeplus$-cooled region near the base. As we move the planet out in distance from the star, Lyman-$\alpha$ cooling becomes a more significant cooling source. We model the outflow along the substellar ray. As long as the wind is not magnetically confined, the energy will be redistributed around to the nightside of the planet.

We discuss the methods in Section \ref{sec:hydro}, where we describe our model. Additional details of the implementation are in Appendix \ref{sec:hydro_solver}. Section \ref{sec:results} contains the results for our fiducial hot Jupiter at 0.05 au, as well as a grid of models with varying distance from the star. There we discuss the structure of the atmosphere, including the molecular layer and the outflow, and which parts of the spectrum dominate in the atmosphere at different heights. For the grid, we discuss the mass-loss rates and compare them to the energy limited regime. We comment on differences for low-mass planets in Section \ref{sec:se}. We provide a discussion in Section \ref{sec:disc} and conclude with Section \ref{sec:summary}.

\section{Methods} \label{sec:hydro}

To address the question of efficiency of atmospheric escape, we developed a photochemical-hydrodynamic model that allows us to simulate a planetary outflow from a hydrogen- and helium-dominated atmosphere. We begin by focusing on the hot- and warm-Jupiter regime. The atmospheric composition is initialized to a gas consisting of molecular hydrogen and atomic helium in ratios corresponding to Solar abundance (the mass fraction X=0.75 for H, and Y=0.25 for He \citep{lodders2020solar}).  For Jupiter-sized planets with deep potential wells, the profiles need to be initialized with an initial velocity greater than zero for stability purposes. For our fiducial planet with parameters corresponding to hot Jupiter HD 209458b, we used as a starting point the low-flux solution of \citet{rmc2009}, modified to allocate the mass into H$_2$ and He instead of H and with the density increased by an overall multiplicative factor so that the density at the lower boundary was larger. The choice of initial conditions can be relatively arbitrary--the initial profiles do not have to be self-consistent to start, just enough to be able to evolve and avoid situations that code might have difficulty with, such as a snowplow phase \citep{mccann2019morphology}. For smaller-sized planets such as Neptune and Earth analogs, which typically have lower surface gravities, one may simply initialize the profiles to isothermal hydrostatic atmospheres. 
We solve the coupled equations of hydrodynamics and of our chemical network for a planet being irradiated by extreme ultraviolet (EUV) photons in the 248-12 eV range by time-evolving the species' number density, bulk velocity, and specific internal energy of the gas to steady state. The input spectrum used for ionization and subsequent heating is from \citet{richards1994}, a solar EUV flux model that we scale for the corresponding distance from the star. This spectrum is appropriate for modeling planets around old main-sequence stars, as X-rays do not dominate the outflow. Our code is flexible: an alternate spectrum may be used to model planets around stars early in their evolution, as well as expanded into the far-ultraviolet range to include molecular photochemistry that may be especially important in smaller-sized planet atmospheres.

A planet atmosphere can be modeled as a hydrodynamic outflow only if the gas remains collisional up to the sonic point, which is the distance from the planet where the gas velocity becomes supersonic. In other words, the gas can be treated as a fluid. In a gas consisting of several different species $s$, each with number density $n_s$, $n = \sum_{s} n_s$ is the total number density of the gas. For the gas to be collisional, its mean free path $\lambda = 1/(n\sigma$), with $\sigma$ the mutual cross section for collision of the gas particles, must be less than the scale height of the atmosphere $H = k_B T/(\mu g)$. Here, $k_B$ is the Boltzmann constant, $T$ is the gas temperature (we assume the same temperature for all species $s$), and $\mu$ is the mean molecular weight of the gas. The mean molecular weight $\mu = \sum_{s} n_s m_s /n$, where $m_s$ is the molecular weight of species $s$. The gravitational acceleration $g = GM_p/{r}^2$ is written for a planet with mass $M_p$ at radial distance of interest, $r$, and $G$ is the universal gravitational constant. For hydrogen particles of the same type, the collisional cross section $\sigma$ is either the Coulomb cross section $\sigma_C = 3.9 \times 10^{-14}(T/10^4)^{-2}$ cm$^2$ (for ion-ion scattering) or the hard body cross section $\sigma_{hb} = 3.5\times 10^{-16}$ cm$^2$ (for neutral-neutral scattering), depending on whether the gas is mostly ionized (in the wind, for a high stellar flux) or mostly neutral (at the base). For neutral-ion scattering, the dominant cross section is the charge exchange cross section, $\sigma_{ce} = 4.746 \times 10^{-15} (T/10^4 \mathrm{K})^{-0.1311} \, \mathrm{cm^{2}}$ (e.g., \citet{osterbrock1961ambipolar}). We confirm that our fiducial case presented in Section \ref{sec:hd209} is collisional throughout. We discuss collisionality in detail in Section \ref{sec:collisionality}.

The methods section is structured as follows: the equations of hydrodynamics are desribed in Sections \ref{subsec:hydro}--\ref{sec:lyalpha}, the numerical method used for solving them in Sections \ref{subsec:hydrosolver}--\ref{subsec:boundary}, and the chemical reactions and solver in Sections \ref{sec:chemnet} and Appendix \ref{appendix:chemsolver}. The spectrum and a discussion of the photoionization processes are given in Section \ref{subsec:spectrum}.

\subsection{Hydrodynamics}\label{subsec:hydro}

The hydrodynamic equations, also known as the Euler equations, encapsulate conservation of mass, momentum, and energy. In Lagrangian form, the equations for mass, momentum, and energy conservation, respectively, can be written as (e.g., \citealt{clarke2007principles}):
\begin{equation}\label{eq:mass_lagrangian}
    \frac{\mathrm{D}\rho}{\mathrm{D} t}+\rho \vv{\nabla} \cdot \vv{u} = 0,
\end{equation}

\begin{equation}\label{eq:momentum_lagrangian}
    \rho \frac{\mathrm{D}\vv{u}}{\mathrm{D} t} =  -\vv{\nabla} P -\rho \vv{\nabla}\Psi, \, \mathrm{ and}
\end{equation}

\begin{equation}\label{eq:energy_lagrangian}
    \frac{\mathrm{D}e}{\mathrm{D}t} = \frac{P}{\rho^2}\frac{\mathrm{D}\rho}{\mathrm{D}t} + \Gamma_m - \Lambda_m +\vv{\nabla} \cdot (\kappa_{\rm{th}} \vv{\nabla} T).
\end{equation}

\noindent Here, $\rho$ is the mass density of all of the species of the atmosphere, and $\vv{u}$ is their bulk velocity. The pressure is denoted by $P$, and the external gravitational and tidal forces are written in terms of the gradient of a scalar potential, $\Psi$. The radial coordinate $r$ is the distance from the center of the planet. The term $e$ in Equation \ref{eq:energy_lagrangian} is the specific internal energy, i.e. the internal energy per unit mass ($\rho e$ is the internal energy density, i.e. the internal energy per unit volume). 
The thermal conductivity is computed following the implementation of \citet{munoz2007physical}:
\begin{equation}
    \kappa_{\rm{th}} = \frac{1}{n}\sum_s n_s \kappa_s,
\end{equation}
with $\kappa_s = A_s T^{\alpha_s} \, \mathrm{erg} \,  {\mathrm{cm}^{-1}} \, \mathrm{s}^{-1} \, \mathrm{K}^{-1}$, with the coefficients as given in Table 1 of \citet{munoz2007physical}.

Finally, $\Gamma_m$ and $\Lambda_m$ are the radiative heating and cooling source terms, respectively, per unit mass. The heating comes from the EUV spectrum (Section \ref{subsec:spectrum}), and the cooling included is Lyman-$\alpha$ cooling (Section \ref{sec:lyalpha}) and infrared cooling due to $\hthreeplus$ emission (Section \ref{sec:hthreeplus})\footnote{Recombination cooling contributes to radiative energy loss, with the rate $\Lambda_{rr}\simeq 6.11\times 10^{-10}k_{B}T^{0.11}n_{H^+}n_e/\rho$ erg s${^{-1}}$ \citep{osterbrock2006astrophysics}. We confirm in post-processing that it is negligible for the parameter space of our models, so it is turned off in our models.}.
Bolometric heating and cooling determine the temperature in the lower boundary region, below the heights where substantial energy is deposited by high-energy photons.  While we do not model the structure of this region in detail, we provide this gas with sufficient energy to maintain a temperature which we interpret to be determined by bolometric heating and cooling, as described in Section \ref{sec:bolo}.


We write Equation \ref{eq:mass_lagrangian} in one-dimensional spherically symmetric coordinates, after employing the definition of the Lagrangian derivative $\mathrm{D}/\mathrm{D}t = \partial /\partial t + \vv{u} \cdot \vv{\nabla}$:

\begin{equation}\label{mass_spherical2}
    \frac{\partial \rho}{\partial t} + u\frac{\partial \rho}{\partial r} + \frac{\rho}{r^2} \frac{\partial (u r^2)}{\partial r} = \frac{\partial \rho}{\partial t} + \frac{1}{r^2}\frac{\partial}{\partial r} (\rho u r^2) = 0.
\end{equation}

\noindent An individual mass conservation equation can be written for each species $s$ of our atmosphere, with corresponding density $\rho_s$. We add a chemical source term $S_s$ on the right-hand side of the mass conservation equation to denote the creation and destruction of species $s$ in our chemical network, which we solve for in a separate substep of our routine. 
For each species $s$,

\begin{equation}\label{mass_species}
    \frac{\partial \rho_s}{\partial t} + \frac{1}{r^2}\frac{\partial}{\partial r} (\rho_s u r^2) = S_s
\end{equation}

\noindent We note that while diffusion is implemented in our code, we do not expect it to significantly affect the purely hydrogen and helium models presented here, so we turn it off and leave further discussion for future work.
The bulk density $\rho =  \sum_{s} \rho_s$ satisfies Equation \ref{eq:mass_lagrangian}. Similarly, we rewrite Equation \ref{eq:momentum_lagrangian}, including gravity and tidal gravity as external forces:

\begin{equation}\label{momentum_spherical}
    \frac{\partial u}{\partial t} + u \frac{\partial u}{\partial r} = -\frac{G M}{r^2} - \frac{1}{\rho} \frac{\partial P}{\partial r} + \frac{3GM_{\star } r}{a^3},
\end{equation}

\noindent where $a$ is the distance between the planet and the star.
Finally, we note that in steady state ($\partial/\partial t= 0$), Equation \ref{eq:energy_lagrangian} may be re-written as:
\begin{equation} \label{eq:steadystate_e}
    \rho u \frac{\partial e}{\partial r} = \frac{P}{\rho}u\frac{\partial \rho}{\partial r} + \rho (\Gamma_m - \Lambda_m) + \frac{\partial}{\partial r} \left( \kappa_{\rm{th}} r^2 \frac{\partial T}{\partial r}\right),
\end{equation}
which is a form of Equation (3) from \citet{rmc2009} with the addition of conduction.
More generally, employing mass conservation (Equation \ref{eq:mass_lagrangian}) and expanding the total derivative in Equation \ref{eq:energy_lagrangian} gives:
\begin{equation}
    \frac{\mathrm{\partial}e}{\partial t} + (\vv{u} \cdot \vv{\nabla}) e = - \frac{P}{\rho} \vv{\nabla} \cdot \vv{u} + \Gamma_m - \Lambda_m +\vv{\nabla} \cdot (\kappa_{\rm{th}} \vv{\nabla} T).
\end{equation}

\noindent In one-dimensional spherical coordinates, this is:
\begin{equation}\label{energy_spherical}
    \frac{\partial e}{\partial t} + u \frac{\partial e}{\partial r} = -\frac{P}{\rho r^2}\frac{\partial(u r^2)}{\partial r} + \Gamma_m - \Lambda_m + \frac{\partial}{\partial r} \left( \kappa_{\rm{th}} r^2 \frac{\partial T}{\partial r}\right).
\end{equation}

\noindent The ideal gas law relates the pressure to the density:

\begin{equation}\label{idealgas}
    P = n \kb T = \rho \frac{\kb T}{\mu} = \rho e (\gamma-1).
\end{equation}

\noindent Thus, the specific internal energy is related to the pressure via: $P/\rho = e (\gamma-1)$. For the calculation of the adiabatic index $\gamma$, we include the 3 translational degrees of freedom, as well as the contribution of the rotation of $\htwo$; thus, $\gamma = 5/3$ where $\htwo$ is largely dissociated and $\gamma$ approaches 7/5 in the $\htwo$ layer. We omit vibration, as vibrationally-excited $\htwo$ is not expected to be present in significant amounts in our model due to its rapid dissociation at higher temperatures: $<10\%$ of the number density is in $\htwo$ at 2000 K, for example. We leave detailed estimates of the vibrational level populations for future work, as the nonthermal electrons and excitation by the stellar irradiation field may be a source of additional vibrational excitations of $\htwo$.\footnote{We follow \citealt{d2013three} for the implementation of the rotational degrees of freedom for the $\htwo$. We use the contributions of the main species to estimate $\gamma = 1 + 1/E_{\mathrm{tot}}$, where $E_{tot} = (E_{\h}+E_{\he}+ E_{\htwo}) \mu$, with the mean molecular weight $\mu$ as defined in Equation 19 of \citet{d2013three}.} Equations \ref{mass_species}, \ref{momentum_spherical}, and \ref{energy_spherical} are solved by our code. 

\subsection{Bolometric heating and cooling}\label{sec:bolo}

The base of our atmosphere is optically thick to incoming EUV radiation, but optically thin to  optical and infrared radiation. We implement a bolometric heating and cooling term in this region, as the EUV heating is insufficient to prevent the gas from cooling to unphysically low temperatures.  Our atmosphere is least opaque to photons in the highest-energy bin in our spectrum (124--248 eV). Thus, our calculation is only strictly physical at high enough altitudes that the gas is optically thin for these photons. 

In the bolometric region, the temperature is set by the balance between heating by the entire spectrum of the star (primarily in the optical) and blackbody radiation by the atmospheric gas (primarily in the infrared, IR). The heating and cooling rates depend on the optical and IR opacities of the gas, which we do not model.  Instead, we force this region to be approximately isothermal at the planet's equilibrium temperature by employing an artificially high value of $\kappa$ for both opacities, as detailed below. A more accurate model would include dust sinking as well as contributions to the opacity from molecular lines and atoms. Thus, a detailed calculation of the transition between the bolometric region and the region optically thin to the EUV, implementing appropriate opacities for the optical and IR photons, is left for future work. The bolometric heating and cooling term is restricted to heating the layer below $\tau=3$ for photons in the highest-energy bin of our spectrum, where $\tau$ is the optical depth to incoming radiation.  

For a planetary equilibrium temperature, $T_{\mathrm{eq}}$ at a given distance from the star, the effective heating rate per mass at the substellar point is:  $\Gamma_{\rm{bol}} = F_* n_{\rm{abs}} \sigma_{\rm{abs}} /\rho = F_* \kappa_{\rm{opt}}$, where $n_{\rm abs}$ is the volumetric number density of optical absorbers, and the absorption opacity, $\kappa_{\rm opt}$, is determined by absorbers with optical cross-section, $\sigma_{\rm abs}$. The incoming flux sets the equilibrium temperature of the planet, which we obtain by assuming the incoming flux in the optical is radiated away from the planet uniformly as a blackbody: $F_{\star} = 4 \sigma_{\rm{SB}}T_{\mathrm{eq}}^4$, where $F_*$ is the incoming stellar flux and $\sigma_{\rm{SB}}$ is the Stefan-Boltzmann constant. Thus, we can set the heating rate in terms of the target equilibrium temperature: $\Gamma_{\rm{bol}}  = 4\kappa_{\rm{opt}} \sigma_{\rm{SB}} T_{eq}^4$.  

The cooling term radiates as a blackbody at the local atmospheric gas temperature $T$ with a cooling flux per unit area $F_{\rm{cool}} =   \sigma_{\rm{SB}} T^4$. The cooling rate per mass can be written as: $\Lambda_{IR} = F_{\rm{cool}} n_{\rm{em}} \sigma_{\rm{em}} /\rho =  \kappa_{\rm{IR}} \sigma_{\rm{SB}} T^4$, where IR emitters with emission cross-section $\sigma_{\rm {em}}$ and volumetric number density $n_{\rm em}$ determine the IR opacity $\kappa_{\rm IR}$. The net bolometric heating and infrared cooling term, per mass, is then written as:
\begin{equation}\label{eq:newtonian_cooling}
    \Gamma_{\rm{bol}} - \Lambda_{IR} = \kappa \sigma_{\rm{SB}} (4T_{eq}^4 -  T^4) = \kappa \sigma_{\rm{SB}} (T_{0}^4 - T^4) \mathrm {\, erg\, g^{-1}\, s^{-1}},
\end{equation}
where we have set $\kappa_{\rm opt}$ and $\kappa_{\rm IR}$ to a single inflated value $\kappa = 1$ cm$^2$ g$^{-1}$. In addition, in our implementation we set $T_{0}^4 = 4 T_{\mathrm{eq}}^4$, such that Equation \eqref{eq:newtonian_cooling} has the form of a Newtonian cooling term. The exact correspondence of the local base temperature to the planetary equilibrium temperature depends on the redistribution of heat from the dayside to the entire planet, the details of which we do not model. 
We note that while we have written this bolometric heating and cooling term in a form that references the physics that sets the temperature, leaving room for more detailed modeling in future work, our choices effectively force the bolometrically heated layer to be isothermal at the temperature given by the input parameter $T_{0}$.

\subsection{H3+} \label{sec:hthreeplus}

The $\hthreeplus$ ion, also known as trihydrium, is a significant coolant in gas giant planet thermospheres. In the astronomical context, it was first detected in Jupiter's auroral ionosphere by infrared emission \citep{drossart1989detection}. It has also been observed to be present in the atmospheres of the Solar System giants Saturn (\citealt{geballe1993detection}) and Uranus (\citealt{trafton1993detection}), but has not been detected in Neptune (\citealt{moore2020atmospheric}). It has not yet been detected in an exoplanet.

It is primarily produced from the hydrogen molecule by a chain of two chemical reactions. First, molecular hydrogen is ionized by a photon\footnote{In Jupiter's auroral region the dominant pathway for $\htwoplus$ formation is election impacts with $\htwo$, as opposed to EUV ionization \citep{drossart1989detection}. In this case, the bulk of the $\hthreeplus$ emission is not determined by the incoming EUV flux.} with energy above the ionization threshold to produce the dihydrogen cation: $\htwo$ + h$\nu \rightarrow \htwoplus + \e$. This ionized molecular hydrogen then reacts with neutral molecular hydrogen to form $\hthreeplus$: $\htwoplus + \htwo \rightarrow \hthreeplus$ + H, as first proposed by \citealt{hogness1925ionization}. The main destruction pathway is dissociative recombination, the outcomes of which are twofold: $\hthreeplus + \e \rightarrow \htwo$ + H and $\hthreeplus + \e \rightarrow$ H + H +H.  We do not assume that these pathways dominate (see Section \ref{sec:chemnet} for the list of reactions included in our model), but we verify post-facto that they do indeed dominate in our results.

Due to its structure, the $\hthreeplus$ molecule is an efficient emitter in the infrared. For a detailed review of the vibrational modes contributing to the infrared spectrum of $\hthreeplus$, see \citet{miller2020} and references therein. We implement the $\hthreeplus$ cooling function calculated by \citet{miller2013}:


\begin{equation} \label{eq:h3plus3}
\begin{split}
E(800<T<1800K) = & -62.7016 + 0.0526104 T - 7.22431\times 10^{-5} T^2 + 5.93118\times 10^{-8} T^3 - 2.83755\times 10^{-11} T^4 +\\
 & 7.35415\times 10^{-15} T^5 - 8.01994\times 10^{-19} T^6 \mathrm{\,W  \, sr^{-1} \,molecule^{-1}}
\end{split}
\end{equation}

\begin{equation}
\begin{split}
E(1800<T<5000K) = & -55.7672 + 0.0162530 T - 7.68583\times 10^{-6} T^2 + 1.98412\times 10^{-9} T^3 - 2.68044\times 10^{-13} T^4 +\\
 & 1.47026\times 10^{-17} T^5 \mathrm{\,W  \, sr^{-1} \,molecule^{-1}} \label{eq:h3plus4}
\end{split}
\end{equation}

The temperature-dependent cooling function allows us to express the $\hthreeplus$ contribution to the cooling term in our code as: $\Gamma_{\hthreeplus} = 4 \pi e^{E(T)} n_{\hthreeplus} \times 10^7 / \rho \, \mathrm{erg\, cm^{-3}\, s^{-1}}$. We note that the cooling term is zero for temperatures above the range provided in Equations \eqref{eq:h3plus3} and \eqref{eq:h3plus4}, as we do not expect $\hthreeplus$ to exist there in any significant amounts due to its source $\htwo$ being thermally and/or photo-dissociated in the regions of the atmosphere where these higher temperatures occur.  We demonstrate in Section \ref{sec:h3pemission} that $\hthreeplus$ emission from the photoionization-heated gas is generally optically thin. We note that emission by a given $\hthreeplus$ molecule can proceed in any direction. As this radiation is optically thin, it will not be re-absorbed locally in any significant amount. We treat the radiation that is emitted downward (i.e. toward the lower simulation boundary) as having left the domain. The energetics at the lower boundary are dominated by the bolometric heating and cooling terms, and as such only impact the solution by acting as a lower boundary condition.

\subsection[Lyman-alpha cooling]{Lyman-\(\mathbb{\alpha}\) cooling}\label{sec:lyalpha}

The other significant radiative cooling mechanism in our model is Lyman-$\alpha$ cooling, which occurs due to radiative de-excitation of neutral hydrogen atoms that are collisionally excited to the n=2 state. This cooling converts thermal energy into radiation via impacts. As our gas is partially ionized, the excitation will be primarily due to electron impacts, as electron collisions are frequent due to their higher thermal velocities. We use the expression used by \citet{rmc2009} from \citet{black1981physical}:
\begin{equation}
    \Lambda_{\mathrm{Ly}\alpha} = -7.5\times 10^{-19} n_{\h} n_\mathrm{e} e^{-118348/T} \mathrm{erg}\,\mathrm{cm}^{-3}\,\mathrm{s}^{-1},
\end{equation}\label{eq:lyman_alpha_cooling_rate}
where the density of electrons $n_\mathrm{e} = n_{\hplus}+n_{\htwoplus}+n_{\hthreeplus}+n_{\heplus}+n_{\hehplus}$ is the sum of the number densities of the ionized species (we enforce charge neutrality). We note that the Lyman-$\alpha$ photons must be able to escape in order for the cooling to be efficient. The gas is not optically thin to the radiation at this energy; however, \citet{rmc2009} show that these photons can freely escape after scattering into the Lyman-$\alpha$ line wings.  In Appendix \ref{appendix:lyaesc}, we conduct a Monte Carlo radiative transfer calculation for our fiducial planet to verify that a large fraction of Lyman-$\alpha$ photons escape the outflow.

\subsection{Hydrodynamic solver} \label{subsec:hydrosolver}
We adopt the CIP (constrained interpolation profile) scheme \citep{yabe1991universal} for solving the hydrodynamics equations. It minimizes numerical diffusion by keeping track of not only the cell-centered quantities, but also of their spatial gradients. As suggested by \citep{yabe1991universal}, we also adapt a staggered mesh (Section \ref{sec:grid}) to avoid the problem of damping oscillations caused by the decoupling of odd and even grid cells. Avoiding unnecessary dissipation improves mass and energy conservation, which is important especially for low scale heights at the base of the atmosphere, where spatial concentration gradients are steep. This scheme was tested in the context of atmospheric escape by \citet{kuramoto2013effective}, who found that it reproduces hydrodynamic outflow solutions with relatively high accuracy.

For each hydrodynamics timestep, we split the fluid equations (Equations \ref{mass_spherical2}, \ref{momentum_spherical}, and \ref{energy_spherical}) into an advective part on the left, and source terms on the right (as documented in detail in Appendix \ref{sec:hydro_solver}). The advective part is solved via the CIP method, as described in  Appendix \ref{sec:hydro_solver_ad}. Then, the source terms on the right are explicitly marched forward in time with a finite differencing method, as described in Appendix \ref{sec:nonadvectpart}. We implement a number density floor of zero for all species. 
An artificial viscosity is added for stability (Section \ref{sec:visc}), and the chemical network is solved for via the semi-implicit method StepperSie from Numerical Recipes \citep{press2007numerical} (Section \ref{sec:chemnet} and Appendix \ref{appendix:chemsolver}).

\subsection{Grid}\label{sec:grid}
\begin{figure}[ht] 
  \includegraphics[width=\textwidth]{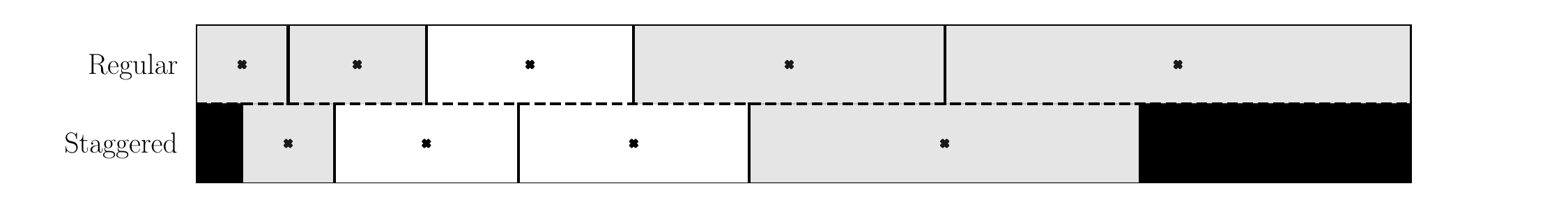}
  \caption{Illustration of a staggered grid with uneven spacing. The staggered grid cells are centered at the regular grid cell boundaries. The ghost cells are shaded in gray.}
  \label{fig:grid}
\end{figure}

We solve Equations \ref{mass_species}, \ref{momentum_spherical}, and \ref{energy_spherical} by time marching the density, velocity, and specific internal energy to a steady state on a one-dimensional non-uniform grid in spherical coordinates. We use a non-uniform grid to allow for a higher uniform resolution at the base, increasing the cell size with increasing altitude. For our fiducial case, the planetary radius $R_p$ is set to 1.35 $R_{\rm{J}}$. The size of the base cell is set to $10^5$ cm, and the grid cell size is further set by a geometric progression, $\Delta r _{i} = 1.014 \, \Delta r _{i-1}$, where $i$ is the cell index on the grid (called the ``regular" grid in Figure 1). The grid spans 1 $R_p$ to $\sim 8.8$ $R_p$. This grid is used for the density and energy profiles---the variables are calculated at the cell centers. Figure \ref{fig:grid} illustrates our setup for a regular grid with a single cell (white) bounded by ghost cells (gray). The velocity grid is staggered from the regular grid such that the cell-centered velocity values coincide with the cell boundaries on the regular grid. The size of the first cell of the staggered grid is set by $\Delta r_{1/2} = \Delta r_0$, where the indeces of the staggered grid are offset from the regular grid by 1/2. The sizes of the following cells are calculated via: $\Delta r_{i+1/2} = 2 \, \Delta r_i - \Delta r_{i-1/2}$. On the regular grid, there are 2 ghost cells at the base, and 2 ghost cells at the top, for a total of 504 cells. On the staggered grid, there is 1 ghost cell at the base, and 1 ghost cell at the top, for a total of 503 cells.

Staggering the grid this way allows for the exact computation of $\partial (ur^2)/\partial r$ on the right-hand side of the mass and energy conservation equations using adjacent cells. In addition, it prevents oscillations arising from odd-even decoupling present for a non-staggered grid (see Appendix \ref{ap:sec:grid}).

\subsection{Artificial viscosity}\label{sec:visc}
For stability, we add a numerical viscosity term $Q_i$ to the pressure term, analagous to that used by \citet{yabe1991universal} for the shock-tube problem. First, we calculate the change in velocity across cell $i$ using the divergence equation: $(\vv{\nabla}\cdot \vv{u})_i\Delta r_i = ((r^2 u)_{i+1/2}-(r^2 u)_{i-1/2})/r^2_i$, as suggested by \citet{yabe1991universal}. Then, the artificial viscosity is given by a combination of two terms \citep{landshoff1955numerical}, linear and quadratic in the divergence of velocity:
\begin{equation}
    Q_i  = 
    \begin{cases} 
    0.75 \left[ -\rho_i \left(\frac{\gamma P_i}{\rho_i} \right)^{1/2} (\vv{\nabla}\cdot \vv{u})_i \Delta r_i + \frac{\gamma+1}{2}\rho (\vv{\nabla}\cdot \vv{u})^2_i\Delta r_i^2 \right] & \text{, if } (\vv{\nabla} \cdot\vv{u})_i < 0 \\
         0         & \text{, if }(\vv{\nabla} \cdot \vv{u})_i \geq 0.
    \end{cases}
\end{equation}
The first term, linear in $(\vv{\nabla}\cdot \vv{u})$, is required to stabilize shock waves that might arise, since the scheme reduces numerical dissipation \citep{nishiguchi1983second}. It is the product of the density, adiabatic sound speed, and the divergence of the velocity. This term reduces short-wavelength numerical oscillations. The second term is a von-Neumann viscosity term \citep{vonneumann1950method}---it supresses longer-wavelength numerical oscillations. The tunable constant of order 1 in front of the viscosity term controls the strength of the artificial viscosity, and we choose 0.75 following \citet{yabe1991universal}. We note that our implementation is similar to that of the tensor artificial viscosity of \citet{stone1992zeus}, where instead of the standard gradient implementation the divergence of velocity is used.

We note that in our final steady-state solutions, the velocity of the outflowing wind increases monotonically with radius, so that $(\vv{\nabla} \cdot \vv{u})_i > 0$ and $Q_i = 0$. Thus, the artificial viscosity helps our code evolve stably to a steady-state solution but does not impact the final result.

\subsection{Boundary conditions}\label{subsec:boundary}
As the flow is supersonic, no boundary conditions for density, pressure, and velocity have to be specified at the outer boundary, as waves cannot enter the domain from the outer boundary.  Both outflow and Neumann boundary conditions work at the outer boundary.  We ensure that the the grid's outer boundary is positioned well away from the sonic point to prevent it from influencing the solution as it evolves to steady state. For simulations beyond 0.05 au, the grid is extended accordingly. We note that conduction makes the choice of outer boundary conditions more important, as the conduction term does not explicitly include a flux limiter. At the inner boundary, the flow is subsonic, so we specify the boundary conditions for the number densities of each species and the pressure. We extrapolate the velocity from the grid: $u_0 = u_1 - (r_1-r_0)(u_2-u_1)/(r_2-r_1)$: a wave can travel from the domain to the inner boundary, as the flow is subsonic, so the value of the velocity is part of the solution and cannot be imposed as a boundary condition.
\subsection{Chemical network} \label{sec:chemnet}
The chemical network of reactions (Table \ref{tab:reactions}) includes the ionization processes that heat the gas and drive a wind. They also include the reactions that generate $\hthreeplus$, an important infrared coolant in the molecular layer. In addition, they determine how the wind transitions from the molecular to the atomic layer. The reactions were compiled by \citet{yelle2004aeronomy} and \citet{munoz2007physical}. The reaction rates per volume are computed by: $r_j = k_j \prod_{s, j} n_{s, j} $, where $n_{s, j}$ are the number densities of the reactants in reaction $j$ and $k_j$ is the corresponding rate coefficient. The rate coefficients for the photoionization reactions are determined as in Section \ref{subsec:spectrum}. The evolution of the concentration of the 7 atomic/molecular species (+electrons) is given by Equation \ref{mass_species}, where the 22 reactions of our chemical network are encapsulated by the source terms, $S_s$. In terms of the reation rates, the source term for a given species $s$ can be written as $S_s = m_s \Sigma_j r_j$, where the sum is performed over the set of reactions $j$ that species $s$ (with molecular mass $m_s$) is involved in, either as a reactant or a product. We note that if the species $s$ is a reactant for a given reaction, the term $r_j$ takes on a negative sign in the sum. As the rates of the reactions vary by several orders of magnitude, the reaction network cannot be solved via a straightforward explicit method---the set of equations is said to be ``stiff". We implement the semi-implicit method of solving stiff systems of equations from \citet{press2007numerical}, the details of which are given in Appendix \ref{appendix:chemsolver}.

\begin{deluxetable}{cccc}
\tabletypesize{\footnotesize}
\tablecolumns{4}
\tablecaption{\label{tab:reactions} Chemical network}
\tablehead{
\colhead{Label} & \colhead{Reaction}& \colhead{Rate coefficient$^a$}& \colhead{Reference}
}
\startdata
\hline
k1 & H + h$\nu \rightarrow \hplus$ &  & \citet{schunk2009ionospheres}\\ 
k2 & $\hplus + \e \rightarrow$ H & $4.0 \times 10^{-12} (300 \,\rm{K} /T)^{0.64}$ & \citet{storey1995recombination}\\ 
k3  & $\htwo$ + h$\nu \rightarrow \htwoplus + \e$  &   &\citet{schunk2009ionospheres} \\
k4  & $\htwo$ + h$\nu \rightarrow \rm{H} + \hplus + \e$  &   &\citet{schunk2009ionospheres} \\
k5  & $\htwoplus + \e \rightarrow$ H + H  & $2.3\times 10^{-8} (300\,\rm{K}/T)^{0.4}$  & \citet{auerbach1977merged}\\
k6  & $\hthreeplus + \e \rightarrow \htwo$ + H  & $2.9\times 10^{-8} (300\,\rm{K}/T)^{0.65}$  &\citet{sundstrom1994destruction} \\
k7  & $\hthreeplus + \e \rightarrow$ H + H +H  &  $8.6\times 10^{-8} (300\,\rm{K}/T)^{0.65}$ & \citet{datz1995branching}\\
k8  & $\htwoplus + \htwo \rightarrow \hthreeplus$ + H  &  $2.0\times 10^{-9}$ & \citet{theard1974ion} \\
k9  & $\htwoplus + \rm{H} \rightarrow \hplus + \htwo$  &  $6.4\times 10^{-10}$ & \citet{karpasb1979ion}\\
k10  & $\hplus + \htwo\rightarrow \htwoplus + \rm{H} $  &  $1.0\times 10^{-9} e^{-2.19\times 10^4/T}$ & \citet{yelle2004aeronomy}\\
k11  & $\hthreeplus + \rm{H} \rightarrow \htwoplus + \htwo $  &  $2.0 \times 10^{-9}$ & \citet{yelle2004aeronomy}\\
k12$^b$  & $\htwo + \rm{M} \rightarrow \rm{H} + \rm{H} + \rm{M} $  &  $1.5\times 10^{-9} e^{-4.8\times 10^4/T}$ & \citet{baulch1992evaluated}\\
k13$^b$  & $\rm{H} + \rm{H} + \rm{M} \rightarrow \htwo + \rm{M} $  &  $8.0\times 10^{-33} (300\,\rm{K}/T)^{0.6}$ & \citet{ham1970gas}\\
k14 & $\he$ + h$\nu \rightarrow \heplus$ &  & \citet{schunk2009ionospheres} \\ 
k15 & $\heplus + \htwo \rightarrow \hehplus + \rm{H}$ & $4.2 \times 10^{-13}$ & \citet{schauer1989reactions} \\
k16 & $\heplus + \htwo \rightarrow \hplus + \rm{H} +\he$ & $8.8 \times 10^{-14}$ & \citet{schauer1989reactions} \\
k17 & $\hehplus + \htwo \rightarrow \hthreeplus + \he$ & $1.5 \times 10^{-9}$ & \citet{bohme1980determination} \\
k18 & $\hehplus + \rm{H} \rightarrow \htwoplus + \he$ & $9.1 \times 10^{-10}$ & \citet{karpasb1979ion} \\
k19 & $\heplus + \e \rightarrow \he$ & $4.6\times 10^{-12} (300\,\rm{K}/T)^{0.64}$ & \citet{storey1995recombination} \\ 
k20 & $\hehplus + \e \rightarrow \he + \rm{H}$ & $1.0\times 10^{-8} (300\,\rm{K}/T)^{0.6}$ & \citet{yousif1989recombination} \\
k21$^b$ & $\hplus + \htwo + \rm{M} \rightarrow \hthreeplus + \rm{M}$ & $3.2\times 10^{-29}$ & \citet{miller1968reactions} \\%
k22 & $\htwo$ + h$\nu \rightarrow \rm{H} + \rm{H}$ &  & \citet{schunk2009ionospheres} \\%
\enddata
\footnotesize{$^a$ Rate coefficients are given in units of cm$^3$ s $^{-1}$ for two-body reactions, and cm$^6$ s $^{-1}$ for three-body reactions. Photoionization and dissociation rates are as calculated in Section \ref{subsec:spectrum}.}\\
\footnotesize{$^b$ In the regions of interest in this work, M is $\htwo$, $\h$.}\\
\end{deluxetable}

\subsection{Spectrum} \label{subsec:spectrum}
Here we describe the EUV spectrum of the star. We use the binned spectrum of \citet{richards1994}, which provides the solar flux in 37 wavelength bins corresponding to a high level of stellar activity.

Consider a ray originating at the upper boundary of our atmosphere, which we define to be optical depth zero. We can integrate along the distance $l$ of the ray, moving inward toward the planet, to calculate a wavelength-dependent optical depth for solar radiation penetrating the atmosphere, $\tau_{\nu}(l)$, for each spectral bin:
\begin{equation}
    \tau_{\nu}(l) = \sum_{s} \int_{0}^{l} n_{s}(l')\sigma_{\nu,s} dl'.
    \label{eq:tau}
\end{equation}
In terms of the radial coordinate $r$, $l = r_{\rm{out}} - r$, where $r_{\rm{out}}$ is the radius of the simulation upper boundary. Here $\sigma_{\nu,s}$ is the total photoabsorption cross section of species $s$ per wavelength $\nu$, and $n_{s}(l)$ is the number density of species $s$ at a distance $r = r_{\rm{out}} - l$.

\begin{figure}[ht]\includegraphics[width=0.5\textwidth]{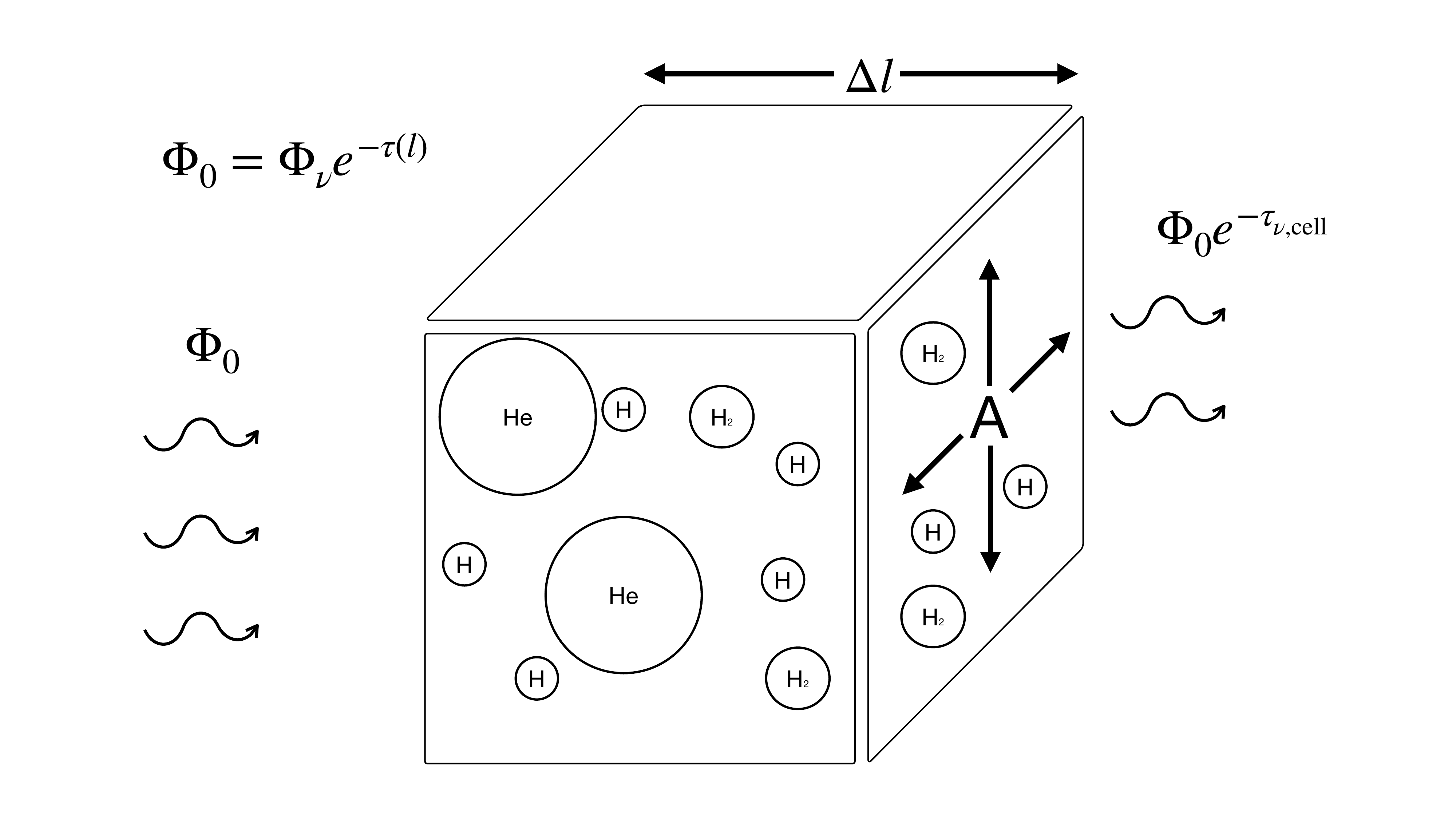}
\caption{Grid cell of length $\Delta l$ and cross-sectional area $A$ containing absorbing species $\h$, $\htwo$, and $\he$. Ionizing stellar flux of wavelength $\nu$ impacts the cell on the left. After being absorbed in the cell, the flux is reduced to $e^{-\tau_{\nu, \rm{cell}}}$ times the original flux, $\Phi_0$, where $\tau_{\nu, \rm{cell}}$ is the optical depth through the cell for the particular wavelength of interest.}\label{fig:photon_alloc}
\end{figure}

We then calculate the photoionization rate for species $s$ by photons of wavelength $\nu$ and energy $E_{\nu}$. The photons may be absorbed by species other than species $s$ if the cross sections for absorption $\sigma_{abs, \nu}$ for these other species are non-zero. Therefore, we must allocate the ionizing flux in each wavelength bin between the species. Suppose the absorbing species in our gas are $\htwo$, $\h$, and $\he$. Then, the fraction $x_{s}$ absorbed by species $s$ can be written as \citep{osterbrock2006astrophysics}:
\begin{equation}
    x_{s} = \frac{n_{s}\sigma_{abs, \nu, s}}{n_{\htwo}\sigma_{abs, \nu, \htwo} + n_{\h}\sigma_{abs, \nu, \h} + n_{\he}\sigma_{abs, \nu, \he}},
    \label{eq:abs_fraction}
\end{equation}
where species $s$ is either $\h$, $\htwo$, or $\he$. Here, $n_{s}\sigma_{abs, \nu, s}$ is the number density of species $s$, and $\sigma_{abs, \nu, s}$ is the cross section for absorption of species $s$ for photons of wavelength $\nu$. We note that Equation \eqref{eq:abs_fraction} is valid in both the optically thick and optically thin case. To calculate the photoionization rate for species $s$ in each cell, consider a cell of width $\Delta l$ and cross-sectional area $A$ (Figure \ref{fig:photon_alloc}) located at a depth $l$ from the upper boundary of the grid.

Suppose that the incoming photon number flux $\Phi_0$ from the star is coming in from the left. This incoming flux $\Phi_0 = \Phi_{\nu} e^{-\tau_{\nu} (l)}$, where $\tau_{\nu} (l)$ is the optical depth at distance along the ray $l$ given by Equation \ref{eq:tau}. The total number of photons that gets absorbed per volume in the cell is $\Phi_0 (1-e^{-\tau_{\nu, \rm{cell}}})/\Delta l$, where $\tau_{\nu, \rm{cell}}  = ({n_{\htwo}\sigma_{abs, \nu, \htwo} + n_{\h}\sigma_{abs, \nu, \h} + n_{\he}\sigma_{abs, \nu, \he}}) \Delta l$ for small $\Delta l$ such that the number densities of all species are approximately constant within each cell. Then, of the photons that get absorbed, species $s$ absorbs a fraction $x_{s} = \tau_{\nu, s, cell}/\tau_{\nu, \rm{cell}}$ (Equation \ref{eq:abs_fraction}), where $\tau_{\nu, s, cell} = n_{s}\sigma_{abs, \nu, s} \Delta l$ is the contribution to the optical depth in the cell due to absorbing species $s$. 

For the chemical network presented in Table \ref{tab:reactions}, the reaction rates per volume involving photons with frequency $\nu$ are calculated by multiplying the rate coefficient $k_{j, \nu}$ by the concentration of the absorbing species $n_s$, where the index $j$ is the label of the corresponding chemical reaction. As such, the rate at which species $s$ is ionized by a photon with frequency $\nu$ per volume by in the cell of width $\Delta l$ is
\begin{equation}
    r_{\textrm{ion}, \nu, s} = x_{s, ion} \frac{ \Phi_0(1-e^{-\tau_{\nu, \textrm{cell}}})}{\Delta l}\textrm{cm}^{-3} \textrm{s}^{-1} = \frac{n_s \sigma_{\textrm{ion}, \nu, s} \Delta l}{\tau_{\nu, cell}} \frac{ \Phi_0(1-e^{-\tau_{\nu, \textrm{cell}}})}{\Delta l}\textrm{cm}^{-3} \textrm{s}^{-1},
\end{equation}
where $x_{s, ion} = n_{s}\sigma_{ion, \nu, s}/(n_{\htwo}\sigma_{abs, \nu, \htwo} + n_{\h}\sigma_{abs, \nu, \h} + n_{\he}\sigma_{abs, \nu, \he})$ is the fraction of the incoming flux that ionizes species $s$. We note that, for computational purposes, we implement the following expression for the optically thin limit ($\tau_{\nu, \rm{cell}} < 10^{-9}$): as $\tau_{\nu, \rm{cell}} \ll 1$, $1-e^{-\tau_{\nu, \rm{cell}}} \approx \tau_{\nu, \rm{cell}}$, reducing the rate expression to:
\begin{equation}\label{eq:optically_thin}
    r_{\textrm{ion}, \nu, s} = \frac{n_s \sigma_{\textrm{ion}, \nu, s} \Delta l}{\tau_{\nu, \textrm{cell}}} \Phi_0 \frac{\tau_{\nu, \textrm{cell}}}{\Delta l}  = n_{s}\sigma_{\textrm{ion}, \nu, s} \Phi_0 \, \textrm{cm}^{-3} = n_{s} \sigma_{\textrm{ion}, \nu, s} \frac{F_{\nu} e^{-\tau_{\nu}(l) }}{E_{\nu}} \, \textrm{cm}^{-3}\textrm{s}^{-1}. 
\end{equation}
Then, the rate coefficient for ionization reaction $j$ involving any absorbing species $s$ is:
\begin{equation}\label{eq:ks}
    k_{j, \nu} = \frac{r_{\textrm{ion}, \nu, s}}{n_s} =  \sigma_{\textrm{ion}, \nu, s} \frac{F_{\nu} e^{-\tau_{\nu}(l) }}{E_{\nu}} \frac{(1-e^{-\tau_{\nu, \rm{cell}}})}{\tau_{\nu, \rm{cell}}} \rm{s}^{-1}.
\end{equation} 
Here, $F_{\nu}$ is the energy flux from the star per bin of wavelength $\nu$, $E_{\nu}$ is the average photon energy in the bin (we assume all photons in a bin have the same energy), and $\sigma_{\textrm{ion}, \nu, s}$ is the photoionization cross section for the specific photoionization reaction, as for example $\htwo$ photoionization can proceed via Reaction 3 or 4.

By analogy, we can calculate the heating rate due to the photoionization of H, $\htwo$, and He, as well as the dissociative photoionization and dissociation of $\htwo$. To calculate the heating rate per mass by photons of energy $h \nu$, we add all of the heating term contributions from the photoreactions $j$:
\begin{equation}\label{eq:heating_rate}
    \Gamma_{\nu} = \frac{1}{\rho}\sum_j n_s k_{j, \nu} \epsilon_j E_{\nu} = \frac{1}{\rho}\sum_j n_s \sigma_j \epsilon_j F_{\nu} e^{-\tau_{\nu}(l)} \frac{(1-\tau_{\nu, \rm{cell}})}{\tau_{\nu, \rm{cell}}} \mathrm{\, erg \, g^{-1} \, s^{-1}},
\end{equation}
where the heating efficiency $\epsilon_j$ is given by:
\begin{equation}
    \epsilon_j = \frac{(h \nu - E_{lim})}{h\nu},
\end{equation}
where $E_{lim}$ is the threshold energy for the ionization of H (13.6 eV), He (24.6 eV), $\htwo$ (15.4 eV), the disssociative ionization of $\htwo$ (18.08 eV, \citet{ford1975photoionization}), and the dissociation of $\htwo$ (4.74 eV, \citet{herzberg1961dissociation}). We can obtain the total heating rate per mass, $\Gamma_m$ in Equation \eqref{energy_spherical}, by integrating the heating rate over the entire EUV energy range (12-248 eV) of our input stellar spectrum, i.e. $\Gamma_m = \sum_{\nu} \Gamma_{\nu}$.

\subsection{Secondary ionizations due to high-energy electrons}\label{sec:secondarye}
Photons with higher energies are deposited deeper in our model of the upper atmosphere, where the ionization fraction is low. Electrons released from ionization by these high-energy photons will have energies exceeding the thermal energy of the gas. Their interactions with the gas leads to the further ionization and excitation of the components. It is important to include the results of detailed models of photoelectron energy deposition and loss, as they can affect the efficiency of atmospheric escape (e.g.   \citet{shematovich2014heating}, \citet{chadney2016euv}, \citet{munoz2023efficient}, \citet{gillet2023self}, \citet{garcia2024heating}).

We adapt a simplified treatment by using the results of the model of \citep{dalgarno1999electron}, which provides estimates that we incorporate in our code. We neglect the motion of electrons along magnetic field lines, although it is important in smaller-mass planets, e.g. Venus and Mars \citep{liu2024parameterization}.

The initial energy of the photoelectron produced produced by a photon of energy $E_{\nu}$ interacting with the gas is $E = E_{\nu} - I$. Through successive interactions with the gas particles, the high-energy electron loses energy. The mean number of ionizations produced by a photoelectron of initial energy $E$ is given by $N_{\mathrm{ion}} = E/W$. Simulations indicate that the mean photoelectron energy per ionization event, $W$, is roughly constant for photoelectron energies much larger than the energy required for ionization, $I$. For example, ionization of $\h$ requires $I = 13.6$ eV, and for photoelectrons with energy $E \gtrsim 100$ eV, $W \approx 35-40$ eV \citep{dalgarno1999electron}. Then, the number of $\h$ photoionizations is $N_{\mathrm{ion}} = (E_{\nu} - I)/W_{H^+}$, and similarly for $\htwoplus$. The ionization of $\he$ and $\htwoplus$ is negligible through this process in a mixture with $\h$ or $\htwo$ (\citet{dalgarno1999electron}, Table 4) for the electron energies present in our simulation.

To implement this, we include an additional ionization reaction in our chemical network to account for the secondary ionization rate. The rate constant $k1_{\mathrm{sec}}$ becomes $k1_{\mathrm{sec}} =  \sum_{\nu} (N_{\mathrm{ion}, \nu}) k_{\hplus, \nu} + \sum_{\nu} (N_{\mathrm{ion}, \nu}) k_{\htwoplus, \nu} + \sum_{\nu}  (N_{\mathrm{ion}, \nu}) k_{\heplus, \nu}$, where the secondary ionization terms in the sum apply for energy bins that generate photoelectrons with sufficient energy to ionize $\h$. The rate coefficients $k_{s, \nu}$ for each ionizing species $s$ are computed using Equation \ref{eq:ks}. The photoionization rate of $\htwo$, corresponding to $k3_{\mathrm{sec}}$, is computed in a similar manner.


The heating efficiency $\eta$ of the electron gives the fraction of the initial photoelectron energy that is deposited as heat, and it depends on depends on ionization fraction. Per photoelectron generated from the initial ionization, the heating rate is $\eta (E_{ \nu} - I)$. The initial photoionization rate gives the rate of photoelectron production. Then, the heating rate is the photoelectron production rate times the heating rate per photoelectron.





We note that the approximations presented in \citet{dalgarno1999electron} are only valid for a gas with ionization fraction $<0.1$. Photons with energies $\gtrsim 30+13.6$ eV are capable of generating photoelectrons with enough energy to further ionize the gas. The $\tau_{\nu} = 1$ surfaces for photons with these energies lie in the mostly neutral layer of the gas, with ionization fractions of $<0.1$, justifying the use of this approximation. Thus, for photons with high enough energies we modify our heating rate by including the factor $\eta$, as given by \citet{dalgarno1999electron}.


We compute the mean energy for production of $\hplus$ and $\htwoplus$ using
\begin{equation}
    W = W_0 (1 + C x^{\alpha}),
\end{equation}
where we use the parameters for $\hplus$ from Table 4 for the $\h$ and $\he$ mixture, and the parameters for $\htwoplus$ from to $\htwo$ and $\he$ mixture rows of the table. Thus, we have the yield $N_{\mathrm{\hplus}} = E/W_{\hplus}$ and $N_{\mathrm{\htwoplus}} = E/W_{\htwoplus}$. 
\citet{dalgarno1999electron} provide a heating efficiency $\eta$, the fraction of the electron energy that is deposited as heat due to collisions of the electron with the gas components, as well as the rotational excitations of $\htwo$ that are thermalized. To compute the actual electron heating efficiency in our gas, we need to account for the heating due to the thermalization of the vibrational excitations of $\htwo$, as well as the dissociation heat input due to the triplet and singlet electronic excitations of $\htwo$.

Leftover energy from the process of $\htwo$ dissociation also contributes a significant heating source that needs to be added back to compute the total heating efficiency. $\htwo$ can be excited into triplet states that dissociate almost 100\% of the time with a mean heating rate of 5.4 eV per dissociation (which is just the energy of the dissociating triplet state minus the $\htwo$ bond energy). The exact breakdown of the triplet state yields (and the higher-order triplet states denoted ``pseudostates") is not stated, but using the mean energy is sufficient for our calculations. In a neutral mixuture of $\htwo$ and $\he$, we can use Table 5 of \citet{dalgarno1999electron} to compute the heating rate per dissociation:
\begin{equation}
    Q = 5.4 N_{diss} = 5.4 \frac{E}{W_0} \mathrm{eV}.
\end{equation}
This can be modified to take into account a gas with an ionization fraction $x$:
\begin{equation}
    Q = 5.4 \frac{E}{W_0(1+Cx^{\alpha})} \mathrm{eV},
\end{equation}
where we've used Equation (13) of \citet{dalgarno1999electron} to account for the change in mean energy per dissociation as a function of ionization fraction, using the values of $C$ and $\alpha$ provided in Table 5. In a mixture of $\htwo$, $\h$. and $\he$, the dissociation yield will be smaller with an increasing fraction of $\h$: this is due to the excitation of $\h$ competing with the excitation of $\htwo$ that leads to dissociation. As suggested in \citet{dalgarno1999electron} and explicitly stated in Equation (11) of \citet{glassgold2012cosmic}, the heating rate (per initial photoelectron of energy $E$) can be calculated via:
\begin{equation}
    Q =  5.4 \frac{x(\htwo)}{x(\h)+x(\htwo)} \frac{E}{W_0(1+Cx^{\alpha})} \mathrm{eV},
\end{equation}
where $x(\h) = n(\h)/(n (\h)+2n(\htwo))$ and $x(\htwo) = n(\htwo)/(n(\h)+2n(\htwo))$. Finally, the contribution to the efficiency is then:
\begin{equation}
   \eta_{diss} =  5.4 \frac{n(\htwo)}{n(\h)+n(\htwo)} \frac{1}{W_0(1+Cx^{\alpha})}.
\end{equation}

Electrons directly excite the $\nu=1$ and $\nu=2$ vibrational levels of the $\htwo$ ground state. The number density in our simulation far exceeds the critical density for the excitations to be radiated away -- thus, the energy is thermalized and added back to the heating efficiency. The energy to excite $\htwo$ to the $\nu=1$ vibrational level is 0.516 eV, and to the $\nu=2$ level is 1.032 eV.
We can again use the values in Table 5 to compute the energy required to excite these levels in a mixture of $\htwo$ and $\he$:
\begin{equation}
    Q = 0.516 \frac{E}{W_{0, \, \nu=1}(1 +  C_{\nu=1} x^{\alpha_{\nu=1}})} + 1.032 \frac{E}{W_{0, \, \nu=2}(1 +  C_{\nu=2} x^{\alpha_{\nu=2}})} \mathrm{\, eV}.
\end{equation}
This translates to a heating efficiency of:
\begin{equation}
    \eta_{vib} =0.516 \frac{1}{W_{0, \, \nu=1}(1 +  C_{\nu=1} x^{\alpha_{\nu=1}})} + 1.032 \frac{1}{W_{0, \, \nu=2}(1 +  C_{\nu=2} x^{\alpha_{\nu=2}})}.
\end{equation}
To compute the equivalent in a mixture of $\htwo$, $\h$. and $\he$, for $x\geq 10^{-4}$ we weight by the number densities:
\begin{equation}
    \eta_{vib}  = \eta_{vib} \frac{2 n(\htwo)}{n(\h) + 2 n(\htwo)},
\end{equation}
and for $10^{-4}>x\geq10^{-7}$:
\begin{equation}
    \eta_{vib}  = \eta_{vib} \frac{ n(\htwo)/n(\h)}{n(\htwo)/n(\h) + 0.5 (x/10^4)^{0.15}},
\end{equation}
and for $x<10^{-7}$
\begin{equation}
    \eta_{vib}  = \eta_{vib} \frac{ n(\htwo)/n(\h)}{n(\htwo)/n(\h) + 0.177},
\end{equation}
corresponding to Equation (11) of \citet{dalgarno1999electron}.

The electronic states $B^1\Sigma_u^+$ and $C^1\Pi_u$ are denoted the Lyman and Werner $\htwo$ electronic states. These states radiatively decay to the vibrational ground states of $\htwo$ with a probability of $\sim90 \%$, with the other $\sim 10 \%$ resulting in dissociation. We adapt an average 2 eV level for the vibrational states, and a 0.4 eV dissociation heat input. The heating as a result of this is then:
\begin{equation}
    Q = (0.9 \times 2. + 0.1 \times 0.4)  \frac{E}{W_{0,B}(1+C_Bx^{\alpha_B})} + \frac{E}{W_{0,C}(1+C_Cx^{\alpha_C})} \mathrm{eV},
\end{equation}
where the $B$ and $C$ represent the Lyman and Werner states of $\htwo$. Similar to the dissociation of $\htwo$, the yield in a mixture of $\htwo$, $\h$. and $\he$ is calculated by weighting by the relative abundances:
\begin{equation}
   \eta_{fl} =  (0.9 \times 2. + 0.1 \times 0.4) \frac{x(\htwo)}{x(\h)+x(\htwo)} \left(\frac{1}{W_{0,B}(1+C_Bx^{\alpha_B})} + \frac{1}{W_{0,C}(1+C_Cx^{\alpha_C})}\right).
\end{equation}

In summary, we use the efficiencies computed by \citet{dalgarno1999electron} to compute an initial heating rate due to collisions of the photoelectrons with the gas compononents, as well as the thermalization of the rotationally excited $\htwo$. We use the expressions provided by \citet{dalgarno1999electron} to compute the additional heating due to $\htwo$ dissociation following triplet excitation, the thermalization of the collisionally-excited $\nu=1$ and $\nu=2$ levels of $\htwo$, and the dissociation of $\htwo$ and thermalization of the vibrational ground states of $\htwo$ following singlet excitation. The approximations from \citet{dalgarno1999electron} have stated errors of, at most, 15\%. The remaining energy goes into secondary ionization, the energy required to dissociate $\htwo$, and Lyman-$\alpha$ radiation following collisional excitation. Any energy otherwise not explicitly accounted for is assumed to be radiatively lost.

\section{Results: Hot and Warm Jupiters}\label{sec:results}
We discuss the results for our fiducial hot Jupiter at 0.05 au in Section \ref{sec:hd209}. We show how our model produces three layers. Key reaction rates involving $\h$, $\htwo$, and $\he$ in the molecular layer are highlighted. In Section \ref{sec:energy_allocation}, we illustrate how the cooling terms are sourced from the heating terms at each altitude. Then, we evaluate how each cooling term contributes to the energy balance within each energy bin of the spectrum. In Section \ref{sec:epsilon} we compare our results to the energy-limited mass-loss approximation. 
Section \ref{sec:grid_of_models} contains the results of our grid of models, varying the semimajor axis of the planet from 0.015 to 0.3 au.
\subsection{HD209458b}\label{sec:hd209}
\begin{figure}[ht] 
  \includegraphics[width=\textwidth]{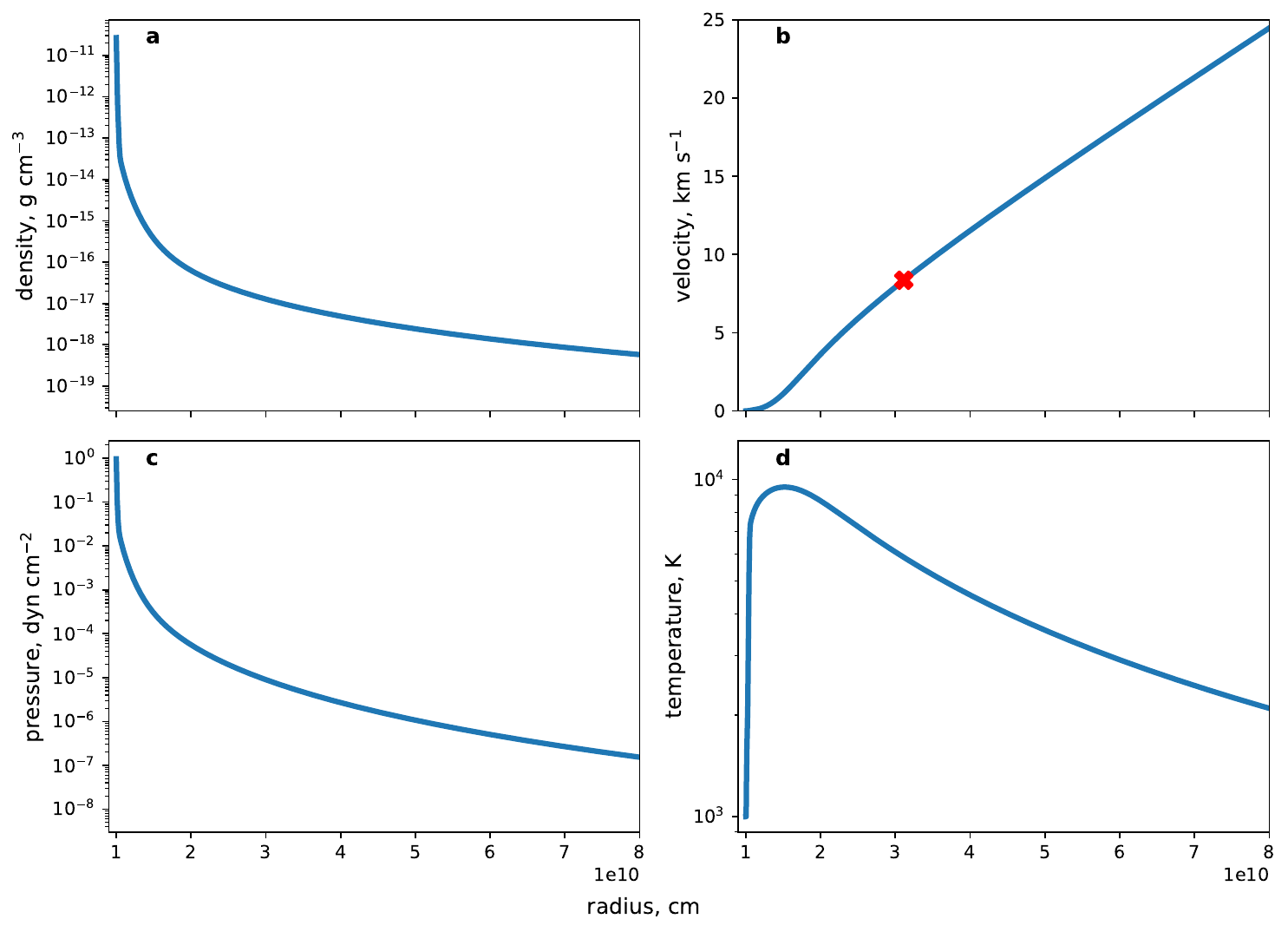}
  \caption{Profiles of wind density (a), velocity (b), pressure (c), and temperature (d) with radius for our fiducial hot Jupiter at 0.05 au (Table \ref{tab:params}).  The red ``X" marks the sonic point of the wind.
  }
  \label{fig:rhovpt_fiducial}
\end{figure}

We model the upper atmosphere of our fiducial planet with parameters similar to that of the hot Jupiter HD209458b (Table \ref{tab:params}). The steady-state profiles of our model are shown in Figure \ref{fig:rhovpt_fiducial}.
\begin{figure}[ht] 
  \includegraphics[width=\textwidth]{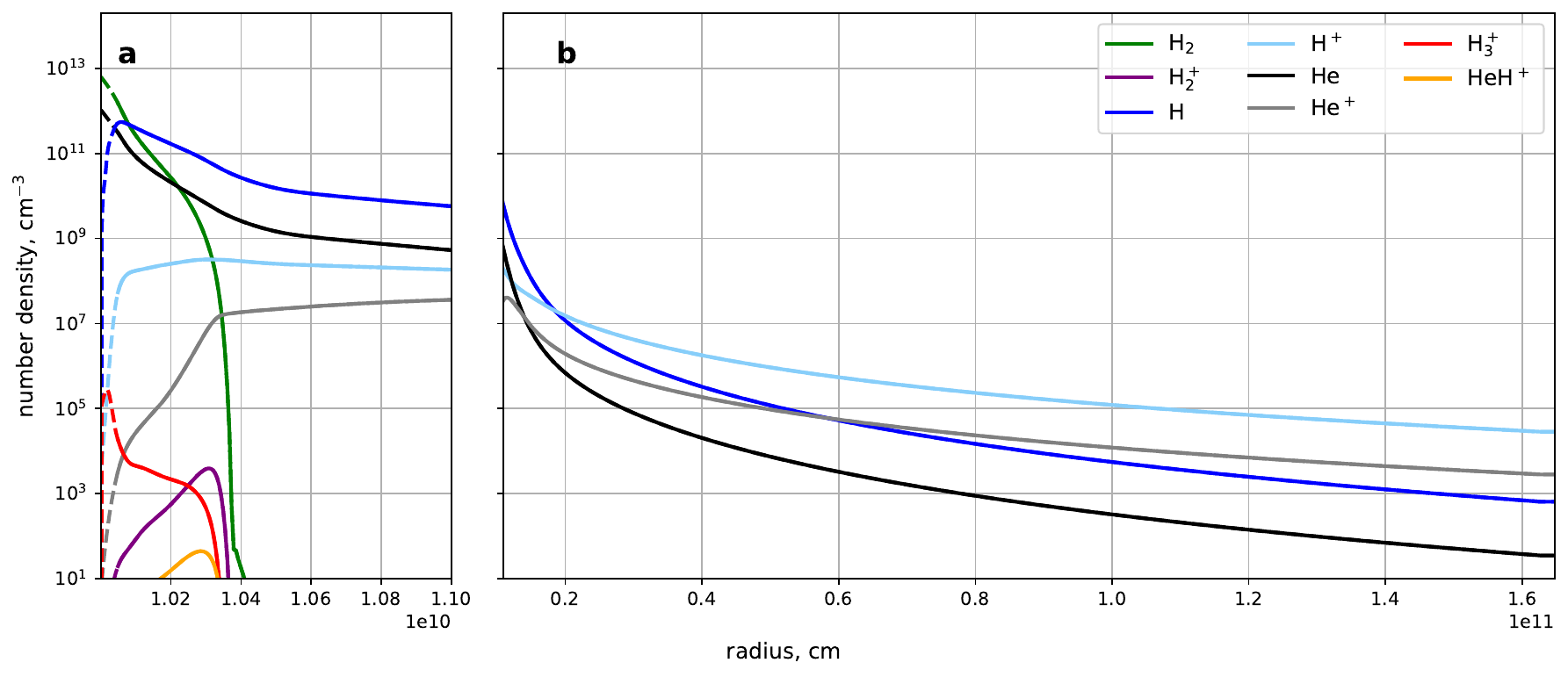}
  \caption{Atmospheric composition as a function of radius, zoomed in near the wind base (panel a) and at larger radii (panel b), for the same outflow shown in Figure \ref{fig:rhovpt_fiducial}. The dashed lines in panel (a) denote the region where bolometric heating and cooling dominate. The transition from: $\htwo$ (green)$ \rightarrow \h$ (dark blue) occurs at $1.01 R_{\mathrm{P}}$, $ \h $ (dark blue) $ \rightarrow \hplus$ (light blue) at $1.8 R_{\mathrm{P}}$, and $\he$ (black)  $\rightarrow \heplus$ (gray) at $1.4 R_{\mathrm{P}}$. The molecular layer drops off sharply at $1.04 R_{\mathrm{P}}$. The temperature rises from 2000K to 4000K in the layer where $\htwo$ thermal dissociation dominates. The kink in the $\htwo$ density profile does not affect the result, and is likely a result of applying the density floor in a region where the profile drops off steeply. The molecules  H$_3^+$ (red), H$_2^+$ (purple), and HeH$^+$ (orange) are present only where H$_2$ survives.}
  \label{fig:composition}
\end{figure}

\begin{figure}[ht] 
  \includegraphics[width=\textwidth]{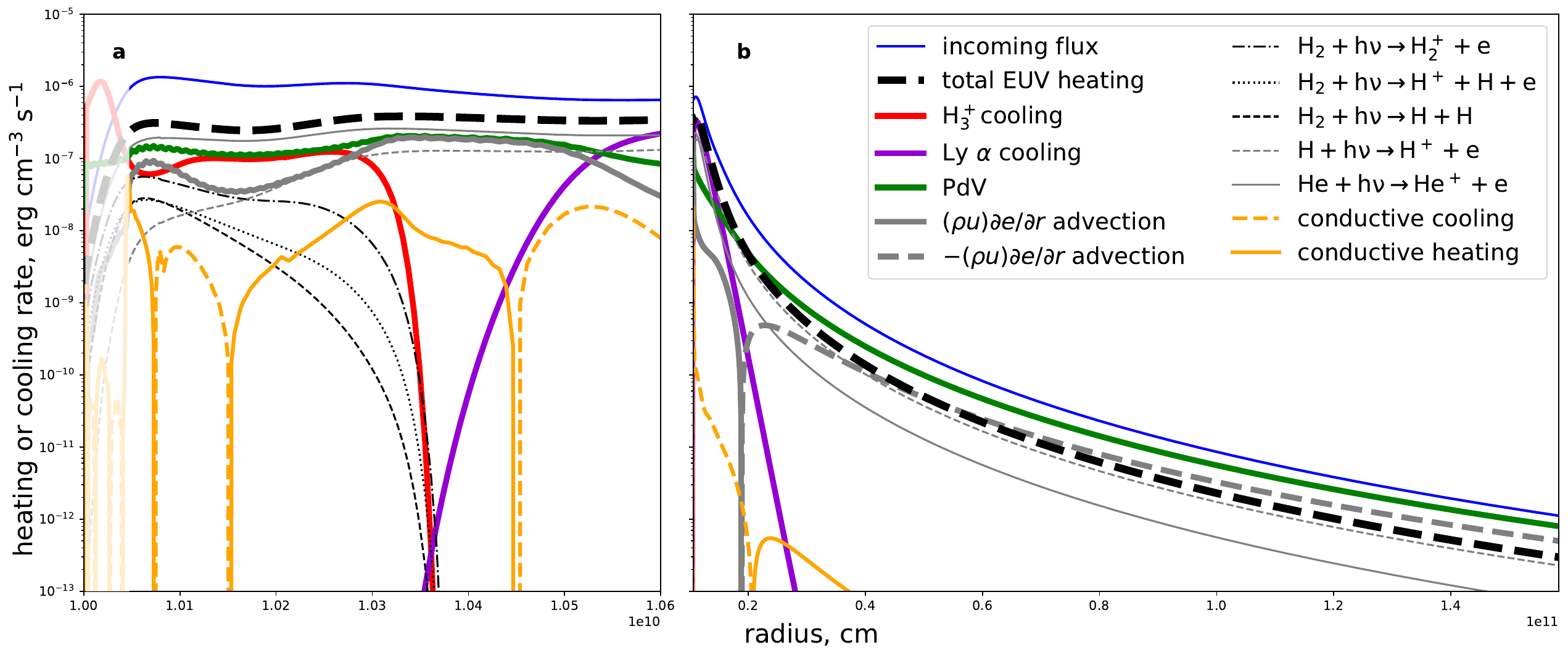}
  \caption{Heating and cooling rates.  Panel (a): The molecular hydrogen layer ($ r < 1.04 \times 10^{10}$ cm) is primarily heated by photoionization of He (thin gray solid), with contributions from photoionization of $\htwo$ (black dashed-dotted) deeper in the layer, and photoionization of $\h$ (thin gray dashed) higher up. Most of the total energy deposited in this region (thick black dashed) is radiated away by the $\hthreeplus$ (red). In the atomic hydrogen layer ($1.04 \times 10^{10} < r < 1.5 \times 10^{10}$ cm), Lyman-$\alpha$ (purple) is the primary net coolant, balancing comparable heating contributions from the photoionization of H and He. Bolometric heating and cooling (not shown) dominate for $r < 1.005 \times 10^{10}$ cm (de-emphasized faint region). Panel (b): 
  In the larger-scale outflow ($r > 1.5 \times 10^{10}$ cm), EUV heating is predominately due to the ionization of atomic hydrogen and is primarily balanced by $PdV$ cooling (green). We note that the sum of the EUV heating and $PdV$ work terms is equal to the advection curve; in other words, advection is transporting thermal energy from below up into the wind. In all regions, photodissociation (thin dashed black), ionizing photodissociation (dotted black), and conduction (yellow) are energetically sub-dominant.  Note that the thick black dashed line indicating total EUV heating is equal to the sum of the photoionization reaction heating terms.}
  \label{fig:energy_balance_reactions}
\end{figure}

\begin{figure}[ht] 
  \includegraphics[width=\textwidth]{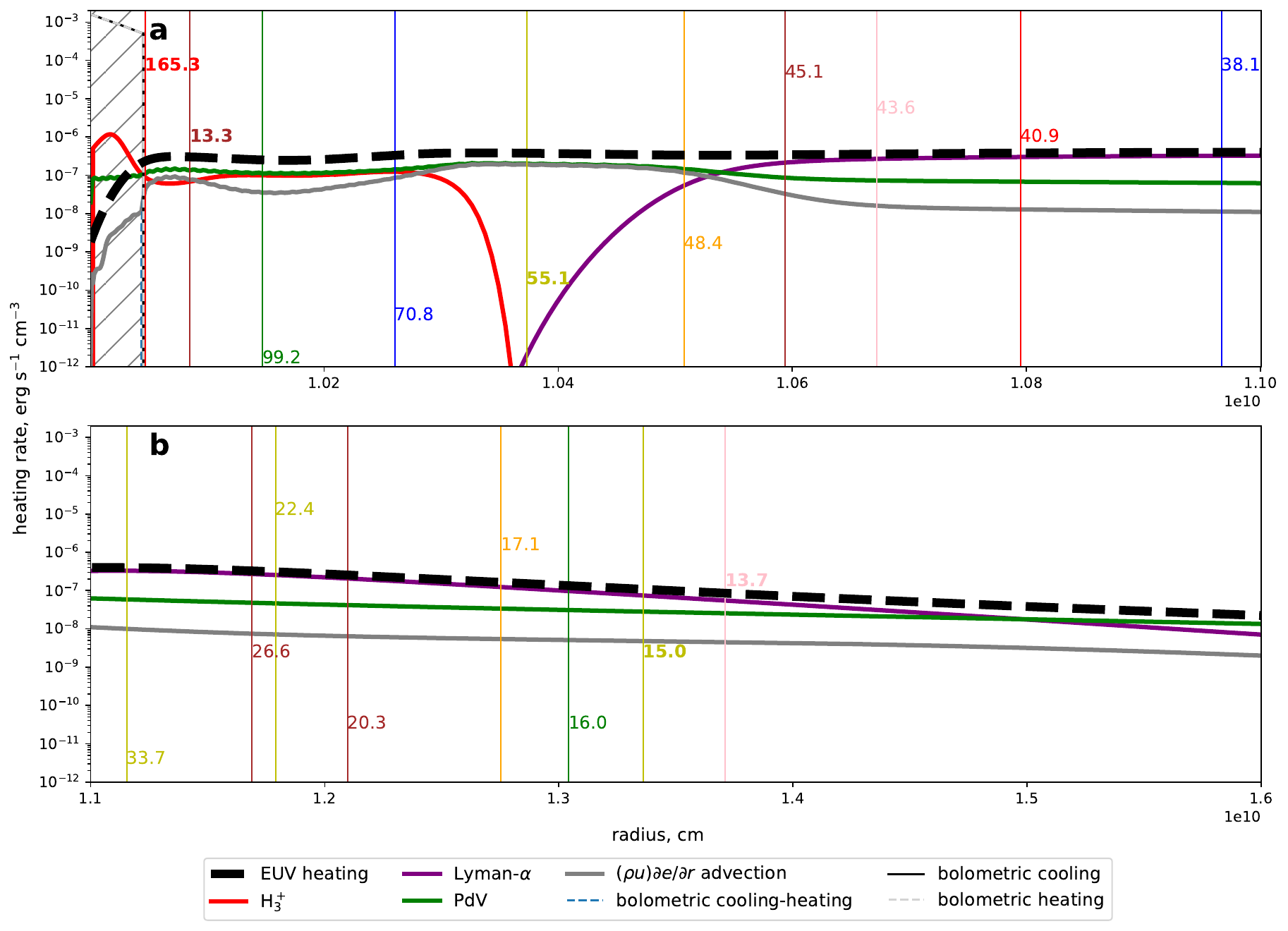}
  \caption{Zoomed-in view of Figure \ref{fig:energy_balance_reactions}, with $\tau_{\nu}=1$ surfaces for photons labeled in eV. The hatched region ($r < 1.005 \times 10^{10}$), de-emphasized in Figure \ref{fig:energy_balance_reactions}, contains the bolometric heating and cooling terms. As illustrated in Panel (a), $\tau_{\nu}=1$ surfaces for the highest-energy photons (248-55 eV) lie within the $\htwo$ molecular layer at the base. Photons in the 55-15 eV energy range are deposited at altitudes directly above the $\htwo$ layer, where gas is primarily $\h$ and Lyman-$\alpha$ cooling dominates. Panel (b) shows how the Lyman-$\alpha$-cooled region transitions into the outflow layer. Photons in the 15-13.6 eV energy range ionize neutral hydrogen and are more efficient at powering the outflow. The simulation region is largely optically thin to photons with energies $< 13.6$ eV; $\htwo$ dissociation is a source of opacity in the far-ultraviolet (e.g. see the 13.3 eV $\tau_{\nu}=1$ surface marked at the base).}
  \label{fig:energy_taus}
\end{figure}

The wind reaches a velocity of $\sim$ 8.4 km s$^{-1}$ at the sonic point distance of 3.1 $R_{\mathrm{P}}$. The maximum temperature reached is $\sim 9500$ K, as Lyman-$\alpha$ cooling prevents further heating. The mass-loss rate per steradian, $\dot{M} = \rho u r^2 = 9 \times 10^9$ g s$^{-1}$ sr$^{-1}$, is evaluated along the ray connecting the planet to the star on the dayside. Integrating over $4\pi$ steradians may lead to a mass-loss overestimate of the true mass-loss rate, as the mass loss is expected to peak along the direction of the substellar ray, where UV irradiation and tidal gravity are the strongest. \citet{rmc2009} estimated that the directional dependence of the UV irradiance reduced the mass-loss rate by a factor of $\sim 0.3$, compared to mass-loss rate obtained by integrating over $4 \pi$, with an additional factor of at most $\sim 0.8$ to account for the directional variation in tidal gravity. Thus, the total mass loss may be reduced to $\sim 0.24$ as compared to the maximal possible rate in the case of HD209458b, with the tidal variability factor declining in importance for modeling planets further from the star. 

Figure \ref{fig:composition} shows the composition of the atmosphere, zoomed in on the base in panel (a), and depicting the outflow in panel (b). $\htwo$ and $\he$ are the major constituents of the atmosphere at the base of our simulation, which we denote the molecular hydrogen layer. This layer also contains the important coolant $\hthreeplus$. The composition transitions to $\h$ and $\he$ in the intermediate layer between the base and the outflow layer. Finally, the outflow is dominated by $\hplus$.

\subsubsection{Heating and cooling}
The heating and cooling rates are presented in Figure \ref{fig:energy_balance_reactions}. The net incoming flux deposited in the atmosphere is plotted in blue. The total EUV heating (thick black dashed line) is the heating of the gas as a result of photoionization (and photodissociation, though this process is always sub-dominant). The gas is then cooled by adiabatic expansion ($PdV$, green solid line) and radiative cooling. Two significant sources of radiative cooling are included in our model: Lyman-$\alpha$ cooling (purple solid line) and $\hthreeplus$ cooling (red solid line). Panel (a) is zoomed in to the base, the composition of which is depicted in panel (a) of Figure \ref{fig:composition}. $\hthreeplus$ cooling is a significant cooling source in the molecular hydrogen layer. The energy deposited here is largely lost to ionization and radiative cooling, and thus does not substantially drive the outflow. As the $\hthreeplus$ is confined to this molecular hydrogen layer, just above the layer it is no longer available to cool the gas, and the temperature rises steeply (panel (d) in Figure \ref{fig:rhovpt_fiducial}). In order to escape the gravitational well of the hot Jupiter, the gas must reach a temperature that allows the thermal pressure to accelerate it to a significant fraction of the escape velocity. The gas heats up enough for substantial Lyman-$\alpha$ cooling (purple solid line) to set in: the electrons\footnote{We note that Lyman-$\alpha$ cooling occurs via collisional excitation, and electron impacts are most efficient due to their high velocities relative to the other gas species in energy equipartition.} have high enough energies to excite the neutral $\h$, and cooling is significant. Panel (b) depicts the larger-scale outflow. Here, the Lyman-$\alpha$ cooling term drops off as the $PdV$ term starts to dominate and cool the outflow, which is accelerated to velocities of km s$^{-1}$ to tens of km s$^{-1}$ (panel (b) in Figure \ref{fig:rhovpt_fiducial}). The deposition of energy throughout the wind by EUV heating (thick black dashed line) and advection  of internal energy from below (gray thick dashed line) are the sources of energy for the wind---they balance the $PdV$ term arising from the adiabatic expansion of the gas (green solid line). The advection term is the left-hand side of the energy equation in steady-state form (Equation \ref{eq:steadystate_e}). When it is negative (i.e., $\partial e/\partial r<0$, since $\rho u>0$ in the outflow), this term acts as an additional source of energy, advecting energy from lower in the atmosphere (labeled ``$-(\rho u)\partial e/\partial r$ advection" in Figure \ref{fig:energy_balance_reactions}). In contrast, at the base as the temperature rises with distance, $\partial e/\partial r>0$, the advective term actually acts as an energy sink, transporting energy up (labeled ``$(\rho u)\partial e/\partial r$ advection" in Figure \ref{fig:energy_balance_reactions}). Consistent with previous work \citep{munoz2007physical,rmc2009}, we find that conduction does not dominate for our fiducial planet.  In summary, the cooling (after accounting for ionization): has a substantial $\hthreeplus$ component at the base, is Lyman-$\alpha$-dominated higher up, and is $PdV$-dominated in the outflow. 

In Section \ref{subsec:spectrum}, we defined optical depth $\tau_{\nu}$ for each bin of the spectrum and the heating rate due to ionization. The presence of multiple ionizing species and, for the case of molecular hydrogen, ionization outcomes, means that the incoming photon flux must be allocated according to each process' respective cross section. The flux declines as $e^{-\tau_{\nu}}$ as we traverse the atmosphere from the top to the base. $\tau_{\nu}=1$ is where the atmosphere transitions from optically thin to optically thick to photons of a given wavelength. Zooming in to the base of the atmosphere, Figure \ref{fig:energy_taus} displays the heating and cooling rates, annotated by the locations of the $\tau_{\nu}=1$ surfaces for selected energy bins.

\subsubsection{Photochemistry}
We now discuss the key processes of our chemical network. We will step through the composition in Figure \ref{fig:composition} from the base, above which EUV-driven ionization and heating drive chemical reactions. 

As an aside, before we focus on the photochemistry, we derive the advective term $n_{tot} u \, \partial f_s/\partial r$, that transports a species $s$ with volumetric number density $n_s$ and number fraction $f_s = n_s/n_{tot}$, where $n_{tot}=\Sigma_s n_s$ is the total number density of nucleii. In steady state, the advective term for the total number density $n_{tot}$, is
\begin{equation}
    \frac{1}{r^2}\frac{\partial}{\partial r}\left( r^2 n_{tot} u\right) = u\frac{\partial n_{tot}}{\partial r} +  \frac{n_{tot}}{r^2}\frac{\partial}{\partial r}\left( ur^2\right) =0.
\end{equation}

\noindent Therefore, we can write the advection term for an individual species $s$:
\begin{eqnarray}\label{eq:advection_term}
    \nonumber \frac{1}{r^2}\frac{\partial}{\partial r}\left( r^2 n_s u\right) &=& u\frac{\partial n_s}{\partial r} + n_s \frac{1}{r^2}\frac{\partial}{\partial r}\left( ur^2\right) = u\frac{\partial n_s}{\partial r} - u\frac{n_s}{n_{tot}}\frac{\partial n_{tot}}{\partial r} \\  &=& n_{tot}u\left( n_{tot}\frac{\partial n_s}{\partial r}-n_s \frac{\partial n_{tot}}{\partial r} \right)/n_{tot}^2 = n_{tot}u \frac{\partial (n_s/n_{tot})}{\partial r} = n_{tot}u \frac{\partial f_s}{\partial r}.
\end{eqnarray}

\begin{figure}[ht] 
  \includegraphics[width=\textwidth]{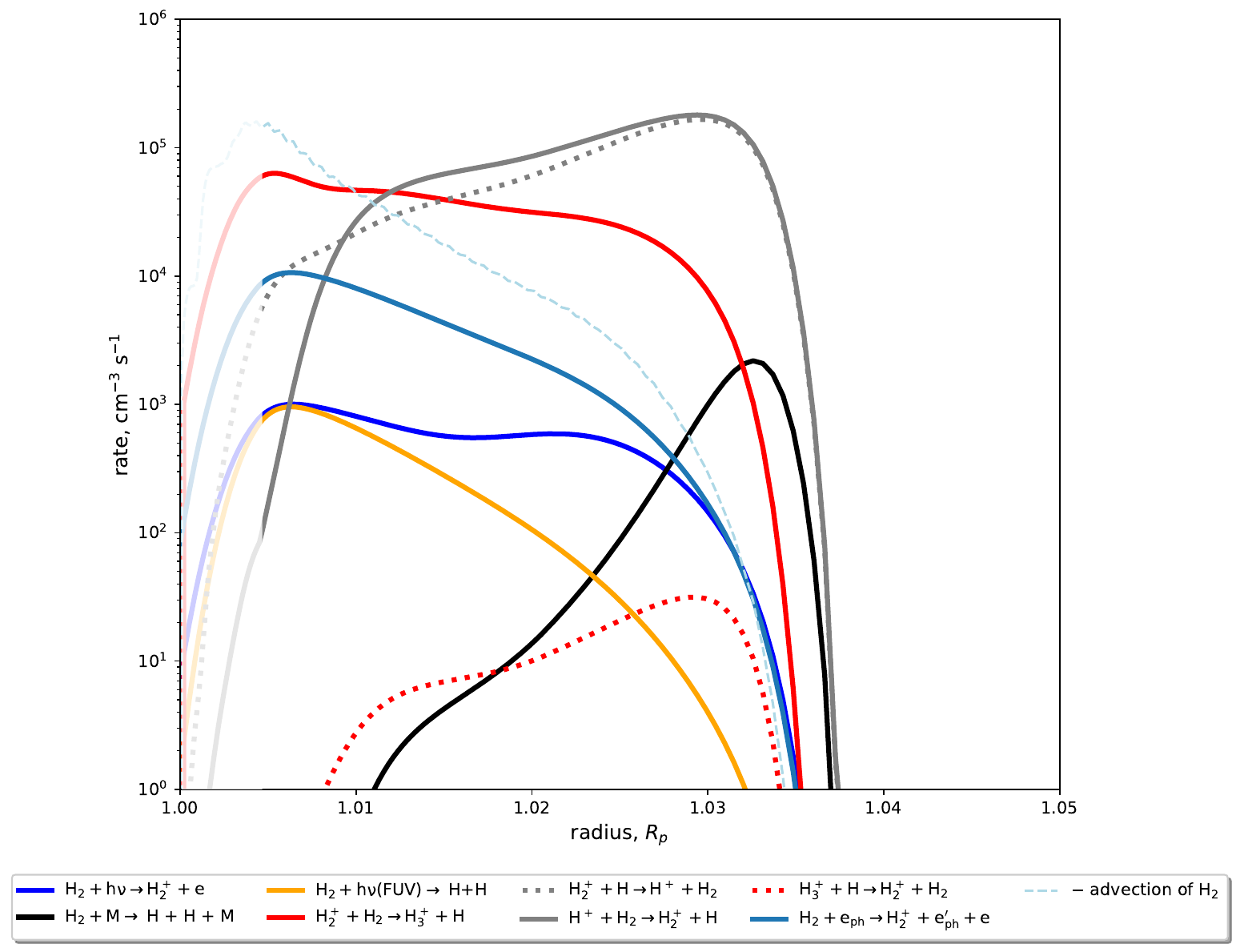}
  \caption{Selected reaction rates involving $\htwo$. The advective term is plotted as derived in Equation \eqref{eq:advection_term}. The gray solid and dotted lines represent the process of $\htwo$ and $\h$ exchanging an electron: this reaction pair does not lead to the destruction of $\htwo$. Instead, $\htwo$ is thermally dissociated (black line) at $\sim 1.04$ planetary radii. The key reaction of $\htwoplus$ with $\htwo$ to form $\hthreeplus$, the most important coolant in this layer, is plotted in red. The photoionization of $\htwo$ is plotted in bright blue and the ionization of $\htwo$ by photoelectrons is plotted in dark blue. The formation of $\htwoplus$ is a crucial step in forming $\hthreeplus$. The far-ultraviolet dissociation of $\htwo$ (orange line) increases with depth in the layer. Advection of $\htwo$ is plotted as a light blue dashed line; its sign follows the convention established in Figure \ref{fig:energy_balance_reactions}: as the sign is negative, the advective term represents an addition of $\htwo$ from the boundary.}
  \label{fig:h2rates_selected_fiducial}
\end{figure}

\begin{figure}[ht] 
  \includegraphics[width=\textwidth]{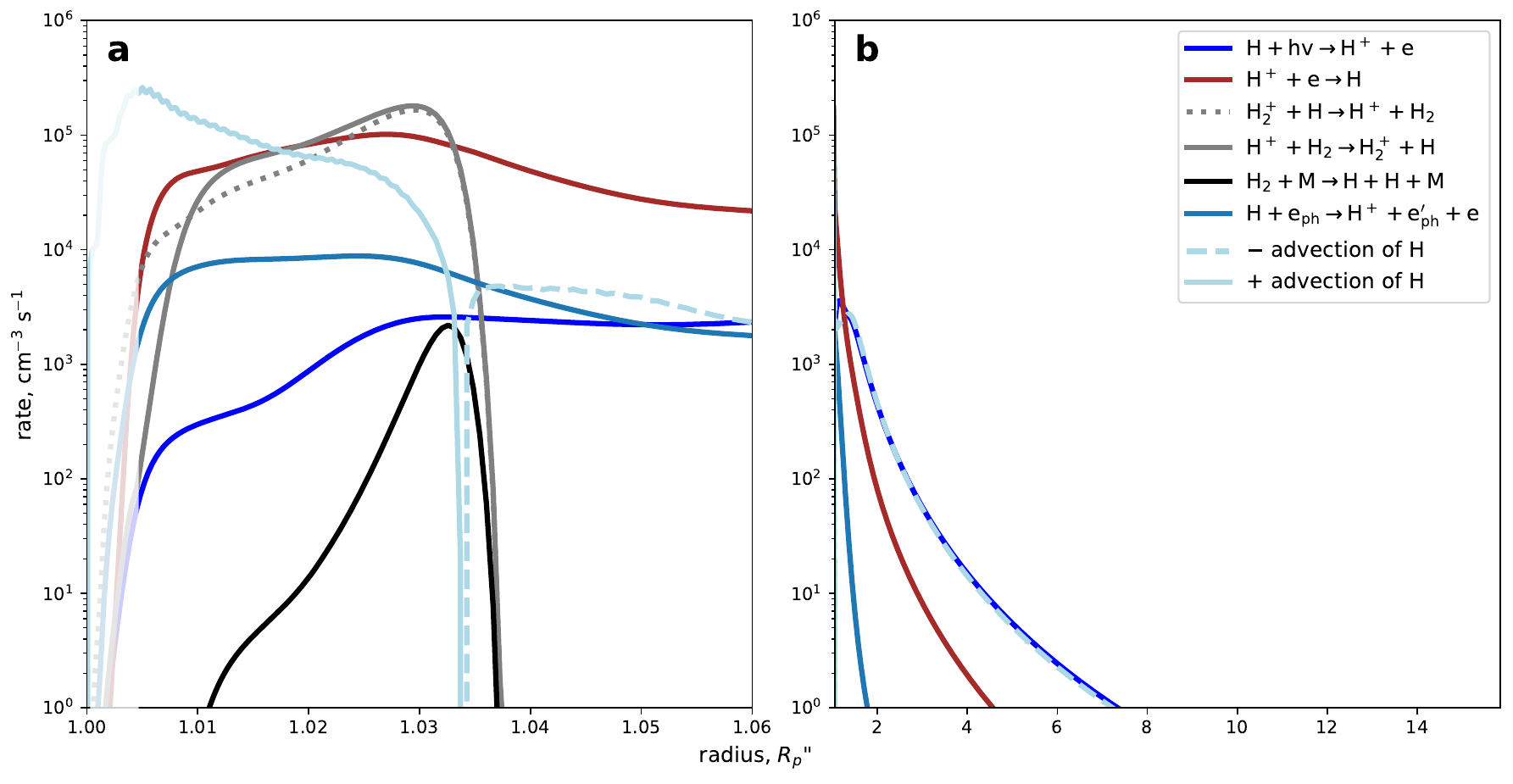}
  \caption{Reaction rates involving $\h$. Panel a: zoomed in to molcular hydrogen layer. Panel b: 1.06 to $\sim 8.8 R_{\mathrm{P}}$. As in Figure \ref{fig:h2rates_selected_fiducial}, the gray solid and dotted lines represent the electron exchange between $\htwo$ and $\h$. The black solid line marks the production of $\h$ as a result of the thermal dissociation of $\htwo$ at the boundary of the molecular layer. The blue solid line is the photoionization of $\h$. In the outflow (Panel b), the ionization of $\h$ is exactly balanced by the advection of $\h$ (light blue dashed line), with a small contribution from recombination (brown solid line). The sign convention for advection is ``$-$" for the advection in of $\h$, and ``+" for the advection of $\h$ away, as in Figures \ref{fig:energy_balance_reactions} and \ref{fig:h2rates_selected_fiducial}.}
  \label{fig:hrates_selected_fiducial}
\end{figure}

\noindent We use this expression to plot the advective terms in Figures \ref{fig:h2rates_selected_fiducial} and \ref{fig:hrates_selected_fiducial}.

We first focus on the conditions required to form $\hthreeplus$, a key coolant present in the base layer of our atmosphere. $\hthreeplus$ forms predominately via $\htwo + \htwoplus \rightarrow \hthreeplus + \h$. As molecular hydrogen is the precursor to $\hthreeplus$, we plot the reactions involving $\htwo$ in Figure \ref{fig:h2rates_selected_fiducial}. The $\hthreeplus$ production rate via its dominant pathway is plotted in red. Here, $\htwoplus$ forms when $\htwo$ is ionized ($\htwo$ + h$\nu \rightarrow \htwoplus + \e$) by photons with energies exceeding the photoionization threshold h$\nu = 15.4$ eV (blue line). The energy deposited by photons in this layer is efficiently radiated away by infrared cooling from the $\hthreeplus$: the ionization of molecular hydrogen ultimately does not drive hydrodynamic escape of the atmosphere.

At the upper edge of the molecular layer, ionizing radiation heats up the gas enough for the molecular hydrogen to dissociate due to collisions: $\htwo + \rm{M} \rightarrow \rm{H} + \rm{H} + \rm{M}$, where M is a species that collisionally dissociates $\htwo$ (black line). In this region, sufficient $\h$ has been produced and the electron fraction is small enough that collisions with $\h$ dominate. We note that although molecular hydrogen can undergo a photoionization/dissociation process ($\htwo$ + h$\nu \rightarrow \rm{H} + \hplus + \e$), thermal dissociation dominates. We also include the dissociation of molecular hydrogen by photons in the far-ultraviolet (FUV) range of the spectrum ($\htwo$ + h$\nu \rightarrow \rm{H} + \rm{H}$) \citep{sternberg1989ultraviolet, jo2017far}, with energies $<13.6$ eV. At energies above 13.6 eV, ionization of $\h$ dominates over this process. The FUV photons resonantly excite $\htwo$, which then decays with a $\sim 10\%$ probability of dissociating into two hydrogen atoms (and the remaining $\sim 90\%$ de-excites via fluorescent emission). The FUV photons probe deeper into the molecular hydrogen layer. This process is the dominant dissociation pathway of $\htwo$ (and therefore the main source of atomic hydrogen - see Figure 7, orange line) further in (as opposed to thermal dissociation (black line on Figure 7), which happens at higher temperatures). The heating from this process is predominately radiated away by $\hthreeplus$ (Figure \ref{fig:energy_taus}, $<13.6$ eV), even though these are the lowest-energy photons in our spectrum. $\hthreeplus$ cooling shuts off at the $\htwo \rightarrow \h$ transition.

In Figure \ref{fig:hrates_selected_fiducial} we show the main reactions involving $\h$. As shown in Figure \ref{fig:h2rates_selected_fiducial}, the thermal dissociation of molecular hydrogen increases as the temperature of the gas starts rising to $\sim 10^4$ K. As molecular hydrogen gives way to atomic hydrogen, some of the atomic hydrogen is photo-ionized (producing electrons capable of exciting the atomic hydrogen) and the gas is heated. Now, the predominant mode of cooling is Lyman-$\alpha$ cooling from neutral atomic hydrogen, which thermostats the temperature to $10^4$ K. Thus, atomic $\h$ can be produced here at a high rate, which further allows for the heating of the gas and the destruction of molecular hydrogen. We note that most of the energy leftover from ionization here in this neutral atomic hydrogen layer above the molecular layer (Figures \ref{fig:composition} and \ref{fig:energy_balance_reactions}) is radiated away by Lyman-$\alpha$ cooling. Some of the heating here is converted to $PdV$ work, partially powering a hydrodynamic outflow.

Further up in the wind (starting at about 1.5$R_{\mathrm{P}}$), the gas becomes mostly ionized (the $\hplus$-dominated region in Figure \ref{fig:composition}).  We refer to this region as the outflow layer. The ionizations here are not balanced by recombinations ($\hplus + \e \rightarrow \h$), but by advection. Figure \ref{fig:hrates_selected_fiducial}, panel b shows the reactions involving $\h$ in the outflow: advection of $\h$ and a minor contribution from recombination exactly balance ionization. As is shown in Figure \ref{fig:energy_balance_reactions}, panel b, the $PdV$ term is the dominant cooling term in the outflow: it balances EUV heating (black dashed line) as well as heat advected from below (dashed gray line). We note that the photons that are absorbed in the outflow layer (in the 13.6-15 eV range) have $\tau=1$ surfaces that all lie outside of the molecular hydrogen layer. The heating from this energy range of photons will be unaffected by cooling by $\hthreeplus$, as they are absorbed in a layer above.

\begin{deluxetable}{cc}
\tabletypesize{\footnotesize}
\tablecolumns{2}
\tablecaption{ Simulation data \label{tab:params}}
\tablehead{
\colhead{Parameter} & \colhead{Value}
}
\startdata
\hline
\rule{0pt}{1.\normalbaselineskip}Planet/star parameters\\
\hline
Planet mass, $M_{\mathrm{P}}$ ($M_{J}$) & 0.7 \\
Stellar mass, $M_\star$ ($M_{\sun}$) & 1 \\
Planet radius, $R_{\mathrm{P}}$ (cm) & $10^{10}$ \\
Semimajor axis, $a$ (au) & 0.05 \\
Temperature at base, $T_0$ (K) & 1000 \\
Pressure at base, $P_0$ (dyn cm$^{-2}$) & 0.96 \\
Adiabatic index, $\gamma$ & 5/3 \\
Spectral range$^b$ (eV) & 248-12 \\
\hline
\rule{0pt}{1.\normalbaselineskip}Photoreaction thresholds (eV)\\
\hline
$\h + h\nu \rightarrow \hplus$  &  13.6\\ 
$\htwo + h \nu \rightarrow \htwoplus$  & 15.4 \\
$\htwo + h \nu \rightarrow \h + \hplus + e$  & 18.08 \\
$\htwo + h\nu \rightarrow \h + \h$  & 4.74 \\
$\he + h\nu \rightarrow \heplus$  & 24.6 \\
\enddata
\footnotesize{$^b$ EUV flux is calculated by scaling \citealt{richards1994} solar spectrum by $(\mathrm{au}/a)^2$ for a given semimajor axis $a$.}\\
\end{deluxetable}

\begin{figure}[ht] 
  \includegraphics[width=\textwidth]{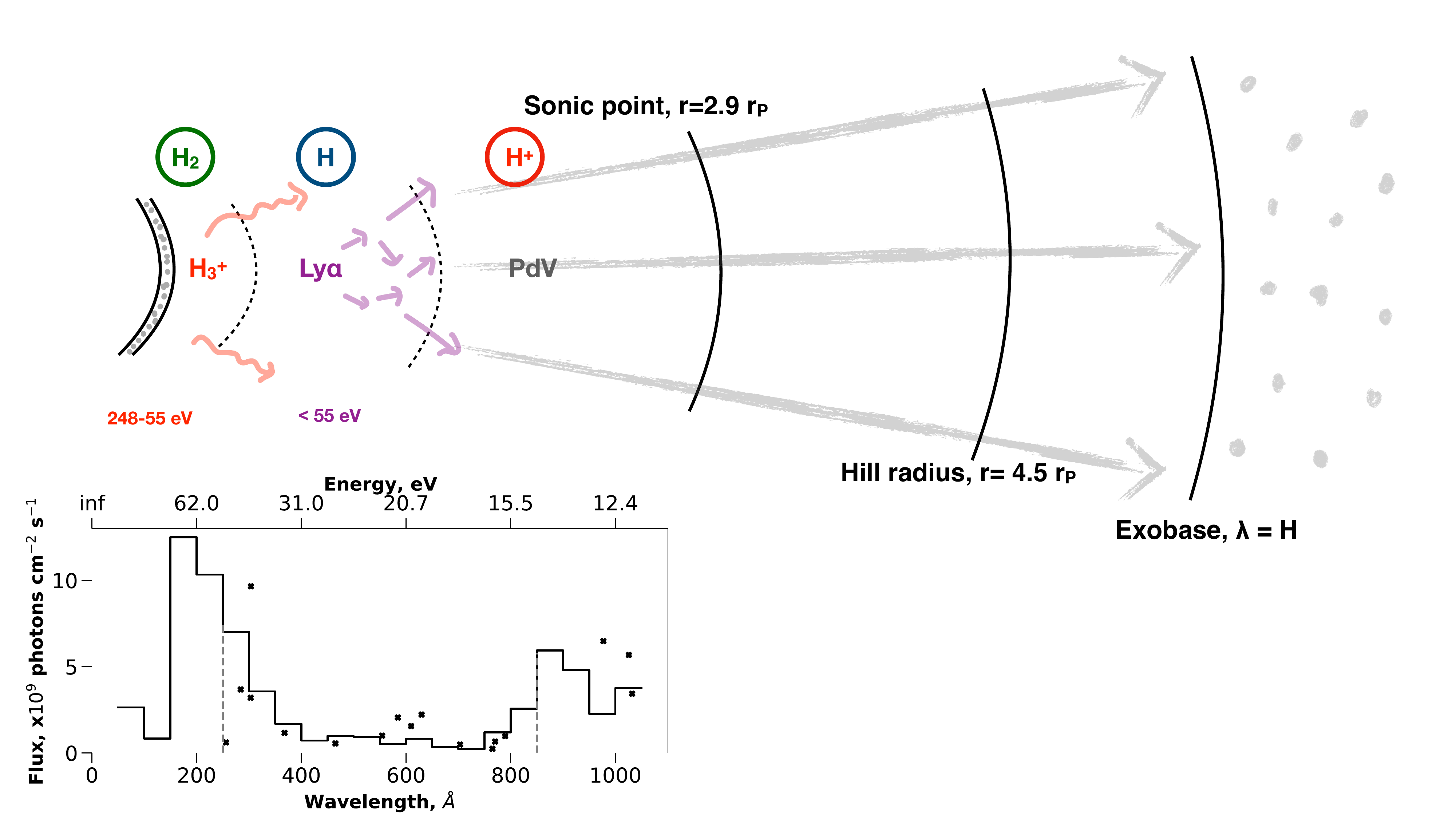}
  \caption{The three-layer diagram of the outcome of our model. The inner layer is the $\htwo$ layer cooled by $\hthreeplus$, where the highest-energy photons are deposited. The neutral atomic hydrogen layer further up is predominately cooled by Lyman-$\alpha$, with a contribution from $PdV$ as well. Intermediate-energy photons are depostied there. The outflow layer is cooled via $PdV$. Its structure is shown as a hydrodynamic outflow out to the sonic point, extending out to the Hill radius of the planet, where the gas becomes unbound from the planet. The exobase, where the outflow can is no longer collisional, is shown. The spectrum below the diagram corresponds to the P200 model from \citet{richards1994} with a high level of steller activity. The dots on the spectrum represent emission lines, and the bins are 20 wavelength bins with widths of 50 \AA. The 248-55 eV photons' $\tau=1$ surfaces lie within the $\hthreeplus$-cooled layer, and the $<55$ eV - within the Lyman-$\alpha$-cooled layer. The $< 13.6$ eV photons responsible for the resonant dissociation of $\htwo$ reach deeper in the $\hthreeplus$ - cooled layer.}
  \label{fig:layer_diagram}
\end{figure}

\begin{figure}[ht] 
  \includegraphics[width=\textwidth]{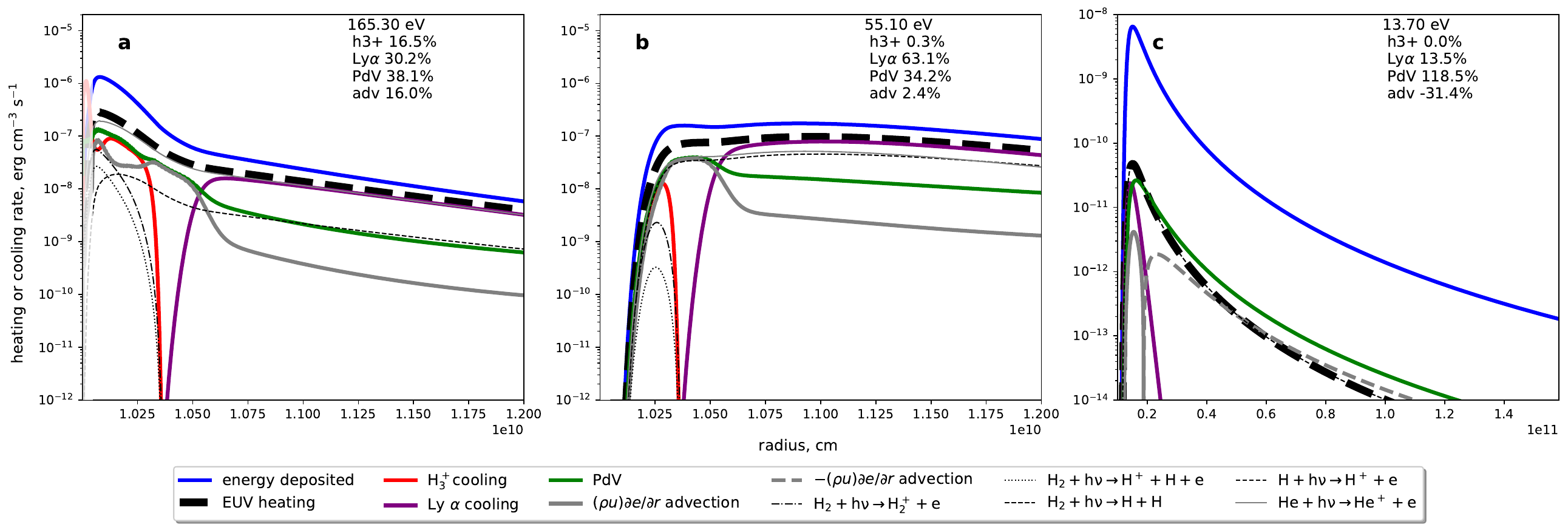}
  \caption{Heating and cooling rates for selected energy bins. The energy is allocated according to the expressions from Section \ref{sec:energy_allocation}. The panels display the breakdown of the heating rate as a result of all of the photoreactions with radius (see thin lines in legend). Panels a and b are zoomed in at the base, where radiative cooling is important. $\hthreeplus$ cooling dominates panel a, whilst Lyman-$\alpha$ is the major cooling source in panel b. $\he$ ionization (approximately coincident with the total EUV heating) is an important heating source in panel a, while $\h$ ionization becomes more important for heating in panels b and c. Although dominated by Lyman-$\alpha$, panel b still displays a significant contribution to the $PdV$ term from this intermediate energy. Finally, panel c displays an energy bin whose heating is primarily goes into driving the $PdV$ outflow (though the advective term here pulls extra energy from the internal energy of the gas in addition to the heating). The heating in this bin is all due to $\h$ ionization. The advective term sign convention follows that of Figure \ref{fig:energy_balance_reactions}.} \label{fig:bins}
\end{figure}

\begin{figure}[ht] 
  \includegraphics[width=0.5\textwidth]{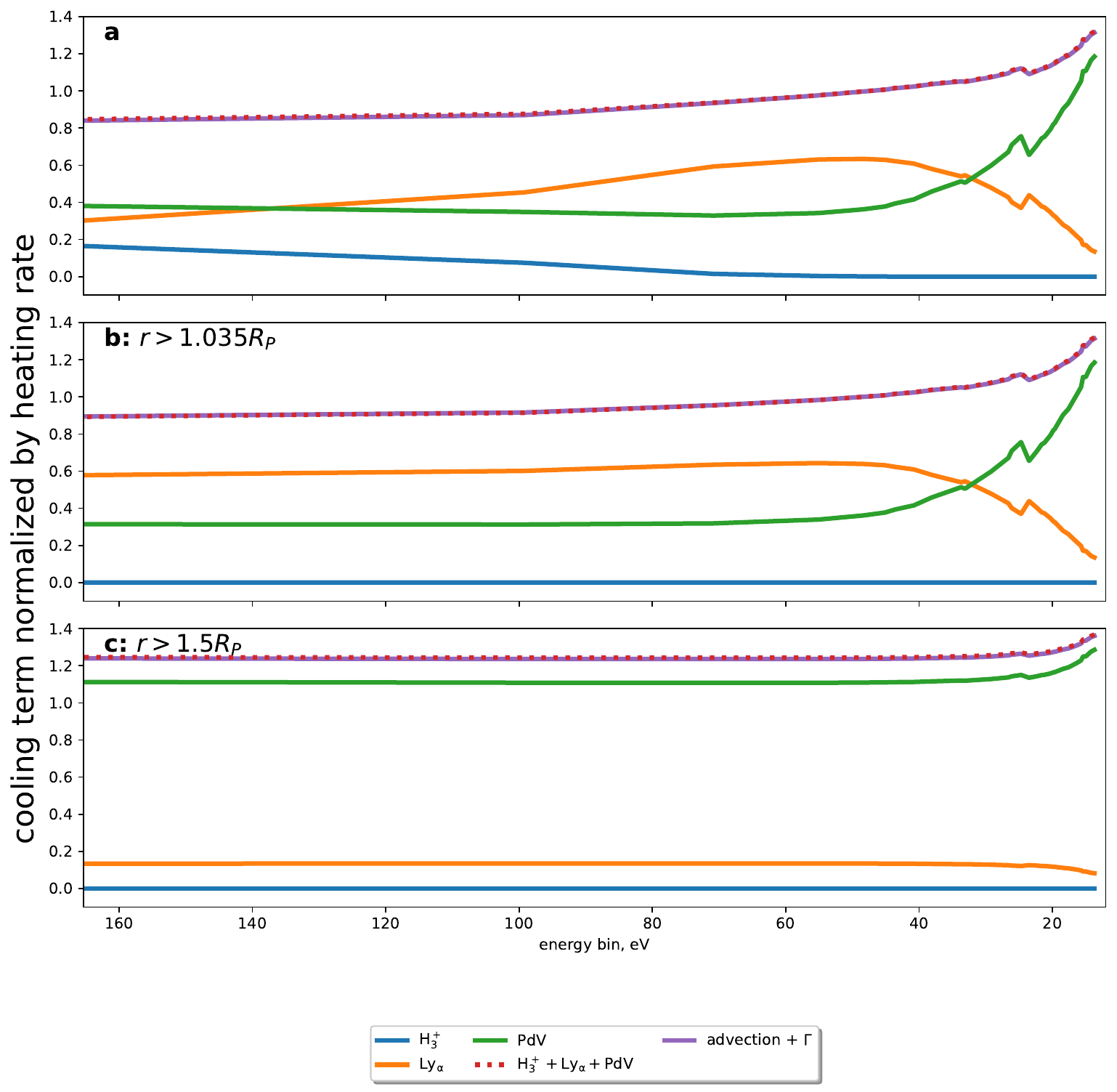}
  \caption{$PdV_{\nu}$, $\Lambda_{H_3^+, \nu}$, $\Lambda_{\rm{Ly}\alpha, \nu}$ normalized by the integrated stellar heating rate per steradian from $R_{\mathrm{P}}$ to $r_{out}$. The cooling terms balance the sum of advection and stellar heating (normalized by the stellar heating) in steady state (Equation \eqref{eq:steadystate_e}). Panel a is shown for the entire radius range (beyond the bolometric region acting as a boundary region). Panel b is shown for $r>1.035 R_{\mathrm{P}}$, the transition between $\hthreeplus$- and Lyman-$\alpha$-dominated cooling (Figure \ref{fig:energy_taus}). Panel c is shown for $r>1.5 R_{\mathrm{P}}$, beyond which the outflow becomes energy-limited.}\label{fig:fraction}
\end{figure}

\begin{figure}[ht] 
  \includegraphics[width=\textwidth]{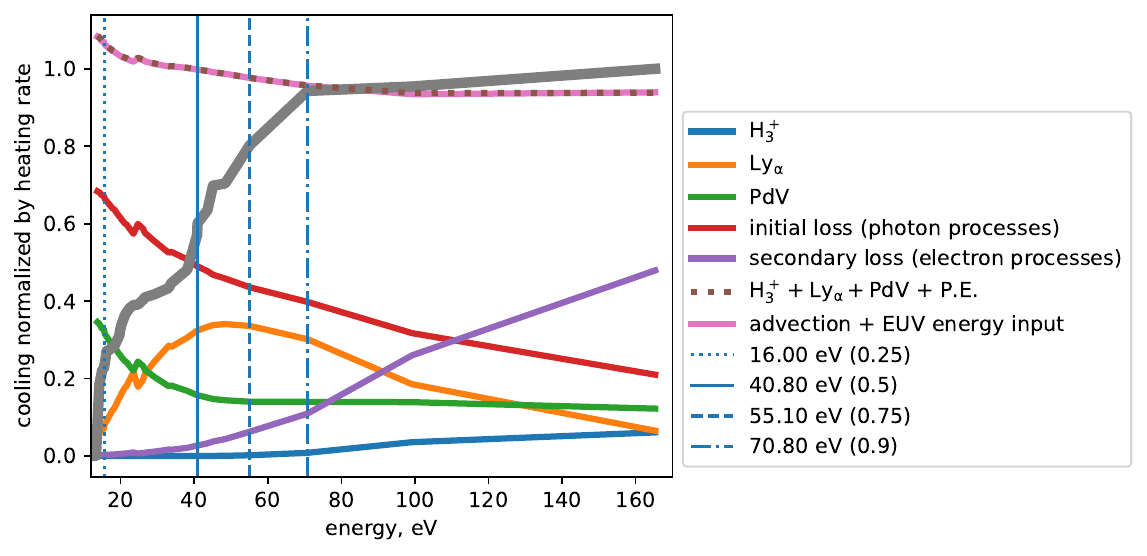}
  \caption{As in Figure \ref{fig:fraction},  $\Lambda_{H_3^+, \nu}$ (orange) and $\Lambda_{\rm{Ly}\alpha, \nu}$ (green) are normalized by the stellar heating rate $\Gamma_{\nu}$. The normalized $PdV$ term (red) is additionally shown as a cumulative fraction from the low- to the high-energy bins (thick blue line). As the low-energy bins are dominated by $PdV$-cooling (red), the cumulative term (thick blue) gives a sense of where in the spectrum most of the $PdV$ outflow power is derived. 0.25 of the $PdV$ comes from bins left of the vertical dotted line, 0.5 - of the solid line, 0.75 - of the dashed line, and 0.9 - of the dashed-dotted line. Beyond $\sim 70$ eV, almost all of the energy is radiated away, predominately by $\hthreeplus$ from the base molecular layer in the atmosphere.}
  \label{fig:csum_lyalpha}
\end{figure}

Figure \ref{fig:layer_diagram} is a summary of the three-layer structure that is the outcome of our model. The $\htwo$-dominated base layer is cooled by $\hthreeplus$, with the highest energies of our spectrum (248-55 eV) deposited there. The Lyman-$\alpha$-cooled layer containing a significant fraction of neutral hydrogen absorbs photons of energies 55-15 eV. We note that although the $PdV$ term has contributions from all altitudes, the 15-13.6 eV photons' $\tau=1$ surfaces lie closer to the layer where $PdV$ dominates over the other cooling sources, making the photons deposited higher-up in the wind an especially efficient driver of hydrodynamic escape compared to the photons deposited at lower altitudes. As an exception, the $<13.6$ eV photons contributing to the far-ultraviolet resonant dissociation of $\htwo$ are deposited in the $\hthreeplus$-cooled layer, not contributing to the outflow. We will show in Section \ref{sec:energy_allocation} that what drives the wind is photons in the 13.6-70 eV energy range.

The distance from the planet where the wind velocity reaches the sound speed is the sonic point, located at 3.1 planetary radii for our fiducial model. The Hill radius $R_{\mathrm{H}} = a (M_{\mathrm{P}}/(3 M_{\star}))^{1/3}$, where the tidal gravity from the star starts exceeding the gravity of the planet, is shown at 4.5 planetary radii. The exobase is the location where the mean-free path to collisions is comparable to the scale height of the atmosphere. Beyond, the particles are no longer collisional, and the atmosphere can no longer be modeled as a hydrodynamic outflow. The neutral species' cross section is smaller than the Coulomb cross section appropriate for ionized particles: they decouple from the outflow before the gas becomes collisionless, starting to free-stream out. Kinetic modeling is appropriate for the decoupled particles, which is beyond the scope of our work. Nevertheless, for our fiducial planet at 0.05 au, the gas remains collisional out to the sonic point, allowing us to model the outflow hydrodynamically. Collisionality is further discussed in Section \ref{sec:collisionality}.

\subsection{Energy balance bin-by-bin}\label{sec:energy_allocation}
In order to evaluate which parts of the spectrum contribute to the outflow and which parts are radiatively cooled (and which location that corresponds to in the atmosphere), we will use Equation \eqref{eq:steadystate_e}. The heating term, $\Gamma_m$, consists of contributions across the spectral energy range. We analyze the output of our simulation by evaluating the heating and cooling terms for different regions in the atmosphere, as well as for each spectral energy bin.

For example, as shown in Figure \ref{fig:energy_taus}, the highest-energy photons from our spectrum tend to be deposited in the molecular layer of the atmosphere at the base (the 55-248 eV range). $\hthreeplus$ emission is important in this layer. Some of the energy deposited in this layer still contributes to $PdV$ work and therefore the outflow (green solid line). We are interested in the contribution of heating in this layer to the total $PdV$ work term. To calculate this, first we define the integrated $PdV$ work term between radii $r_1$ and $r_2$ per unit solid angle, as we are working in one-dimensional spherical coordinates:
\begin{equation}\label{eq:pdv_tot}
    PdV = \int_{r_1}^{r_2} \left(\frac{P}{\rho}u\frac{\partial \rho}{\partial r} \right)r^2 dr.
\end{equation}
Here, the term in parentheses is the $PdV$ term from the steady-state energy equation. 
The total $PdV$ work term, $PdV_{\rm tot}$, can then be calculated from Equation (\ref{eq:pdv_tot}) by setting $r_1 = r_{\rm bol} = 1.005 R_{\mathrm{P}}$, the outer edge of the bolometric region acting as a boundary condition (hatched area in Figure \ref{fig:energy_taus}) and $r_2 = r_{\rm out}$, the outer-most radius in the grid. 

As illustrated in Figure \ref{fig:energy_taus}, the base layer of molecular hydrogen lies between $r_1 = r_{\rm bol}$  and $r_2$ at the outer edge of the layer with significant $\hthreeplus$ cooling ($r = 1.035 \times 10^{10} $ cm)---where the $\hthreeplus$ cooling curve drops sharply. The contribution to the total $PdV$ work from this region, $PdV/PdV_{\rm tot}$, is 5\%. As we go up in the atmosphere, the next primary cooling mechanism is briefly $PdV$ and advection. The contribution of this region ($1.035 \times 10^{10} < r < 1.05 \times 10^{10} $ cm) to the total $PdV$ work is $4\%$, after which the primary cooling mechanism transitions to Lyman-$\alpha$ cooling. The primary cooling mechanism transitions back to $PdV$ work at $r \approx 1.05 \times 10^{10} $ cm, and the contribution to the total $PdV$ from this Lyman-$\alpha$-cooled region is $34\%$. The total contribution to the $PdV$ work from the radiatively-limited region is significant (with most of it coming from the Lyman-$\alpha$ cooled region). Finally, 56\% of the total $PdV$ work comes from the cooling in the $PdV$-dominated upper layer, which transitions at $r = 1.5 \times 10^{10} $ cm.

We can look at the energy balance in this upper layer. The contribution from $PdV$ as a fraction of the EUV heating and advection from below is 90\%, placing the region above approximately 1.5 planetary radii in the energy-limited regime in the sense that most of the input energy after ionization is converted to $PdV$ work. In contrast, the $PdV$ term accounts for only 21\% of the incoming EUV heating in the Lyman-$\alpha$ cooled region. We return to a discussion of the overall energy limit in Section \ref{sec:epsilon}.

To evaluate which parts of the input stellar spectrum contribute the most to the outflow, for each photon energy we determine the range of radii over which the photons deposit energy in the wind.  At each radius in the outflow, we know the fraction of energy going into each cooling process (Figure \ref{fig:energy_balance_reactions}).  Integrating over radius, we then calculate the amount of energy that goes into cooling by each process weighted by the fraction of the flux deposited at that radius.  For example, for a photon energy with optical depth unity in the H$_3^+$-cooled region of the wind, the majority of the deposited energy is lost to H$_3^+$ cooling.  However, some photons of this energy are absorbed in higher layers of the wind, contributing to Lyman-$\alpha$ cooling and $PdV$ work. 

Figure \ref{fig:bins} illustrates how each cooling term balances the heating from a particular wavelength bin as a function of radius. At any given radius, we plot the total heating rate due to photons in a wavelength bin and distribute that heating into $PdV$, Lyman-$\alpha$, $\hthreeplus$ cooling, and advection. The panels of the plot are annotated with percentages that show the integrated fraction of heating from each wavelength bin that is either radiatively cooled, contributes to the outflow via $PdV$ work, or is advected. To calculate these percentages shown in Figure \ref{fig:bins}, we first define the ratio of the heating from a specific energy bin to the total heating at a given radius: $f_{\nu} = \Gamma_{\nu}(r)/\sum_{\nu}\Gamma_{\nu}(r)$. Then, we multiply each cooling rate ($\Lambda_{\hthreeplus}, \Lambda_{\rm{Ly}\alpha}$, and $PdV$) at that radius by that fraction. This way, we estimate how much of each of the cooling processes balance each heating energy bin: $\Lambda_{H_3^+, \nu}(r) = f_{\nu} \Lambda_{\hthreeplus}(r)$, $\Lambda_{\rm{Ly}\alpha, \nu}(r) = f_{\nu} \Lambda_{\rm{Ly}\alpha}(r) $, and $PdV_{\nu}(r) = f_{\nu} PdV(r)$. We can obtain the total contribution to the wavelength bin by integrating over radius\footnote{After multiplying by $\rho$ in the integrals the quantities are in units of energy per volume, not energy per mass.}: $\Lambda_{H_3^+, \nu} = \int_{R_{\mathrm{P}}}^{r_{out}} \Lambda_{H_3^+, \nu}(r) \rho(r) r^2 dr$, $\Lambda_{\rm{Ly}\alpha, \nu} = \int_{R_{\mathrm{P}}}^{r_{out}} \Lambda_{\rm{Ly}\alpha, \nu}(r) \rho(r) r^2 dr$, and $PdV_{\nu} = \int_{R_{\mathrm{P}}}^{r_{out}} PdV_{\nu}(r) r^2 dr$. Finally, we can divide this by the total heating, $\int_{R_{\mathrm{P}}}^{r_{out}} (\sum_{\nu}\Gamma_{\nu}(r)) \rho(r) r^2 dr$, to find out what fraction of the heating goes into what cooling process for each respective energy bin.

Figure \ref{fig:fraction} (top panel) depicts the energy budget broken up into wavelength bins. $\hthreeplus$ cooling affects the highest-energy photons the most, offestting up to $\sim 20\%$ of the available heating. Lyman-$\alpha$ cooling increases in importance and is dominant in the 140-32 eV range. Then, for energies below around 32 eV over half of the heating goes into $PdV$. This may suggest that the outflow is energy-limited from the ionization threshold of H (13.6 eV) to 32 eV. However, the $\tau=1$ surfaces for photons with these energies lie in the region where Lyman-$\alpha$ dominates over the $PdV$ work term (Figure \ref{fig:energy_taus}). This is because the higher-energy photons also contribute significantly to the heating in this Lyman-$\alpha$ cooled region, and their contribution to the heating rate is mostly radiated away. 

If we restrict the radius range to $r>1.5$ planetary radii (Figure \ref{fig:fraction}, bottom panel), we can see that 15\% of the heating goes into Lyman-$\alpha$, and the rest goes into $PdV$ or is advected away over the entire range of the spectrum.  However, we have shown that only about half of the wind's energy is deposited in that upper layer.  The wind's ``base" is thus not simply defined. We find that the heating rate peaks at $\sim 1.1R_{\mathrm{P}} $, which we consider to be a reasonable qualitative choice.

As in Figure \ref{fig:fraction} (top panel), Figure \ref{fig:csum_lyalpha} displays the cooling rates normalized by the energy deposited from the spectrum. However, we plot the $PdV$ term as a cumulative sum from the lower-energy bins of the spectrum (which have almost all the energy after accounting for ionization going into $PdV$ work) to the higher-energy bins (which are relatively less efficient at driving the wind). This allows us to get a quantitative estimate for which spectral energies are driving the wind: most of the spectral energy used to power the outflow is from energies less than $\sim$ 70 eV.

\subsection{Grid of models}\label{sec:grid_of_models}
\begin{figure}[ht] 
  \includegraphics[width=\textwidth]{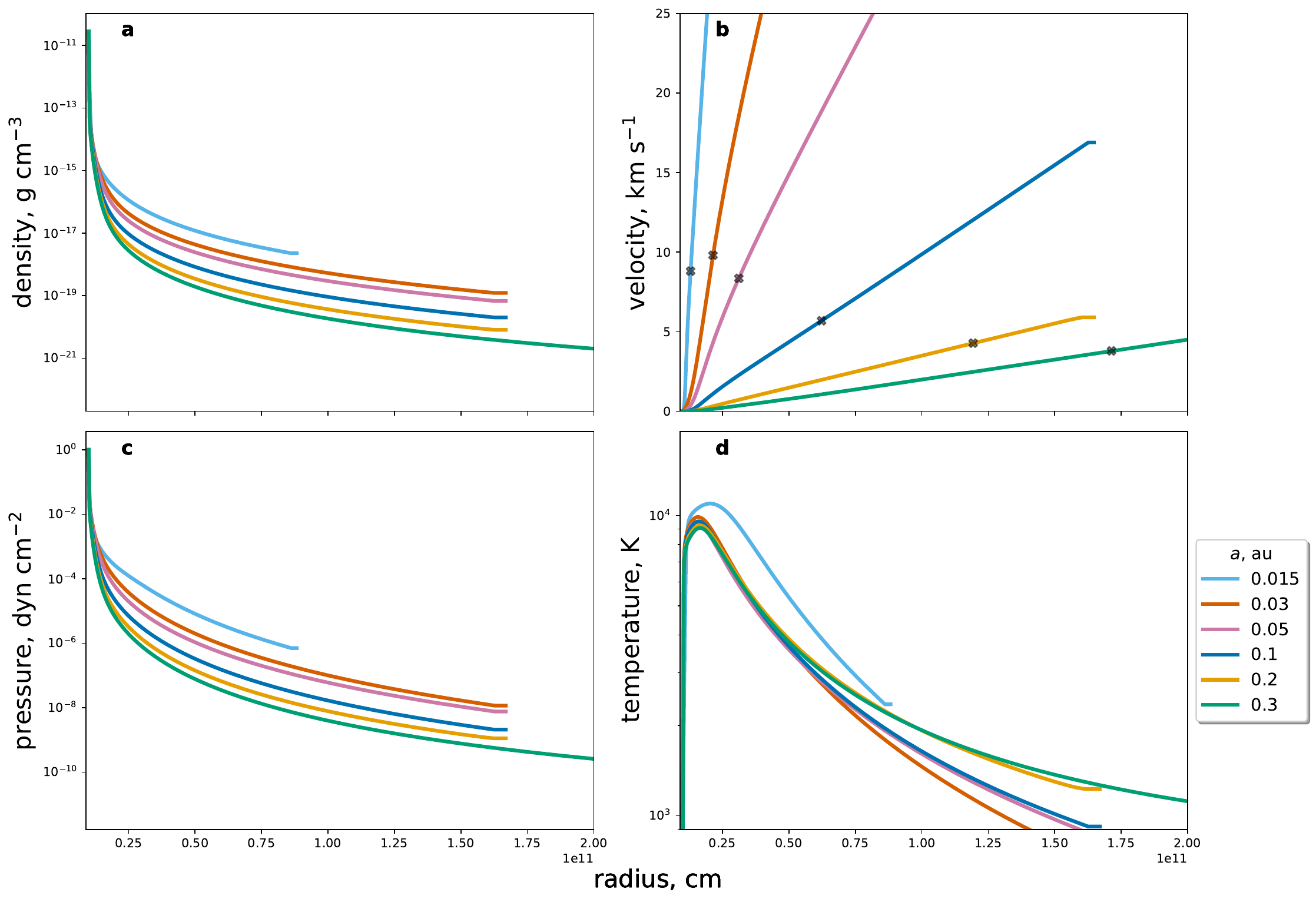}
  \caption{Density (panel a), velocity (panel b), pressure (panel c), and temperature (panel d) profiles for varying semimajor axes in the range 0.015-0.3 au.}
  \label{fig:rhovpt_fiducial_grid}
\end{figure}

We address the question of how the efficiency of atmospheric escape varies with stellocentric distance $a$. For that purpose, we run a grid of models, placing our fiducial hot Jupiter at 0.015-0.3 au from the host star. We scale the EUV flux by $a^{-2}$, effectively varying the distance. Conduction is turned on in the range 0.05-0.3 au. Tidal gravity is present in our model (Appendix \ref{appendix:tidal}), enhancing the mass-loss rate for the closest-in planets. As the term scales as $a^{-3}$, beyond $\sim0.1$ au the effect is negligible. We plot the density (panel a), velocity (panel b), pressure (panel c), and temperature (panel d) profiles in Figure \ref{fig:rhovpt_fiducial_grid}. The sonic point moves outward as the stellar EUV flux decreases with semimajor axis and less energy is available to lift the gas out of the potential well of the planet.

\begin{figure}[ht] 
  \includegraphics[width=\textwidth]{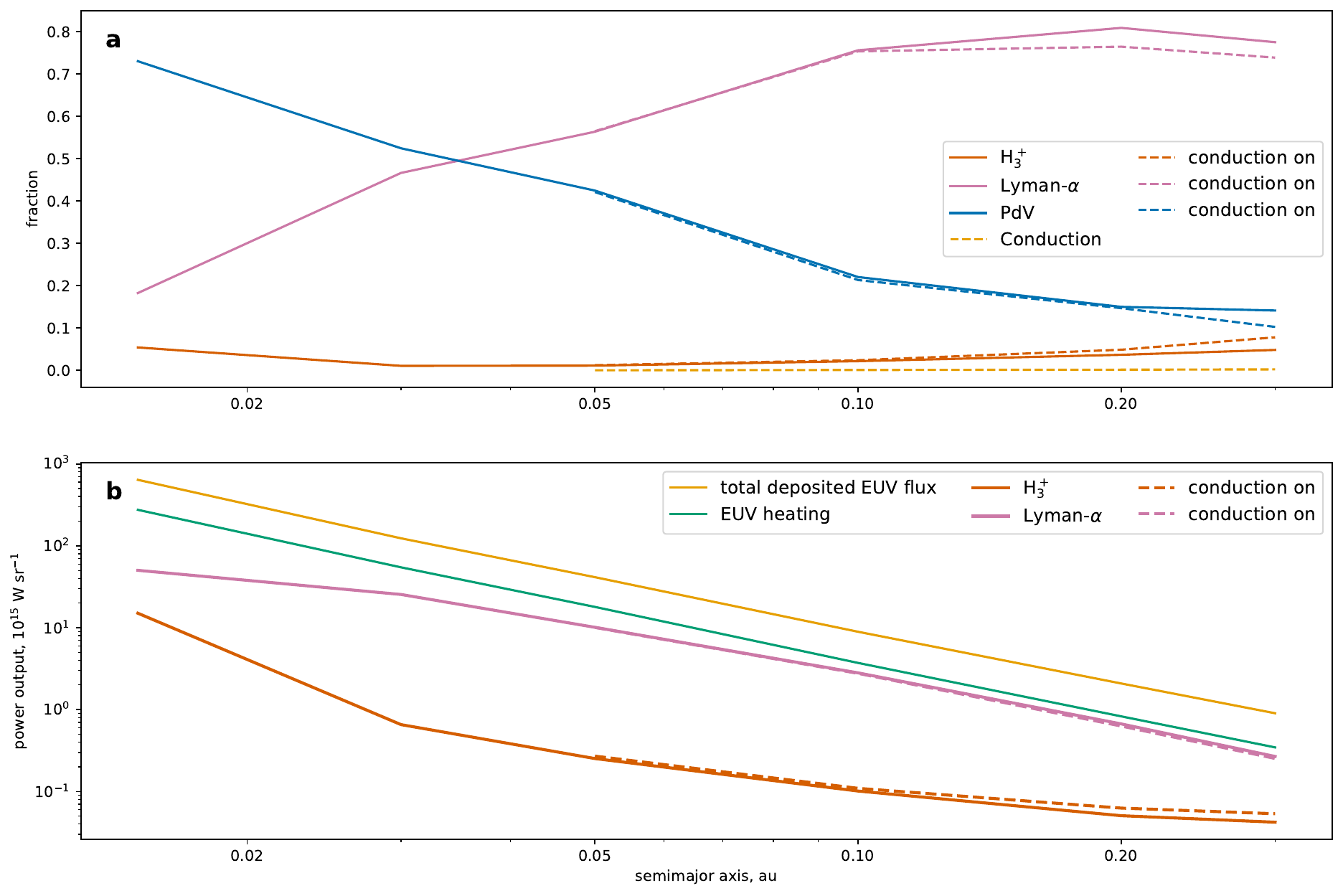}
  \caption{Panel a: fraction of the total heating (after accounting for energy required for ionization) that ultimately is balanced by $PdV$ work (blue), Lyman-$\alpha$ cooling (pink), and $\hthreeplus$ cooling (red). The $PdV$ fraction decreases with distance, implying that the outflow is increasingly not energy-limited with greater distance from the star. The increasing neutral fraction of the outflow causes an increase in the Lyman-$\alpha$ cooling as a fraction of the total available heating in the case where conduction is turned off (pink solid line). However, conduction causes the $\hthreeplus$ to radiate away a greater fraction of the incoming flux, with a decrease in the Lyman-$\alpha$ emission. Conduction through the base is shown (orange dashed). Panel b: the total power output for the model per unit solid angle. The total EUV flux from the star incident on the planet is plotted in orange. The heating rate (after ionization) is plotted in green. The total radative cooling fraction increases with distance.}\label{fig:energy_bin_fractions_out}
\end{figure}

Figure \ref{fig:energy_bin_fractions_out}, panel a displays the fraction of the total EUV heating (after accounting for the energy required for primary and secondary ionization) that goes into $\hthreeplus$ cooling, Lyman-$\alpha$ cooling, and $PdV$ as a function of semimajor axis. Though important for the highest spectral energy bins, $\hthreeplus$ cooling was found to be a relatively minor fraction of the overall energy heating the gas from 0.015 to 0.3 au. We plot the total EUV flux per steradian incident on the planet (yellow) and the total heating rate per steradian (green) in Figure \ref{fig:energy_bin_fractions_out}, panel b. The ratio between these curves is 0.45,  approximately constant over the semimajor axis range.  This ratio comes from the removal of energy required for ionization. We will use this value in Section \ref{sec:epsilon} to compute the energy-limited escape rate and evaluate how energy-limited our wind is, i.e. what fraction of the total available energy that is used to heat the gas goes into powering the wind. We reiterate that the fractions in Panel a are computed with respect to the green curve in Panel b.  Panel b also shows the power output in both $\hthreeplus$ (red) and Lyman-$\alpha$ (pink) cooling. The Lyman-$\alpha$ power output exceeds that of the $\hthreeplus$, comprising a greater fraction of the incoming flux further out from the star. Lyman-$\alpha$ cooling drops in importance once tidal gravity becomes strong for the 0.015au case. 

\begin{figure}[ht] 
  \includegraphics[width=\textwidth]{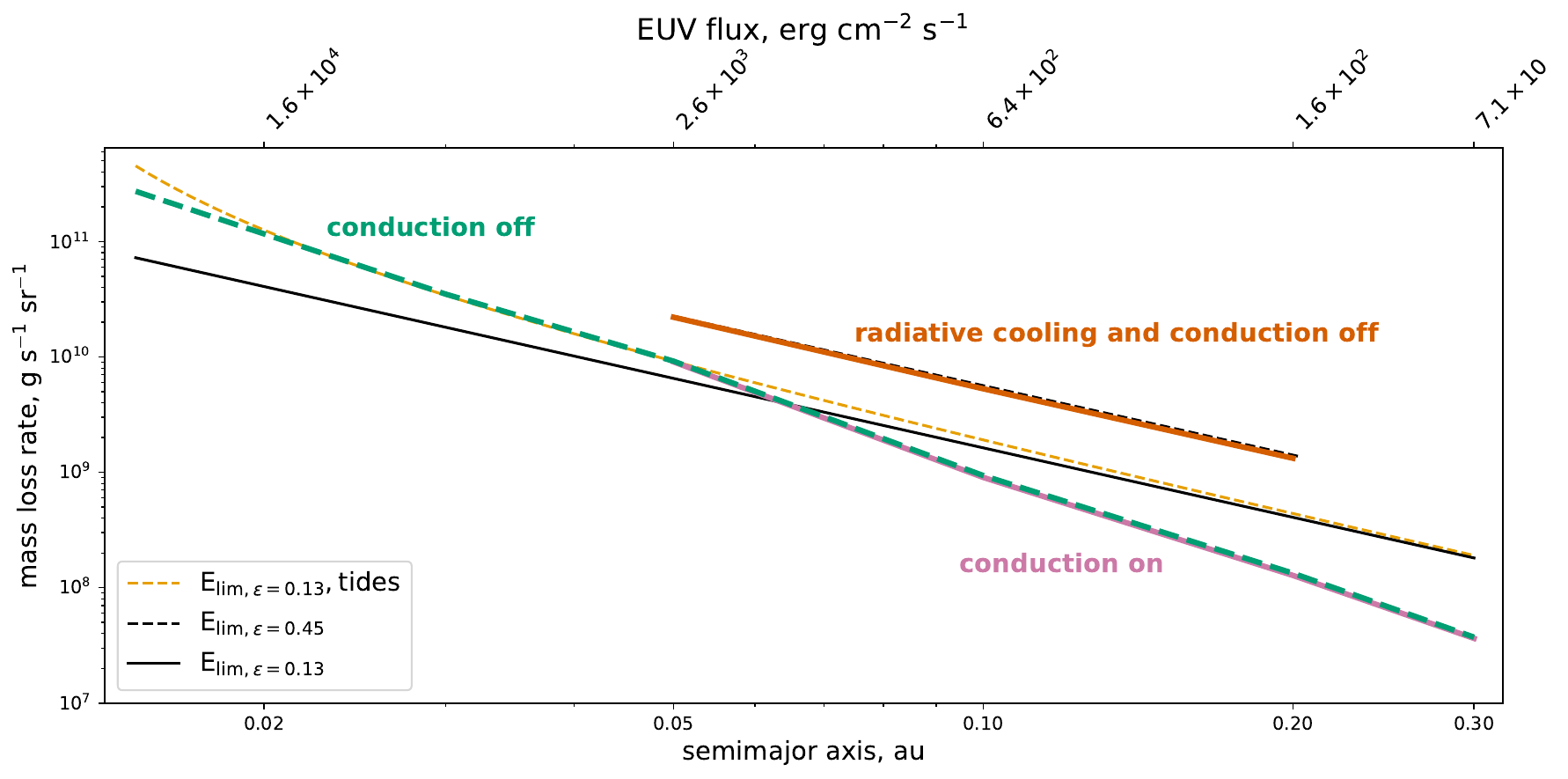}
  \caption{Mass-loss rate as a function of distance. Shown are the mass-loss rates $\dot{M} = \rho u r^2$ per solid angle. The mass-loss rates for planets with radiative cooling and conduction turned off (red solid) approach the energy-limited mass-loss rate (black dashed, plotted for $\epsilon=0.45$ and a wind launch radius $r = 1.2 R_\mathrm{P}$). The cases with conduction off (green dashed) and conduction on (pink solid) are plotted. The energy-limited mass-loss is plotted for a radius $r = 1.2 R_\mathrm{P}$, $\epsilon=0.13$ with a tidal correction factor (orange dashed, see Section \ref{sec:epsilon}) and without (black solid line). For $a=0.3$ au, $\dot{M} = 3.6 \times 10^7$ g s$^{-1}$ for conduction on, and $\dot{M} =  3.7 \times 10^7$ g s$^{-1}$ - for conduction off.}\label{fig:mdot}
\end{figure}
The mass-loss rates for two grids of models are plotted in Figure \ref{fig:mdot}. The red solid line marks the mass-loss rates for a series of runs with all radiative cooling turned off: all of the EUV heating (minus the heating required for ionization) is available to power the outflow. The black dashed line is the energy-limited approximation to the mass-loss rate, with efficiency $\epsilon=0.45$ to account for the energy required to overcome the ionization potentials of the ionizing species. At $1.2 R_P$, we estimated the wind launch radius to be between where the heating rate peaks at $1.1 R_p$ and the average deposition radius $R_{\mathrm{D}} = \int r \Gamma_m(r) \rho(r) r^2 dr /(\int \Gamma_m(r) \rho(r) r^2 dr) \simeq 1.5 R_{\mathrm{P}}$, where $\Gamma_m(r)$ is the heating rate per unit mass of the gas, after accounting for the ionization of the species. We note that this is is an over-simplification, as the deposition radius and efficiency are both dependent on photon energy. Nonetheless, it is useful in providing an approximate scaling for the radius in the energy-limited mass-loss rate.

The mass-loss rate for the non-radiative case in Figure \ref{fig:mdot} goes as $F_{EUV}^1$, which is not necessarily true in cases where radiative cooling is significant. At fluxes greater than what is considered in our work ($\gtrsim 10^4$ erg cm$^{-2}$ s$^{-1}$), \citet{rmc2009} found that the ouflows were not energy-limited, reflecting the increased importance of Lyman-$\alpha$ cooling in their high-flux model. They found that outflows were roughly energy-limited for fluxes in the intermediate range ($10^2$ to $10^4$ erg cm$^{-2}$ s$^{-1}$).

In our simulations, for the case where all radiative cooling is turned on (Figure \ref{fig:mdot}, pink solid line and green dashed line for $a<0.05$ au), the mass-loss rate matches the energy-limited rate well for distances $<0.05$ au, which roughly corresponds to the flux range $>2.6 \times 10^3$ erg cm$^{-2}$ s$^{-1}$. The mass-loss rate beyond 0.05 au has a steeper slope than the energy-limited scaling: radiative cooling is more significant with lower stellar flux. Closer-in, $\hthreeplus$ cooling somewhat offsets the increased tidal forcing.

\subsection[H3+ emission spectrum and optical thinness]{$H_3^+$  emission spectrum and optical thinness}\label{sec:h3pemission}

\begin{figure}[ht] 
  \includegraphics[width=0.5\textwidth]{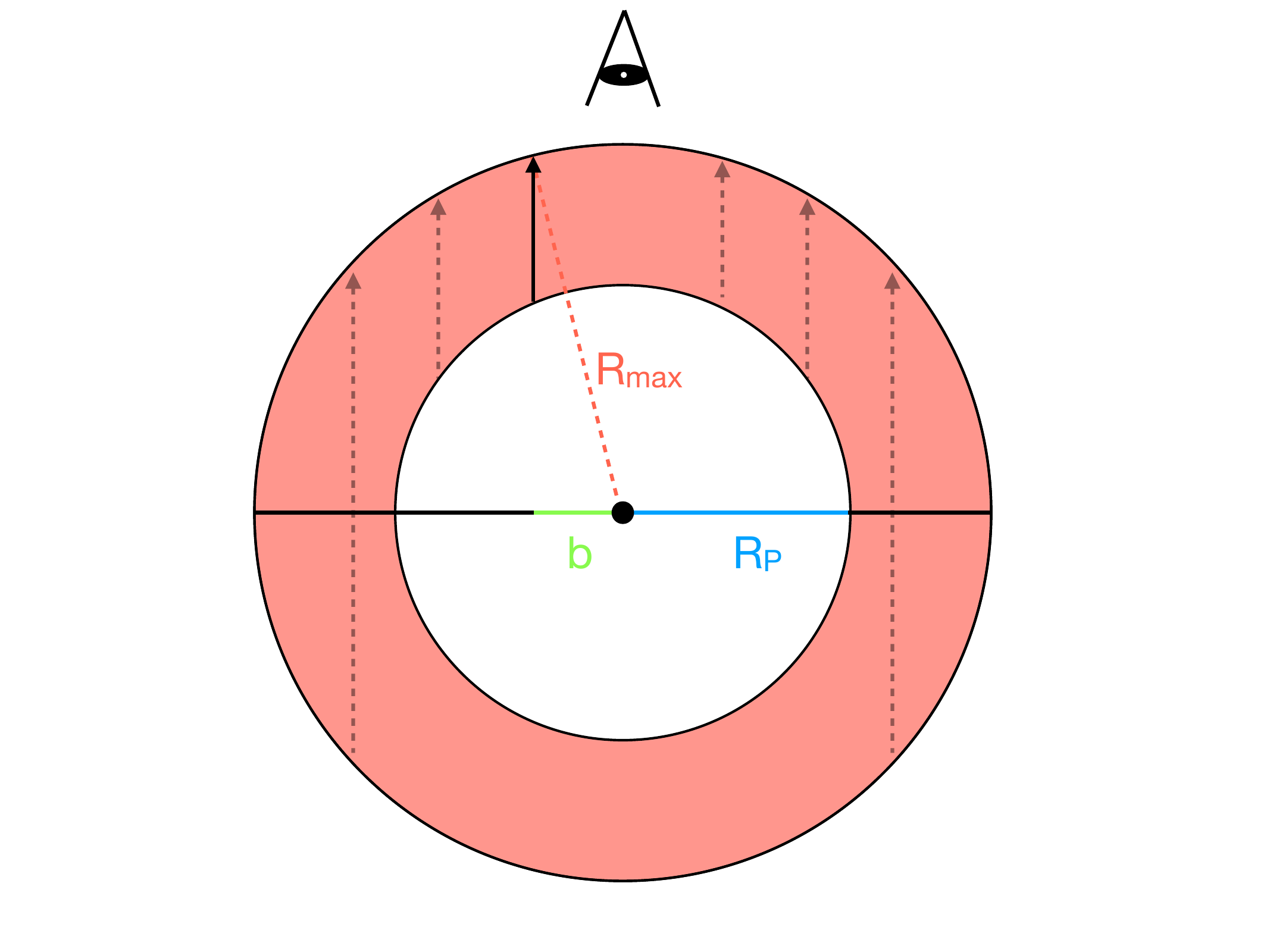}
  \caption{Radiative transfer schematic (not to scale) for computing the $\hthreeplus$ emission spectrum. The $\hthreeplus$ is contained in the layer between $R_{\mathrm{P}}$ and $R_{\mathrm{max}}$. If the impact parameter $b\geq R_{\mathrm{P}}$ we consider a ray originating at the edge of the $\hthreeplus$ emission surface and passing through the limb of the planet. If $b< R_{\mathrm{P}}$, the ray originates inside the planet's disk, at the base of the $\hthreeplus$--containing layer.}
  \label{fig:rad_transfer}
\end{figure}

\begin{figure}[ht] 
  \includegraphics[width=\textwidth]{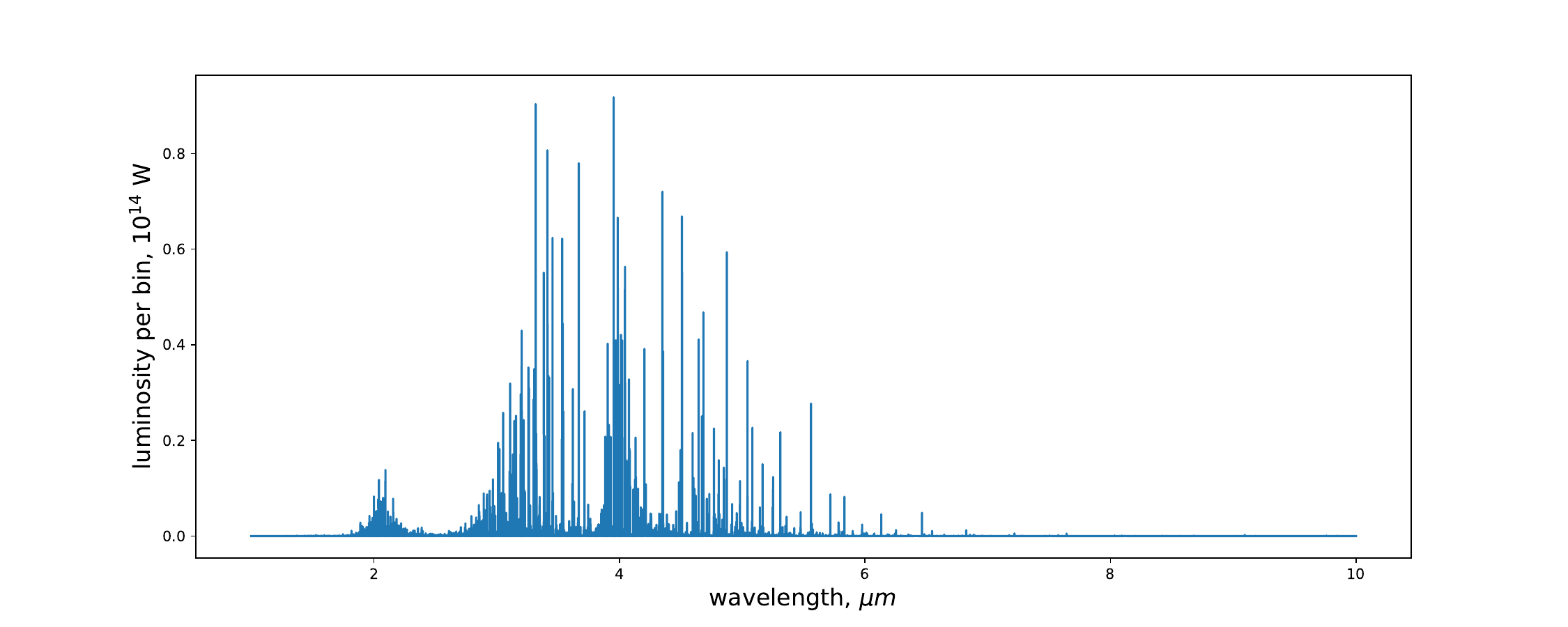}
  \caption{Emission spectrum of $\hthreeplus$ cooling for our fiducial hot Jupiter placed at 0.05 au. Bins are spaced evenly in wavenumber, with 1 bin width = 0.09 cm$^{-1}$.}
  \label{fig:h3p_emission}
\end{figure}

We generate a sample emission spectrum using ExoCross \citep{yurchenko2018exocross}, which uses a line list for $\hthreeplus$ computed by \citet{mizus2017exomol}. To calculate the emission, we integrate the luminosity of each spectral bin along the line of sight for each impact parameter (Figure \ref{fig:rad_transfer}). Then, we sum the contributions to obtain the overall emission at each wavelength bin. We use the interpolated temperature to look up the emissivity at each wavelength, $\epsilon_\lambda$ (units of ergs/s/molecule/sterradian), obtained from ExoCross, multiplied by the interpolated $\hthreeplus$ density values, $n_{\textrm{H}_3^+}$ to calculate the flux per unit length at each point along the ray, $F_\lambda(s) = \pi \varepsilon_\lambda n_{\textrm{H}_3^+}$. We then integrate the luminosity along the ray for each wavelength: $F_{\lambda, \mathrm{ray}} = \int_ {s_{\mathrm{min}}}^{s_{\mathrm{max}}} F_{\lambda}(s) ds$, where $s$ is the coordinate along the ray. Here, $s_{\mathrm{min}}$ is either $-\sqrt{ (R_{\mathrm{max}}^2-b^2)}$ (for $b \geq R_{\mathrm{P}}$) or $\sqrt {(R_{\mathrm{P}}^2-b^2)}$ (for $b < R_{\mathrm{P}}$). The upper bound of the integral $s_{\mathrm{max}} = \sqrt{ (R_{\mathrm{max}}^2-b^2)}$ denotes where the number density of $\hthreeplus$ becomes negligible. We then integrate over the impact parameters $b$: $ \int_0 ^{R_{\mathrm{max}}}2 \pi b F_{\lambda, \mathrm{ray}} db$ to get the total luminosity for a given wavelength.   We plot the emission spectrum for our fiducial hot Jupiter at 0.05au in Figure \ref{fig:h3p_emission}.

We show that $\hthreeplus$ infrared emisison is optically thin, thereby able to escape and cool the gas. 
The optical depth for a given line is:
\begin{equation}
    \tau_{\nu} = N(\hthreeplus) \sigma,
\end{equation}
where $\sigma$ is the column-averaged cross section. The column density $N(\hthreeplus) = \int n_{\hthreeplus} dr$ of $\hthreeplus$ for our fiducial hot Jupiter is $ 10^{13}$ cm$^{-2}$, and, as will be discussed in Section \ref{sec:se} for our super Earth is $1.5\times 10^{13}$ cm$^{-2}$. 
Optical thinness is defined as $\tau_{\nu} \ll 1$. As $\sigma \ll 10^{-14} \, \mathrm{cm}^2$ for all lines considered, $\hthreeplus$ is optically thin.  We have verified using absorption cross-sections from ExoCross \citep{yurchenko2018exocross, mizus2017exomol} that all lines are optically thin for our fiducial planet.  For example, the maximum cross-section in a line at 2000K is $5 \times 10^{-18}$ cm$^2$.

\subsection{Conduction} \label{sec:conduction}
\begin{figure}[ht] 
  \includegraphics[width=\textwidth]{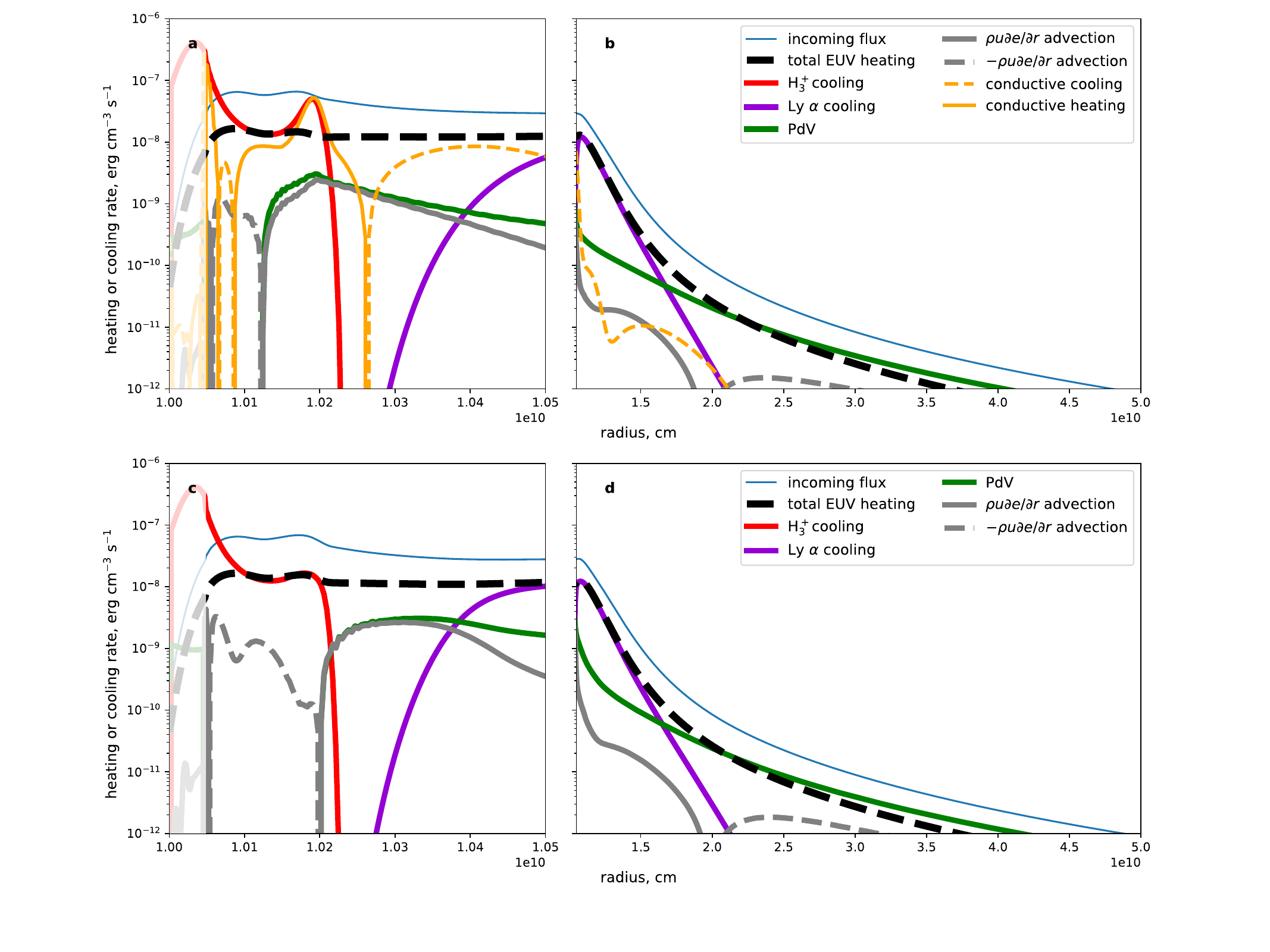}
  \caption{Heating and cooling rates, as described in Figure \ref{fig:energy_balance_reactions}. The top panels correspond to our fiducial hot Jupiter at 0.3 au, with conduction turned on. The bottom panels are the same planet, only with conduction turned off. Conductive heating (orange solid) and cooling (orange dashed) are shown.}
  \label{fig:cond_comp}
\end{figure}

Conduction was found to be subdominant at all fluxes in our grid of models. We present a comparison of the energy distribution with altitude for our fiducial planet placed at 0.3 au in Figure \ref{fig:cond_comp}. The top panels depict the case of conduction turned on, and the bottom -- off. Conduction has a heating effect in the layer containing $\hthreeplus$, which is offset by additional $\hthreeplus$ cooling. Conduction dominates the cooling around 1.04 $R_{\mathrm{P}}$. The mass-loss rate decreases by $\sim 2\%$ for the 0.3au case, as seen in Figure \ref{fig:mdot}. As we expect conduction to matter more for lower fluxes, we cut off our grid at 0.3 au and leave further exploration for future work.

\subsection{Energy-limited approximation and stellar heating efficiency} \label{sec:epsilon}
\begin{figure}[ht] 
  \includegraphics[width=\textwidth]{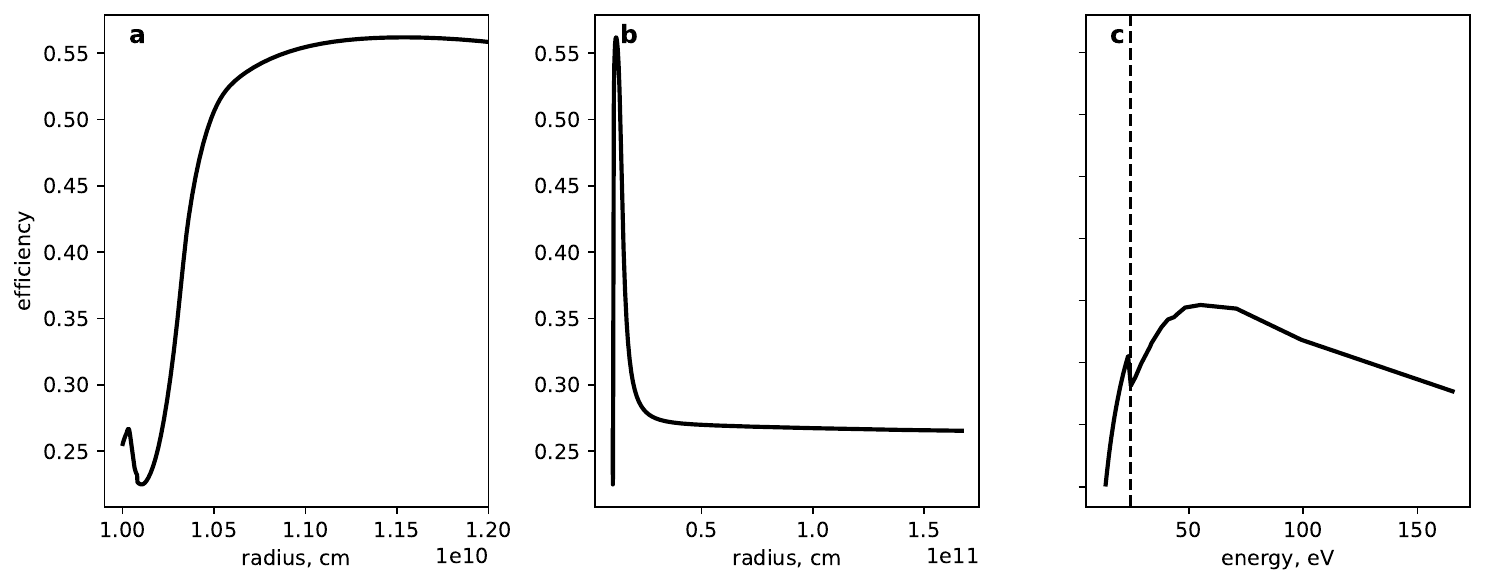}
  \caption{Efficiency of EUV heating, defined as the fraction of total deposited EUV flux that is available to heat the gas. Panel a shows the efficiency with radius, zoomed into the base - heating is less efficient in the base molecular layer, as the higher-energy photons deposited there lose energy to secondary ionization. Panel b shows the larger-scale outflow. Panel c shows the efficiency with energy. The vertical dashed line marks the ionization threshold of $\he$: as the photons rightward are available to ionize, the efficiency drops slightly. We emphasize that this is a deposition efficiency - this does not yet include any losses due to radiative cooling, either $\hthreeplus$ or Lyman-$\alpha$.}
  \label{fig:efficiency_fiducial}
\end{figure}
The energy-limited approximation \citealt{watson1981dynamics} to atmospheric escape is widely used in the literature to estimate planetary mass-loss rates (e.g. \citealt{des2007diagram}; \citealt{erkaev2007roche}; \citealt{sanz2010scenario}; \citealt{ehrenreich2011mass}; \citealt{salz2015high}). This formula relates the energy absorbed in the cross-sectional area $\pi R_{\mathrm{P}}^2$ to the energy going into lifting the gas out of the gravitational potential well, $GM/R_{\mathrm{P}}$:
\begin{equation}\label{eq:energy_lim}
    \dot{M} = \epsilon \frac{R_{\mathrm{P}}^3 F_{XUV} }{GM_{\mathrm{P}}},
\end{equation}
where $\dot{M}$ is the mass-loss rate, and $F_{XUV}$ is the spectral flux in the energy range contributing to escape, commonly the total EUV flux or the sum of the EUV and X-ray flux. The efficiency factor $\epsilon$ (with typical values of 0.1-0.5) encompasses the energy required to ionize the gas and radiative losses that result in not all of the incoming stellar flux being available to power escape. Depending on the geometry of the problem, this formula may often include additional factors of order unity. We choose to write this formula as an escape rate per unit solid angle (i.e. per steradian) along the sub-stellar ray, as this makes it convenient to directly compare to the output of our one-dimensional simulation centered on the substellar ray between the planet and the star.\footnote{The total energy-limited $\dot M$ typically includes a prefactor of $\pi$ to account for the cross-sectional area absorbing the stellar flux.  If this input goes into mass loss over a full $4\pi$ steradians from the planet, one might expect Equation \ref{eq:energy_lim} to require an additional factor of $1/4$.  However, we are calculating loss along the sub-stellar ray.  Translating that into a planet-integrated $\dot M$ would require multiplying by an extra factor of $\sim$$1/4$.  Equation \ref{eq:energy_lim} is appropriate without additional pre-factors along the sub-stellar ray that we compute with our code.}   The radius used in this expression is often the optical radius, but the true absorption radius is higher (and depends on photon energy, see Figure \ref{fig:energy_taus}). 

While the efficiency factor $\epsilon$ is sometimes used to mean the overall efficiency, including the ionization and the radiative cooling, we only consider ionization in this section. 
We may compute the heating efficiency of our model output by defining it as a fraction of the total input EUV flux that goes into heating, after accounting for overcoming the ionization potentials of the species. It may be computed for a given distance from the planet (Figure \ref{fig:efficiency_fiducial}, panel a), or for a given energy bin (Figure \ref{fig:efficiency_fiducial}, panel b). The total efficiency computed over the entire spectral range, integrated over the planetary radius range, is $\epsilon = 0.45$. We note here that although the photons of higher energy are more efficient at heating the gas, they do not necessarily power the outflow more, as they are deposited at altitudes that have significant radiative cooling, both $\hthreeplus$ for the highest-energy photons in our spectrum ($\sim 248-70$ eV), and Lyman-$\alpha$ ($\sim 70-20$ eV).

From our simulations including radiative cooling, the mass-loss rates are lower than this energy-limited rate. The radiative cooling provides an extra factor of loss of efficiency, which depends on distance from the star (and therefore on the EUV flux). We provide a calculation of the output estimate in Section \ref{sec:grid_of_models}.

To account for tides, we can include a correction factor in the energy-limited mass-loss rate formula \citep{erkaev2007roche}:

\begin{equation}\label{eq:tidal_correction}
    \zeta(\chi) = \left(1-\frac{3}{2\chi}+\frac{1}{2 \chi ^3}\right)^{-1},
\end{equation}
\noindent where $1/\chi = (R_{\mathrm{H}}/R_{\mathrm{P}})^{-1}$ is the inverse of the Hill radius normalized by the planetary radius. This produces an over-estimate of the mass-loss rate (Figure \ref{fig:mdot}) for the 0.015 au case. The ratio $1/\chi$ is no longer sufficiently small, so the expansion used to produce Equation \ref{eq:tidal_correction} is no longer valid at this order, and higher-order corrections are needed. Nonetheless, we still use this expression to compare with what's commonly done in the literature.In Figure \ref{fig:mdot}, at distances beyond 0.05 au, the mass-loss diverges from the energy-limited rate due to more effective radiative cooling. For for distances $\lesssim 0.05$ au, tidal gravity enhances the mass-loss rate, though $\hthreeplus$ cooling somewhat decreases this effect (see also Figure \ref{fig:fraction}.

\section{Super Earths/sub-Neptunes} \label{sec:se}
Super-Earths and sub-Neptunes are a category of planets with sizes in between that of the Earth and the $\sim 10 \, M_{\text{E}}$ mass range that characterizes the ice giants. In this study, we will use the terms ``super-Earths" and ``sub-Neptunes" interchangeably to refer to planets smaller than the ice giants that host atmospheres. A more detailed distinction between these categories is beyond the scope of this work. Being larger than Earth, these planets are able to host atmospheres up to order 10\% of their total mass. The super-Earths discovered to date orbit close-in to their host stars, typically with the semimajor axes spanning the range of $\sim 0.01-0.5$ au. In a high-stellar-flux environment, these planets may experience significant mass loss due to stellar irradiation, analogous to the hydrodynamic outflows of the hot Jupiters. 

\begin{figure}[ht] 
  \includegraphics[width=0.9\textwidth]{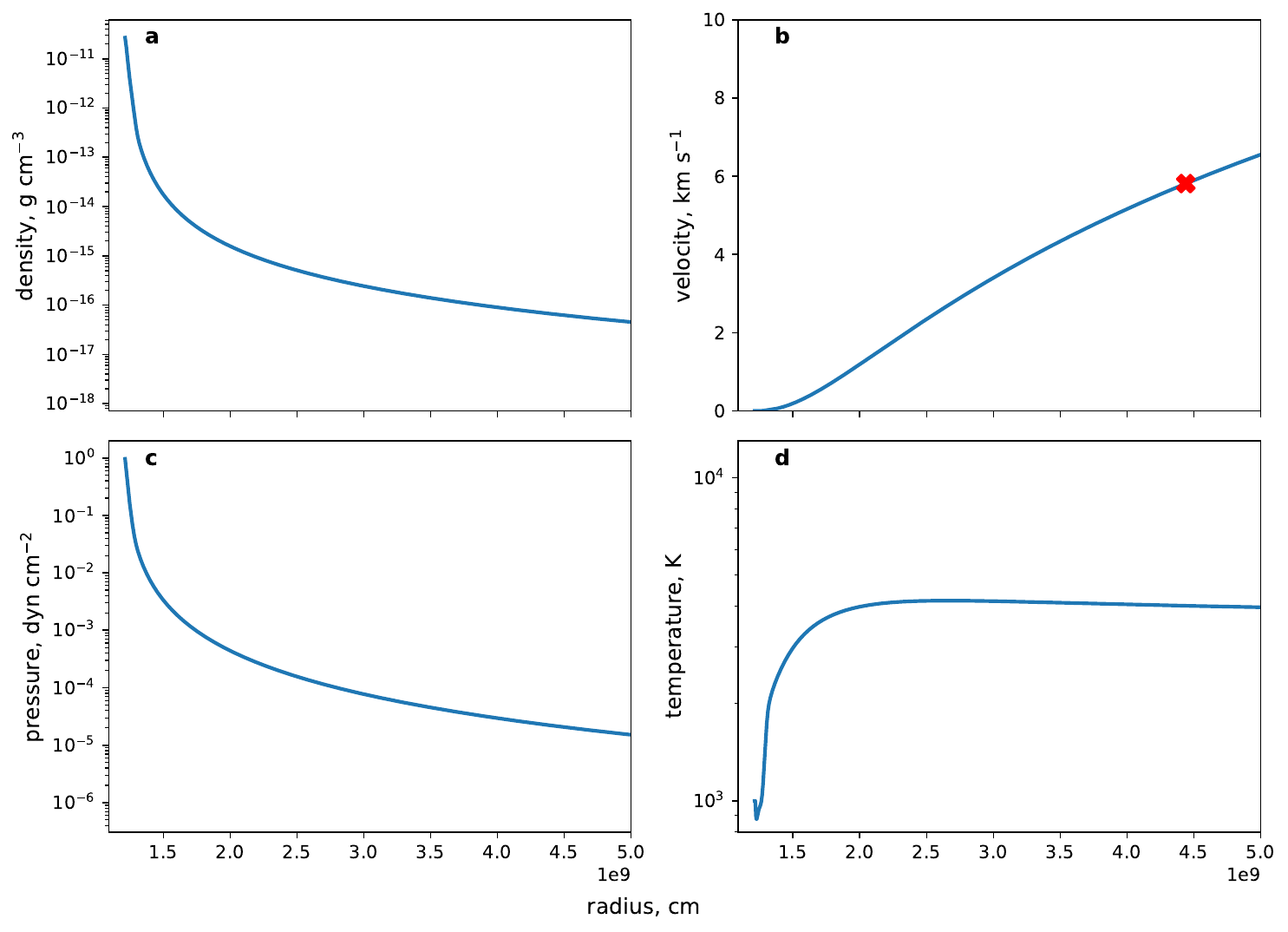}
  \caption{Density (panel a), velocity (panel b), pressure (panel c), and temperature (panel d) profiles for our fiducial super-Earth. The velocity reaches 5.8 km s$^{-2}$ at 4.4 planetary radii. The temperature peaks at 4200 K., and is 4000 K at the sonic point.}\label{fig:rhovpt_se}
\end{figure}

\begin{figure}[ht] 
  \includegraphics[width=0.9\textwidth]{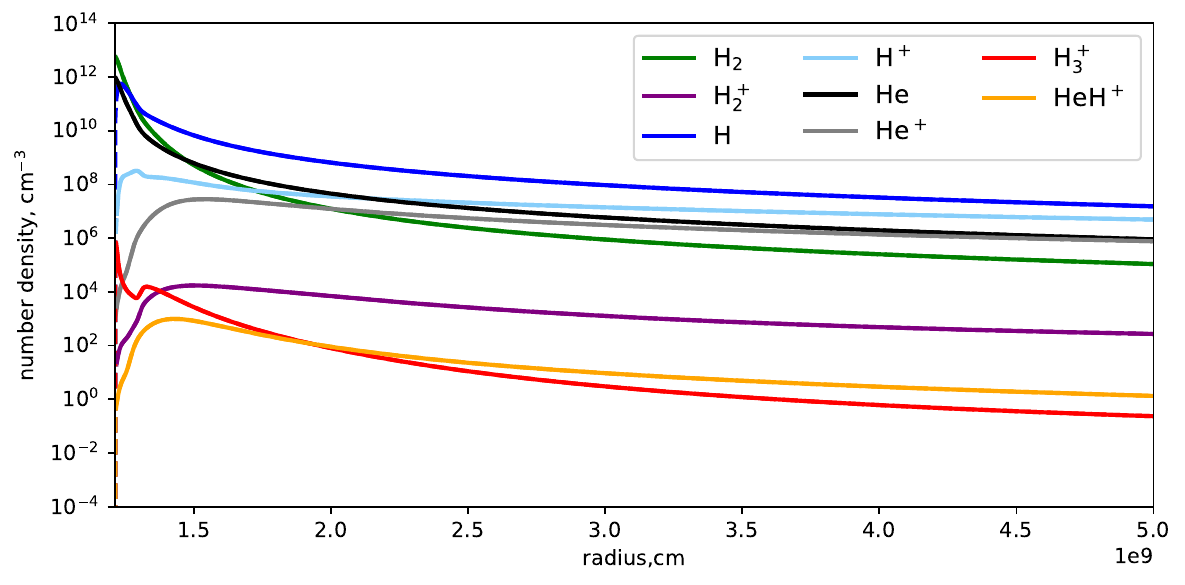}
  \caption{The wind is mostly neutral $\h$ at the sonic point. The $\htwo$ is not cut off sharply as in the hot Jupiter case; its gradual drop off allows $\hthreeplus$ to be present throughout.} \label{fig:comp_se}
\end{figure}

\begin{figure}[ht] 
  \includegraphics[width=0.9\textwidth]{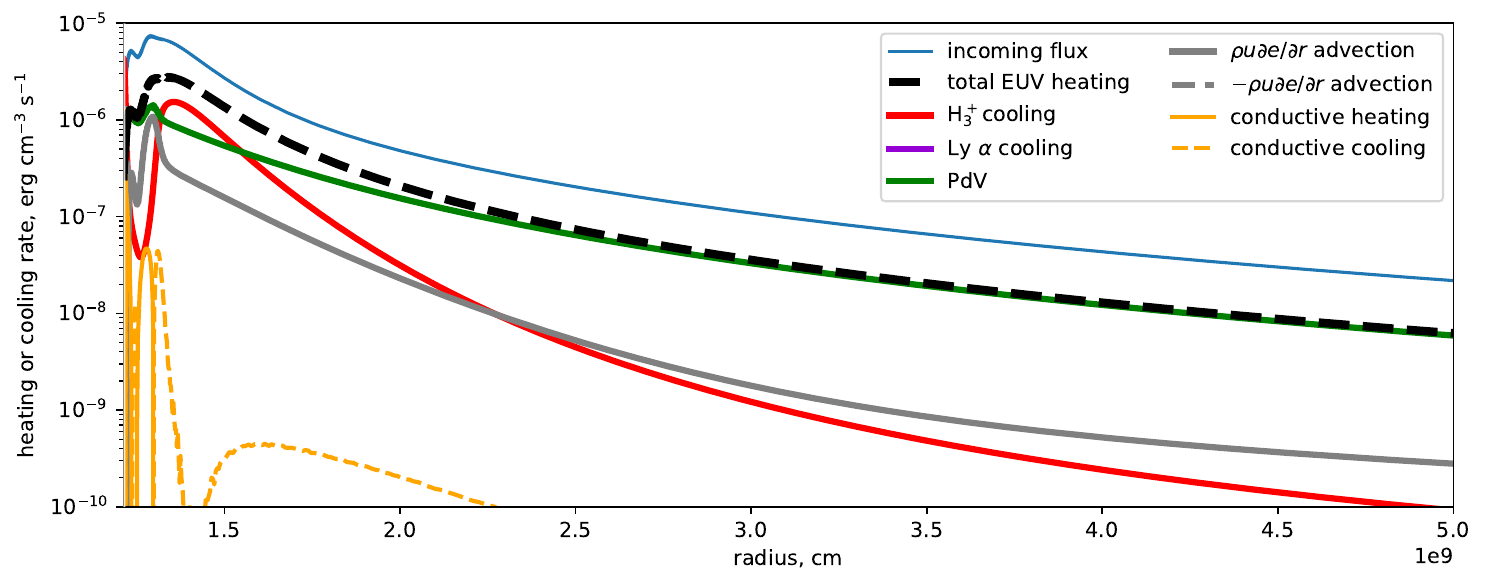}
  \caption{Toward the base, $\hthreeplus$ cooling is a dominant cooling source. Further up, the outflow transitions to energy-limited, where EUV heating is balanced by $PdV$ work. Lyman-$\alpha$ is not an important cooling source due to the lower temperature in the gas. Advection never acts as a heating source: it is not necessary to pull internal energy from deeper in to drive the outflow, as it is in the hot Jupiter case. Conduction is sub-dominant.} \label{fig:energy_se}
\end{figure}

We now apply our model to the super Earth regime, with the goal of obtaining a preliminary understanding of how radiative energy loss affects outflows from these smaller-sized planets. Figures \ref{fig:rhovpt_se}, \ref{fig:comp_se}, and \ref{fig:energy_se} provide model outputs for as fiducial super-Earth with the parameters given in Table \ref{tab:se_data}.

We have shown for the fiducial hot Jupiter case that Lyman-$\alpha$ cooling becomes significant when the gas heats up to $\sim$$10^4$ K, acting as a major radiative coolant. In the predominately atomic hydrogen region between the molecular hydrogen layer and the outflow layer, the gas heats up, allowing Lyman-$\alpha$ cooling to kick in: the $PdV$ term is not yet strong enough to cool the gas and the heating is not yet enough to overcome the massive gravitational potential of the hot Jupiter. However, the surface gravity of super-Earths is typically smaller than that of Jupiter. The energy needed to escape is smaller, meaning that gas does not need to be heated as much to escape the planet's potential well.  For most super-Earths, the atmosphere never gets hot enough for Lyman-$\alpha$ to be an important coolant.

In a hot Jupiter, the primary mode of dissociation that determines the transition from molecular hydrogen to atomic, thus shutting off of $\hthreeplus$ cooling, is thermal dissociation of $\htwo$. As the temperature threshold for significant thermal dissociation is never reached in a super-Earth, molecular hydrogen is present throughout, instead of being confined to a layer at the base. Nevertheless, the atmosphere is quite optically thin to ionizing photons, allowing $\hthreeplus$ to accumulate where $\htwo$ is present. Thus, $\hthreeplus$ is an important radiative coolant for super-Earth-sized planets.  We note that while consideration of other molecules is beyond the scope of this work, we expect other strong molecular coolants to survive in the outflows of super-Earths as well.  Thus, direct application of our gas giant model (with only hydrogen and helium species) to a super-Earth provides an upper limit on the true atmospheric loss rate.

\begin{deluxetable}{cc}
\tabletypesize{\footnotesize}
\tablecolumns{2}
\tablecaption{Simulation data}\label{tab:se_data}
\tablehead{
\colhead{Parameter} & \colhead{Value}
}
\startdata
\hline
\rule{0pt}{1.\normalbaselineskip}Planet parameters\\
\hline
Planet mass, $M_{\mathrm{P}}$ ($M_{\text{E}}$) & 7.2 \\
Planet radius, $R_{\mathrm{P}}$ ($R_{\text{E}}$) & 1.9 \\
Semimajor axis, $a$ (au) & 0.05 \\
Temperature at base, $T_0$ (K) & 1000 \\
Pressure at base, $P_0$ (dyn cm$^{-2}$) &  0.47\\
\enddata
\end{deluxetable}

\begin{figure}[ht] 
  \includegraphics[width=0.5\textwidth]{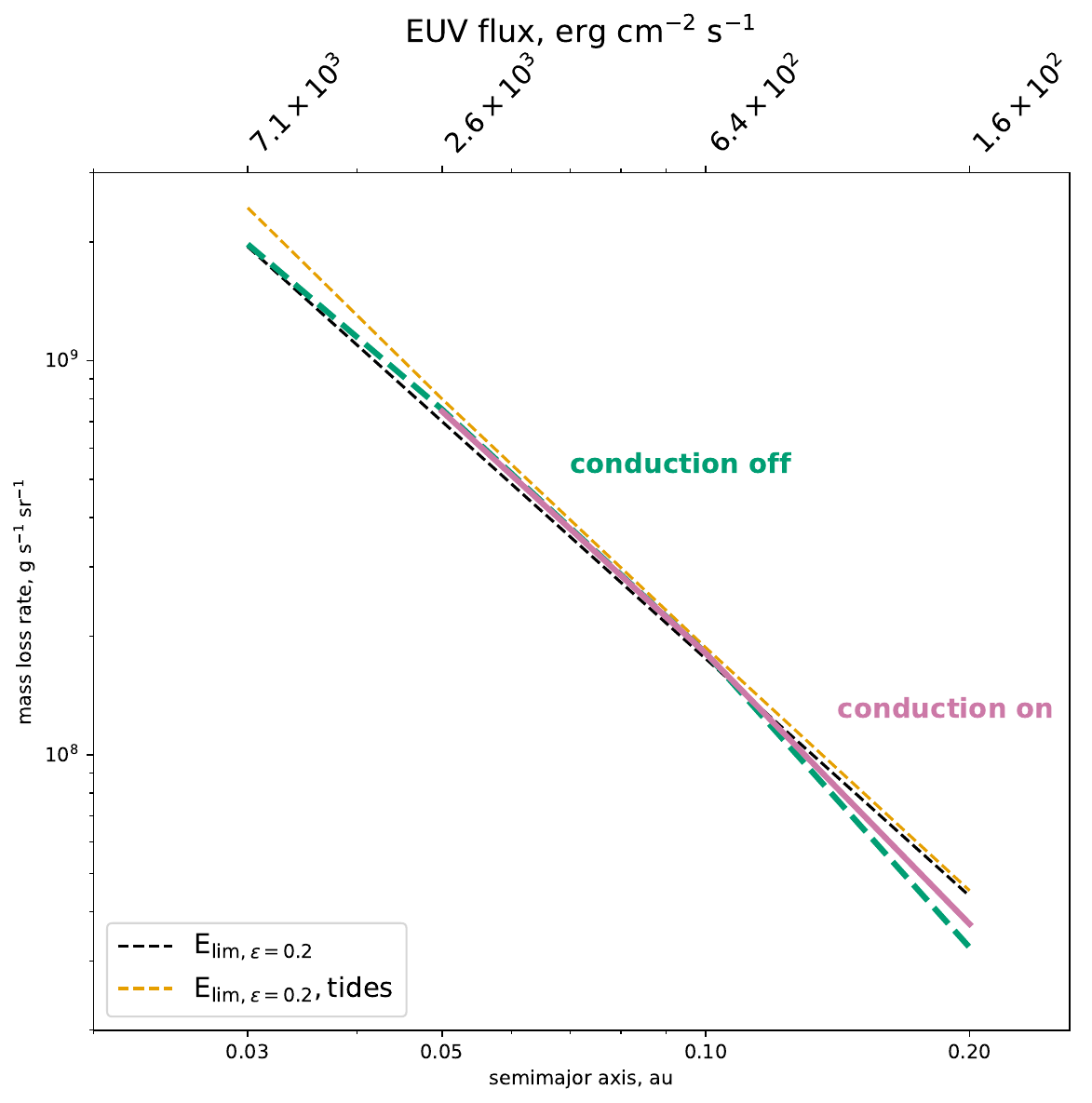}
  \caption{Mass-loss rate as a function of distance for our fiducial super-Earth for a semimajor axis range 0.03-0.2 au. The energy-limited mass-loss approximation with tides off (black dashed) and tides on (orange dashed) is shown for an efficiency $\epsilon = 0.2$ and deposition radius of $1.3 R_{\mathrm{P}}$. Conduction is sub-dominant here.}\label{fig:mdot_se}
\end{figure}

In Figure \ref{fig:mdot_se} we plot the mass-loss rate per solid angle for our grid super-Earths, varying the distance from 0.03 to 0.2 au. For comparison, we plot the energy-limited formula for an efficiency of 0.2 and deposition radius of 1.3 $R_{\mathrm{P}}$, both with the tidal correction factor (orange dashed) and without (black dashed). The efficiency accounts for both the losses due to ionization and $\hthreeplus$ radiative cooling. We emphasize that the efficiency varies with photon energy and deposition radius. In addition, the actual deposition radius is typically $\gtrsim R_{\mathrm{P}}$ and varies depending on the photon energy. 

For $a\lesssim 0.1$ au, we find that the mean-free-path to collisions for ion-ion, neutral-neutral, and neutral-ion collisions is well below the value of the scale height in the atmosphere---the wind is collisional at least throught the sonic point. Beyond $\sim$0.05 au, the wind is predominately neutral. Although the $\hplus$ in the wind is still collisional, the neutral $\h$ becomes collisionless with itself and with the $\hplus$ before reaching the sonic point beyond $\sim$ 0.1 au. We conclude that a kinetic description may be needed for super-Earths beyond 0.1 au, as the winds there are mostly neutral and collisionless.

\section{Discussion}\label{sec:disc}
\subsection{Stellar wind, magnetic field, and radiation pressure} \label{sec:stellar_pressure}
The plantary outflow may be significantly redirected by the stellar wind and/or the stellar magnetic field if the stellar wind ram pressure, $P_{\star}$, the magnetic field pressure, $P_{B, \star}$, and the thermal pressure, $P_k$ exceed the sum of the thermal and ram pressures from the planetary outflow, $P_{k,p}+P_{p}$ (for example, forming a cometary tail, as in \citealt{mccann2019morphology}). The stellar wind ram pressure is $P_{\mathrm{ram}} = \rho_{\star} v_{\star}^2$. We can take the stellar wind velocity as the approximately constant $v \sim 400$ km s$^{-1}$ for our semimajor axis range $a<0.3$, and the number density of protons as approximately $n_{p} \simeq 2.4 \times 10^3 (0.05\mathrm{au}/a)^2$ cm$^{-3}$, as chosen in \citet{rmc2009}. Then, the ram component of the stellar wind pressure can be estimated: $P_{\mathrm{ram}} \simeq m_{\mathrm{H}} n_{p} v^2 \simeq 6.4 \times 10^{-6} (0.05 \mathrm{au}/a)^2$ dyn cm$^{-2}$, or 6.4 picobars at 0.05 au. Taking the proton temperature to be 10$^6$ K, the thermal pressure due to protons in the stellar wind is $P_k = n_pkT_p \simeq 3.3 \times 10^{-7} (0.05 \mathrm{au}/a)^2$ dyn cm$^{-2}$, or 0.3 picobars at 0.05 au.  The pressure due to the solar magnetic field is $B_r^2/(8 \pi)$. If $B_r$ is approximately 0.01 G at 0.05 au \citep{banaszkiewicz1998analytic}, the magnetic field pressure is $\simeq$ 4 picobars. The ram and magnetic pressures dominate, for a total of $\simeq 10$ picobars at 0.05 au. 

At the sonic point distance from the planet, the thermal and ram pressures of the outflow are equal. At 0.05 au for our fiducial hot Jupiter, both components are approximately 7.5 picobars, so $P_{k,p}+P_{p} \simeq 15$ picobars. As this is comparable to the stellar wind pressure in scale, especially at close-in distances we conclude that the outflow shape may be affected by the stellar wind. However, for our super-Earth at 0.05 au, the sum of the thermal and ram pressures of the wind at the sonic point is $\sim 40$ picobars, which is greater than the stellar wind pressure estimated at 0.05 au.

If the pressure of the stellar wind exceeds the pressure of the planetary wind upstream of the sonic point, the wind may be significantly redirected and possibly confined. If the pressure of the stellar wind exceeds the planetary wind pressure downstream of the sonic point, the supersonic solution is valid, as the boundary conditions will not travel upstream.

Radiation pressure is typically considered for interactions with dust grains. However, radiation pressure can also act on hydrogen atoms. The cross section for scattering by ionized hydrogen is negligible compared to the scattering by neutral hydrogen of Lyman-$\alpha$ photons, which can be significant \citep[e.g.,][]{bourrier2013atmospheric, bourrier2015radiative,debrecht2020effects, mccann2021atmospheric}. With increasing distance from the star, our outflow becomes increasingly neutral. While radiation pressure will not affect the strongly ionized outflows close to the star, we expect radiation pressure could play a significant role for the neutral hydrogen farther out, especially as the neutral hydrogen starts to decouple from the ionized outflow (see also Section \ref{sec:collisionality}).

\subsection{Planetary magnetic field confinement}
A planetary outflow may be confined in the presence of a strong planetary magnetic field. We estimate strength of the magnetic field at the surface of the planet, such that the magnetic field pressure $P_B$ becomes equal to the wind pressure $P = P_k + \rho v^2$ - the sum of the thermal and the ram pressures - at the sonic point. The magnetic field pressure of a dipole field goes as:
\begin{equation} \label{eq:b_field}
    P_B(r) = P_B(r_{sonic}) \left(\frac{r}{r_{sonic}}\right)^{-6}.
\end{equation}
Then, the magnetic field at the surface is: $B = (8 \pi P_B (r_P))^{1/2}$, which for our fiducial hot Jupiter is 0.4G. For comparison, the magnitude of Jupiter's magnetic field is 4.17 G. However, the magnitude of the magnetic fields on hot Jupiters is currently unknown. 

Using Equation \ref{eq:b_field}, the magnetic field at the surface of our fiducial super Earth that would produce a magnetic field pressure at the sonic point comparable to the pressure of the planetary outflow is $\sim 1.1$ G. For comparison, the magnetic field of Earth is between 0.25 and 0.65 G, and Uranus has an equatorial magnetic field of 0.23 G. The significantly more neutral outflow for our super-Earth, compared to the hot Jupiter, as well as the expected relatively weaker magnetic field, lead to the super Earth outflow that is less affected by the planetary magnetic field.

\subsection{Collisionality}\label{sec:collisionality}
An outflow may be modeled as a fluid if it is collisional out to the sonic point. In other words, the mean-free path to collisions, $1/(n_{col}\sigma$) is less than the pressure scale height, $H = k_{\rm{B}}T/(\mu g)$, with the gravitational acceleration $g = GM_P/r^2$. There are three cross sections for collision relevant to our problem: the hard-body cross section that is used to approximate the collision cross section for neutral hydrogen, $\sigma_{hb}$, the Coulomb scattering cross section used for ions, $\sigma_C$, and the charge-exchange cross section between ions and neutrals, $\sigma_{ce}$. The gas is mostly ionized in our gas giant models throughout our grid, and the ionized hydrogen is collisional throughout our semimajor axis range out to the sonic point, justifying the use of a fluid model.

Though sub-dominant, the behavior of the neutral component is important because Lyman-$\alpha$ observations probe neutral hydrogen.  At the base, the neutral hydrogen remains collisional with itself, but in the wind as we go out in semimajor axis, the mean free path to collisions calculated using the hardbody cross-section, $\sigma_{hb}$, generally exceeds the scale height for our models. However, neutrals and ions are coupled via charge exchange (cross-section $\sigma_{ce}$), and we find that neutrals are coupled to the outflow at least to the flow sonic point for planetary orbits out to 0.2 au for our Jupiter models.  For super-Earths, the flow remains fully collisional out to 0.1 au. Beyond these distances, the neutral hydrogen stops being collisional with the rest of the outflow below the sonic point, and it starts to free-stream out. 

For Jupiters, this will not impact the actual mass-loss rate, as outflow is mostly ionized at all distances considered in our model. However, it will impact the observed mass-loss rates by decreasing the mass-loss rate of the neutral hydrogen component of the wind, which may lead to an underestimate in the inferred total mass-loss rate. For super-Earths, as the outflow is substantially more neutral and as the neutral hydrogen component of the wind loses pressure support, we will see a decrease in both the actual and observed mass-loss rates.

We note that at very large distances from the planet (not modeled here), the gas reaches ionization/recombination equilibrium and neutral hydrogen in the outflow may actually spends most of its time in the ionized state, where it has plenty of time to thermalize with the surrounding ions,  so that the neutral hydrogen can still be treated as collisional. However, near the sonic point of the outflow, the recombination time is typically longer than the outflow time for the planets in our simulations and and this consideration does not affect their collisionality.

\subsection{Literature context}
There have been many theoretical studies of hydrodynamic escape. Because the focus of this paper is on radiative cooling, in this paper we only focus on papers that have made explicit statements on the effects of radiative cooling on the outflow properties. Here we present a discussion of several studies with the aim of placing our paper in the broader context of the theoretical work. Theoretical work on atmospherical escape includes models of escaping hydrogen and helium atmospheres bathed in EUV radiation, in addition to possible metals entrained in the wind.

\citet{munoz2007physical} and \citet{yelle2004aeronomy} created one-dimensional time-dependent models that both introduce the general structure of hot Jupiter thermospheres. \citet{munoz2007physical} found that the escape can be supersonic. As also discusssed in \citet{yelle2004aeronomy}, the $\hthreeplus$ ion was found at the base of the modeled atmosphere, which is an effective radiative coolant for energy deposited in that region. They found that the escape approaches the energy-limited (\citealt{watson1981dynamics}, \citealt{lammer2003atmospheric}) regime for strongly irradiated hot Jupiter planets. However, Lyman-$\alpha$ cooling was not included as part of the radiative coolants in this study. Nevertheless, our model qualitatively agrees with both models in terms of the structure: a molecular layer at the base cooled by $\hthreeplus$, which is thermally disssociated into atomic hydrogen, which outflows as an ionized wind. Both models found outflows reaching of order $10^4$ K, which were cooled via adiabatic expansion above the $\hthreeplus$ layer, however, \citet{yelle2004aeronomy} did not model the region where the outflow reaches supersonic velocities. \citet{rmc2009} found that Lyman-$\alpha$ cooling can be significant, particularly for higher stellar ionizing fluxes. In addition, \citet{koskinen2007stability} used a coupled thermosphere-ionosphere model to calculate the upper atmospheric temperature without explicitly calculating the outflow structure and thus not including PdV work.  They found that $\hthreeplus$ radiative cooling can significantly limit outflows by keeping these temperatures low.

\citet{caldiroli2022irradiation} focused on the evaporation efficiency of planets in a similar mass range to those in our study, though their fiducial low-gravity planet GJ 3470b is higher in gravity than our fiducial super-Earth. They specifically were interested in how the efficiency scales with flux due to Lyman-$\alpha$ cooling, rather than the molecular coolants. They found that for high-gravity planets, the outflow becomes less energy-limited with decreasing XUV flux, and the reverse trend for the lower-gravity planets. This matches the trend that we observe in Figure \ref{fig:mdot} for our fiducial hot Jupiter: Lyman-$\alpha$ cooling increases in importance with lower stellar flux. However, for our super-Earth (Figure \ref{fig:mdot_se}) the planet gravity is too low for Lyman-$\alpha$ cooling to be significant at any flux level. Instead, the $\hthreeplus$ cooling is more important with lower stellar flux, causing the mass-loss rate to deviate from the energy-limited rate.

Having been detected in the upper atmospheres of giant planets, metals (and their corresponding line cooling) are important to include going forward. For example, \citet{koskinen2013escape} created a hydrodynamic model (incorporating photoelectron heating) that modeled metals entrained in the outflow, demonstrating that they will not necessarily diffusively separate to the base but rather will be transported upwards enough to produce an observable signature.

For this paper, we are interested in the role of molecular coolants and chemistry in the molecular hydrogen layer, and thus chose to focus on the layer deeper in the outflow. We've tested our model and compared it to the hot Jupiter models from the literature. We now discuss theoretical work focusing on the smaller-planet regime, which is our direction for future work with this model.

\citet{koskinen2022mass} included a model of atmospheric escape from a Uranus-sized planet ($\sim14.5 M_{\earth}$) in the context of XUV-driven hydrodynamic escape to the Roche lobe overflow transition at close-in distances. Our super-Earth runs are qualitatively similar to the results of their hot Uranus model, in which instead of the sharp transition between the molecular hydrogen layer and the atomic layer, as seen in our hot Jupiter simulations (and also in \citet{yelle2004aeronomy} and \citet{munoz2007physical}), molecules are present througout the outflow. Particularly interesting is the presence of $\hthreeplus$ throughout the outflow, instead of being confined to a narrow region at the base. In agreement with \citet{koskinen2022mass}, this is largely due to the lower temperature in the wind. Our temperature range is $1000-3800$K, in comparison to their temperature prediction of $4000-5000$ K for the hot Uranus, continuing the trend that a planet with lower gravity (and similar stellar and orbital parameters) will host a lower-temperature outflow. Our cooling term is dominated by $\hthreeplus$ cooling at low altitudes, with the $PdV$ expansion dominating throughout the outflow despite the presence of $\hthreeplus$ thoughout, in agreement with their model. Lyman-$\alpha$ cooling is not significant, as the required temperatures are not reached in the Uranus model either. The comparison to the Uranus-like model is interesting because it suggests that H/He atmospheres are perhaps more similar to super-Earths than to hot Jupiters. In contrast, an upper atmospheric model of hot Neptune HAT-P-11b ($\sim25 M_{\earth}$) is presented in Fig. 5 of the Extended Data of \citet{ben2022signatures}, revealing a layered structure similar to that of the hot Jupiter models. While this hot Neptune exhibits a temperature peak characteristic of the hot Jupiter models, it only reaches a maximum temperature of $\sim 8400$ K, in contrast to the $10^4$ K temperature peak of our hot Jupiter. The outflow is largely neutral for the hot Uranus and super Earth, while the hot Neptune and Jupiter are ionized.

In the super-Earth regime, studies have shown that the impact of radiative cooling by molecules in $\htwo$-dominated atmospheres can be significant (e.g., \citet{yoshida2022less}, \citet{yoshida2024suppression}).

Our model (along with some of the abovementioned hot Jupiter models) suggests that Lyman-alpha cooling is a dominant cooling source for our fiducial hot Jupiter. However, metal-rich atmospheres, especially around lower-mass planets around the terrestrial mass scale, may have metal lines that will dominate the Lyman-alpha cooling. For example \citet{ito2021hydrodynamic} modeled the extreme case of a rocky Earth-size planet orbiting at 0.02 au from its host star. Figure 9 of the study depicts the energy budget, in which most of the heating from the incoming XUV flux is radiated away by radiative cooling of the sodium D line. At $10^{-4} - 10^{-3}$, the heating efficiency in such an atmosphere is greatly reduced compared to the order $0.1$ estimate found here and in other hot Jupiter hydrodynamic escape models. An increased atmospheric metallicity will aid in atmospheric retention, whether through atomic or molecular line coolants. This is something we will apply our model to in future work.

\section{Summary} \label{sec:summary}
Our goal was to evaluate the efficiency of atmospheric escape as a result of EUV heating, while being radiatively cooled by $\hthreeplus$ and Lyman-$\alpha$. For this, we created a one-dimensional hydrodynamic model that solves the fluid equations to steady state, coupled to a chemical network solver that allowed us to include reactions between the various species of $\h$ and $\he$ in our atmosphere. We apply our model to a fiducial hot Jupiter at 0.05au. We model the region of the outflow above the 1 $\mu$bar level, near which molecules such as $\htwo$ are present.

As a result of our simulations, we uncovered a three-layer structure to the atmosphere of our hot Jupiter:
\begin{itemize}
    \item a molecular hydrogen layer at the base (cooled by $\hthreeplus$),
    \item an intermediate layer of predominately neutral atomic hydrogen (cooled by Lyman-$\alpha$), and
    \item an outflow of $\hplus$, in which $PdV$ was the main cooling mode.
\end{itemize}

We then ran a grid of models with increasing distance from the star. We found that the proportion of total EUV flux that is radiatively cooled by $\hthreeplus$ stayed relatively low within the parameter range of our study. In contrast, as the outflow became increasingly neutral with distance, the Lyman-$\alpha$ cooling had an increasingly important effect as a fraction of the total heating. This decreased the mass-loss rate even faster than the na\"ive estimate of $a^{-2}$. Tidal forces enhanced the mass-loss rate close-in to the star, though enhanced $\hthreeplus$ cooling somewhat mitigated this effect.

We have developed a flexible code that can be expanded to include the effects of higher-energy photons, which are important for planets around younger stars. In addition, expanding the spectrum to the far-ultraviolet range would allow us to include molecular dissociation, which is especially important for smaller planets. Diffusion has been implemented in the code, but not turned on, as its effects are negligible for the purposes of this study. However, it is important for planets with metals in the atmosphere.

Including the effects of secondary ionization by high-energy photoelectrons increased the ionization fraction in the molecular layer near the base. The rate of collisional destruction of $\hthreeplus$ by electrons was increased, thereby decreasing the amount of $\hthreeplus$ that could accumulate. This resulted in less $\hthreeplus$ radiative cooling, compared to a model run without secondary ionization turned on. In that case, the base region heating was almost entirely offset by $\hthreeplus$ radiative cooling. After accouting for secondary ionization, the mass-loss rate increased from $6.4 \times 10^9$ to $9 \times 10^9$ $\mathrm{g \, s^{-1} \,sr^{-1}}$ for our fiducial hot Jupiter run, which is counter to the intuition that secondary ionization results in a decreased efficiency of escape.

As we increased the semimajor axis in our grid of models beyond $0.1-0.2$ au, the neutral hydrogen started to potentially decouple from the collisional outflow, free-streaming out. Such an outflow may have a two-component structure, where the ions are able to escape hydrodynamically, while the neutral atoms are in a slower, Jeans-escape regime. As ionized metals are expected to be entrained in the ionized outflow, there may be structure that is observable as a result of this decoupling of the escape regimes. The mass-loss rate produced by our models is an upper bound for outflows where neutral hydrogen is a significant component.

We further expanded our model to include atmospheres of $\h$ and $\he$ in the super-Earth regime. Being significantly cooler, these atmospheres did not have Lyman-$\alpha$ cooling as a significant cooling source. In contrast to the hot Jupiter models, the $\hthreeplus$ molecule was not confined to the layer of molecular hydrogen at the base. The cooler gas was not sufficiently heated for the sharp transition between molecular and atomic hydrogen to occur: $\htwo$ (and therefore $\hthreeplus$) was present throughout the outflow. Peaking near the base of our model, the cooling due to $\hthreeplus$ was a significant energy sink. Moreover, the efficiency of $\hthreeplus$ cooling increased as we moved the planet outward from 0.03 to 0.2 au. This contrasts with the limited radiative losses due to the $\hthreeplus$ in the hot Jupiter case, where only the highest-energy photons, deposited deep in our simulated region, were affected.

The atmospheres of the lower-mass planets may be enhanced in metals, particularly molecules such as, for example, CO, $\text{C}\text{O}_2$, $\text{H}_2\text{O}$, $\text{C}\text{H}_4$, $\text{N}_2$, and OH. $\hthreeplus$ may be destroyed, particularly by reacting with $\text{H}_2\text{O}$ and CO. In addition, molecular hydrogen can be lost by reacting with OH (\citealt{munoz2007physical}), limiting the production of $\hthreeplus$. $\hthreeplus$  may also be destroyed by recombination with electrons, for which the photoionization of metals such as C, O, Si, and N may be an additional source. This may limit the radiative cooling due to the $\hthreeplus$ significantly. Furthermore, metals entrained in the outflow may be a separate source of radiative cooling. Future work is required to determine how the presence of metals in the atmosphere affects the outflow.

\begin{acknowledgements}
This work was supported by funding from the Heising Simons Foundation.
\end{acknowledgements}

\appendix

\section{Hydrodynamic solver}\label{sec:hydro_solver}

We solve the mass conservation equation of each of the 8 component species separately. We assume that the gas is thermalized, i.e. the temperatures of the species are the same. All of the species travel with the bulk velocity $u$, and diffusion is dealt with implicitly in the mass conservation equation.

We keep track of the variables $\rho$, $u$, and $e$, as well as $\rho'$, $u'$, and $e'$, where the primes denote the gradient of the variables $n$, i.e. $n' = \partial n / \partial r$. The initial $n'$ are calculated using a central difference formula, $n{_i}' = (n_{i+1}-n_{i-1}) / (r_{i+1} - r_{i-1})$. A forward difference is used for the cell at the lower boundary, $n{_0}' = (n_{1}-n_{0}) / (r_{1} - r_{0})$, and, similarly, a backward difference is used for the cell at the upper boundary, $n_{-1}' = (n_{-1}-n_{-2}) / (r_{-1} - r_{-2})$.

The hydrodynamic equations, Equations \eqref{mass_species}, \eqref{momentum_spherical}, and \eqref{energy_spherical}, can all be re-cast in the form:
\begin{equation}\label{eq:hydro_generic}
    \frac{\partial X}{\partial t} + u \frac{\partial X}{\partial r} = G_X.
\end{equation}
Here, the left-hand side describes the advection of a quantity $X$ such as density, momentum, or energy, and $u$ is the bulk velocity of the fluid. The right-hand side contains the non-advective term. For the mass conservation equation it is:
\begin{equation}
    G_{\rho} = -\frac{\rho}{r^2} \frac{\partial(u r^2)}{\partial r},
\end{equation}
which describes the compression and/or expansion of the gas. For the momentum equation it is:
\begin{equation}
    G_{u} = -\frac{1}{\rho} \frac{\partial P}{\partial r} -\frac{G M}{r^2} + \frac{3GM_{\star } r}{a^3},
\end{equation}
where right-hand side describes the force due to a pressure gradient, gravity, and tidal gravity. Finally, for the energy equation it is:
\begin{equation}
    G_{e} = -\frac{P}{\rho r^2} \frac{\partial (u r^2)}{\partial r} + \Gamma -\Lambda + \frac{\partial}{\partial r} \left( \kappa_{\rm{th}} r^2 \frac{\partial T}{\partial r}\right),
\end{equation}
which contains the $PdV$ work term, the heating and cooling terms, and the conduction term as described in Section \ref{subsec:hydro}. We can solve Equation \ref{eq:hydro_generic} by separating the advective part from the non-advective part. This method of solving our differential equation is called operator splitting: as the terms represent distinct physical processes, the numerical methods best suited for solving each of them may be different.

We now apply the CIP (Cubic-Interpolated Propagation) method \citep{yabe1991universal, yabe2001exactly} to solve these equations. Going back to Equation \ref{eq:hydro_generic}, we can take the spatial derivative with respect to $r$ of this equation to get:
\begin{equation}\label{eq:hydro:dr0}
    \frac{\partial }{\partial r}\left(\frac{\partial X}{\partial t}\right) + \frac{\partial u}{\partial r} \frac{\partial X}{\partial r} + u \frac{\partial }{\partial r}\left(\frac{\partial X}{\partial r}\right)= \frac{\partial G_X}{\partial r},
\end{equation}
or written in the form of Equation \ref{eq:hydro_generic}
\begin{equation}\label{eq:hydro:dr}
    \frac{\partial X'}{\partial t} + u \frac{\partial X'}{\partial r} = -\frac{\partial u}{\partial r} X' + G'_X,
\end{equation}
where we define $X'=\partial X/ \partial r$ and $G'_X = \partial G'_X / \partial r$. In analogy to Equation \eqref{eq:hydro_generic}, Equation \ref{eq:hydro:dr} can also be split into an advective part on the right-hand side, and a non-advective part on the left.

\subsection{Non-advective part}\label{sec:nonadvectpart}
\subsubsection{Hydrodynamic equations}\label{sec:nonadvect_hydro}
We write the non-advective substep of Equation \ref{eq:hydro_generic} as:
\begin{equation}
    \frac{\partial \xt}{\partial t} = G_{X_n}.
\end{equation}
Given $X_n$, we solve for $\xt$ by explicitly integrating with respect to time using a finite difference:
\begin{equation}
    \xt = X + \Delta t G_X.
\end{equation}
On our staggered grid described in Section \ref{sec:grid}, the mass conservation equation becomes:
\begin{equation}\label{eq:fin_diff}
    \tilde{\rho}_i = \rho_i^n - \frac{\rho_i^n}{r_i^2} \frac{u_{i+1/2}^n-u_{i-1/2}^n}{\Delta r} \Delta t.
\end{equation}
The momentum equation becomes:
\begin{equation}
    \tilde{u}_{i+1/2} = u_{i+1/2}^n +(-\frac{2}{\rho_{i+1}^n+\rho_i^n} \frac{P_{i+1}^n-P_i^n}{\Delta r} +\frac{GM_P}{r_i^2}+\frac{3GM_{\star}a}{r_i^3})\Delta t,
\end{equation}
where we've estimated $\rho_{i+1/2}$ as $(\rho_{i+1}+\rho_{i})/2$. The energy equation becomes:
\begin{equation}
\begin{split}
    \tilde{e}_i =& e_i^n - \frac{P_i^n}{\rho_i^n r_i^2}\frac{r^2_{i+1/2} \tilde{u}_{1+1/2}+r^2_{i+1/2} u_{1+1/2}-r^2_{i-1/2} \tilde{u}_{1-1/2}-r^2_{i-1/2} u_{1-1/2} }{2 \Delta r} + \Gamma_i -\Lambda_i +\\
    &\frac{r^2_{i+1} \kappa_{\rm{th},i+1} (T_{i+2}-T_{i})/(r_{i+2}-r_{i})-r^2_{i-1} \kappa_{\rm{th},i-1}(T_{i}-T_{i-2})/(r_{i}-r_{i-2})}{r_{i+1}-r_{i-1}},
\end{split}
\end{equation}
with the heating and cooling terms $\Gamma_i$ and $\Lambda_i$ as described in Section \ref{sec:hydro}.
\subsubsection{Derivative of the hydrodynamic quantities}
As in Section \ref{sec:nonadvect_hydro}, we can write the non-advective substep of Equation \ref{eq:hydro:dr} as:
\begin{equation}\label{eq:nonadvect_X}
    \frac{\partial \tilde{X}'}{\partial t}  = -\frac{\partial u}{\partial r} X' + G'_X.
\end{equation}
We can write the finite difference form for the non-advection $X'_i$ associated with $G'_i$ by approximating $X'$ with a central difference of $X$:
\begin{equation}\label{eq:X_fin_diff}
    \frac{\tilde{X}'_i-X_i^{\prime n}}{\Delta t} = \frac{1}{\Delta t}\left( \frac{\tilde{X}_{i+1}-\tilde{X}_{i-1}}{2 \Delta r}-\frac{X_{i+1}^n-X_{i-1}^n}{2 \Delta r}\right) = \frac{(\tilde{X}_{i+1}-X_{i+1}^n)+(X_{i-1}^n-\tilde{X}_{i-1})}{2 \Delta r \Delta t}.
\end{equation}
Using Equation \eqref{eq:fin_diff}, we can write the grouped terms of the right-hand side as:
\begin{equation}
    \frac{\tilde{X}_{i+1}-X^n_{i+1}}{\Delta t} = G_{X,i+1}
\end{equation}
\begin{equation}
    \frac{\tilde{X}_{i-1}-X^n_{i-1}}{\Delta t} = G_{X,i-1}
\end{equation}
So Equation \ref{eq:X_fin_diff} becomes:
\begin{equation}
    \frac{\tilde{X}'_i-X_i^{\prime n}}{\Delta t} = \frac{G_{X,i+1}-G_{X,i-1}}{2 \Delta t}
\end{equation}
The term on the right-hand side can be complicated to evaluate, so we will use Equation \eqref{eq:X_fin_diff} instead, as all the terms are known. This allows us to avoid having to evaluate $G$. We can write the finite difference form for the first term on the right-hand side of Equation \eqref{eq:nonadvect_X} as:
\begin{equation}
    -\tilde{X}'_i \frac{u_{i+1}-u_{i-1}}{2 \Delta t},
\end{equation}
so the full equation becomes:
\begin{equation}
    \frac{\tilde{X}'_i-X_i^{\prime n}}{\Delta t}  = \frac{(\tilde{X}_{i+1}-X_{i+1}^n)+(X_{i-1}^n-\tilde{X}_{i-1})}{2 \Delta r \Delta t} -\tilde{X}'_i \frac{u_{i+1}-u_{i-1}}{2 \Delta t}.
\end{equation}
We use this equation to solve for the non-advective part of the time evolution of the gradient of $\rho$, $u$, and $e$.

\subsection{Advective part}\label{sec:hydro_solver_ad}
After solving the non-advective part of the fluid equations, we solve the advection equation:
\begin{equation}\label{eq:advect}
    \frac{\partial X}{\partial t} + u \frac{\partial X}{\partial r} = 0.
\end{equation}
This equation is solved by the cubic interpolating profile method, in which we keep track of the variables $X$ and also of their gradients $X'$. We differentiate \eqref{eq:advect} to obtain:
\begin{equation}
    \frac{\partial X'}{\partial t} + u\frac{\partial X'}{\partial r} = -\frac{\partial u}{\partial r}X' = \frac{\partial Y}{\partial t} + u \frac{\partial Y}{\partial r} = -\frac{\partial u}{\partial r} Y, 
\end{equation}
where we define $Y \equiv X'$. For timestep n, we know $X$ and $Y$ at all grid points. We can write the solution to Equation \eqref{eq:advect} at point $i$ and time $t$ as: $X_i(r_i, t)= X_i(r_i-u\Delta t, t-\Delta t)$: we advect the profiles from the previous timestep at time $t-\Delta t$ to the current time $t$. However, the exact solution is only given at time $t-\Delta t$ at the center of the gridpoints $r_i$. We don't know the profile within the cell, so we must interpolate and we choose a third-order polynomial to do so. The polynomial takes the general form:
\begin{equation}
    F_i(\tilde{r}) = a_i \tilde{r}^3+b_i\tilde{r}^2 + c_i \tilde{r} +d_i
\end{equation}
where $\tilde{r} = r-r_i$. For timestep $n+1$, we shift the profile:
\begin{equation}
    X^{n+1}=F(r-u\Delta t), \mathrm{and}
\end{equation}
\begin{equation}
    Y^{n+1}=\frac{d}{dr}F(r-u\Delta t).
\end{equation}
$dF(r)/dr = 3ar^2+2br+c$. At $r=0$, $dF/dr=c$, so $c_i=Y^n_i.$ Furthermore, at $r=0$, $F=d$, so $d_i=X^n_i$. Then,
we also know $F$ and $dF/dr$ at $r_{i+1}$. $\tilde r = r_{i+1}-r_i=\Delta r$, which is the grid size. We can solve the following two equations for the constants:
\begin{equation}
    F(r_{i+1}) = a_i\Delta r^3 +b_i \Delta r^2 + Y^n_i \Delta r + X^n_i = X^n_{i+1}, \mathrm{and}
\end{equation}
\begin{equation}
    \frac{\partial F(r_{i+1})}{\partial r} = 3a_i\Delta r^2 + 2b_i\Delta r+Y^n_i = Y^n_{i+1}.
\end{equation}
The constants are:
\begin{equation}
    a_i = \frac{Y^n_{i+1}+Y^n_i}{\Delta r^2}-\frac{2(X^n_{i+1}-X^n_i)}{\Delta r^3}, \mathrm{and}
\end{equation}
\begin{equation}
    b_i = \frac{3(X^n_{i+1}-X^n_i)}{\Delta r^2}-\frac{Y^n_{i+1}-2Y^n_i}{\Delta r}.
\end{equation}

\subsection{Grid comments} \label{ap:sec:grid}
Using the same grid for density, velocity, and energy may result in high-frequency oscillations in the solution. The velocity is coupled to the pressure via the momentum equation. Ignoring the advective term on the left and the force terms on the right, the finite-difference form of Equation \eqref{momentum_spherical} for cell $i$ on the grid is:
\begin{equation} \label{eq:oddeven}
    \frac{\partial u_i}{\partial t} = -\frac{1}{\rho_i}\frac{P_{i+1}-P_{i-1}}{r_{i+1}-r_{i-1}},
\end{equation}
where we have used the central difference method for calculating gradients. Note that $P_i$ does not appear in the calculation of $\partial u_i/ \partial t$ - even and odd cells are decoupled, which results in two separate modes in the solution that may appear as a zig-zag pattern (also called odd-even grid oscillation or odd-even decoupling). This is a numerical artifact that may impact the stability of the solution \citep{tang2018phenomenon}. To mitigate this issue, the resolution may be increased (at a computational cost), a filter (which may impact accuracy) or artificial viscosity (which introduces numerical dissipation) may be used; another way is to use a staggered grid. That way, instead of Equation \eqref{eq:oddeven} we have:
\begin{equation}
    \frac{\partial u_{i+1/2}}{\partial t} = -\frac{2}{\rho_i+\rho_{i+1}}\frac{P_{i+1}-P_{i}}{r_{i+1}-r_{i}}
\end{equation}
with no decoupling, where we use half-indeces such as $i+1/2$ to denote cells centered on the grid staggered by half a cell size from the cells $i$ on the regular grid.

\section{Chemical network solver}\label{appendix:chemsolver}
We label each reaction $i$ in Table \ref{tab:reactions} by its rate constant k$_i$, which has units of cm$^3$ s$^{-1}$ for two-body reactions and cm$^6$ s$^{-1}$ for three-body reactions. The system of equations resulting from the chemical network presented in Table \ref{tab:reactions} is stiff. As the rate constants for the reactions can vary by many orders of magnitude, using a single timestep for explicitly integrating the reactions becomes computationally unfeasible, introducing errors and instabilities. Therefore, implicit methods are needed. We use semi-implicit Euler method, described in Section 17.5.2 of \citet{press2007numerical}, to solve the system of equations for the concentrations of the species in our chemical network. We now briefly describe the method.

\subsection{Scheme}\label{chemical_scheme}
The concentrations of the species in our network evolve under a set of differential equations. We can write the set of concentrations of the species as a vector $\vv{y}$. Consider the set of equations for a vector quantity $\vv{y}$ whose time derivative is a function of $y$ that is not necessarily linear:
\begin{equation} \label{eq:nonlineq}
    \vv{y'} = \vv{f(y)}
\end{equation}
We can explicitly advance $\vv{y}$ to the next timestep: $\vv{y_{n+1}} = \vv{y_{n}} + dt \, \vv{y'_n}$ - the derivative is computed at timestep $n$. On the other hand, we can advance the solution implicitly by computing the derivative at timestep $n+1$ instead: $\vv{y_{n+1}} = \vv{y_{n}} + dt \, \vv{y'_{n+1}}$. The advantage of implicit schemes is that they maintain the stability of the solution (generally at a computational cost). We can use the implicit differencing scheme to advance Equation \eqref{eq:nonlineq}:
\begin{equation} \label{eq:impdiff}
    \vv{y_{n+1}} = \vv{y_{n}} + dt \, \vv{f(y_{n+1})}.
\end{equation}
As $\vv{f(y_{n+1})}$ is unknown, we can construct a Jacobian matrix $\partial \vv{f}/\partial \vv{y}$ (evaluated at $\vv{y_n}$) to linearize the equations:
\begin{equation}
    \vv{f(y_{n+1})} = \vv{f(y_{n})} + \frac{\partial \vv{f}}{\partial \vv{y}} \cdot (\vv{y_{n+1}}-\vv{y_n}).
\end{equation}
Together with Equation \eqref{eq:impdiff}, this gives:
\begin{equation}
    \vv{y_{n+1}} = \vv{y_{n}} + dt \, \left(\vv{f(y_{n})} + \frac{\partial \vv{f}}{\partial \vv{y}} \cdot (\vv{y_{n+1}}-\vv{y_n})\right),
\end{equation}
or, after rearranging,
\begin{equation} \label{eq:lineq}
    \vv{y_{n+1}} = \vv{y_{n}} + \left(\textbf{1}-dt \, \frac{\partial \vv{f}}{\partial \vv{y}}\right)^{-1} dt \, \vv{f(y_{n})},
\end{equation}
where $\textbf{1}$ is the identity matrix. We rewrite this as:
\begin{equation} \label{eq:maineq}
    \vv{f(y_{n})} = \left(\frac{1}{dt} \textbf{1}- \, \frac{\partial \vv{f}}{\partial \vv{y}}\right) (\vv{y_{n+1}} - \vv{y_{n}}).
\end{equation}

This equation is solved via LU decomposition with forward substitution, followed by back substitution, as implemented in section (2.3) of \citet{press2007numerical}. 

The Jacobian $\partial \vv{f}/\partial \vv{y}$ is calculated analytically for our problem, as are the time derivatives of the concentrations on the left-hand side of Equation \ref{eq:maineq}. The integration timestep $dt$ must be chosen carefully. Even though we wish to advance the solution by a desired timestep $\Delta t$, we need to cross the interval $\Delta t$ with a series of timesteps $dt_i$, for each of which we solve Equation \ref{eq:maineq}. One cannot always use an arbitrarily large timestep in this method - the time interval must be subdivided into sufficiently small timesteps to obtain an accurate solution. The method attempts to increment the solution by  $dt_i$, after which a polynomial extrapolation is performed as described for the Bulirsch-Stoer Method in section (17.3.2) of \citet{press2007numerical} to check for the suitability of the timestep. The timestep is then adjusted as necessary according to the step size sequence defined in (17.5.41) of \citet{press2007numerical}.

As an example, consider separately a reaction network formed by reactions 8 and 11, with rate constants k$_8$ and k$_{11}$:

\begin{equation}
    \htwoplus + \htwo \xrightarrow{k_8}\hthreeplus + \h
\end{equation}

\begin{equation}
    \hthreeplus + \h \xrightarrow{k_{11}} \htwoplus + \htwo.
\end{equation}

We label the time derivatives of the concentrations ($n_s$) of $\htwoplus$, $\htwo$, $\hthreeplus$, and $\h$ as $f(n_s)$:
\begin{equation} \label{eq:derh2}
    \begin{split}
        f(n_{\htwoplus}) &= \frac{d n_{\htwoplus}}{dt} = -k_8 n_{\htwoplus}n_{\htwo} + k_{11} n_{\hthreeplus} n_{\h}, \\
        f(n_{\htwo}) &= \frac{d n_{\htwo}}{dt} = -k_8 n_{\htwoplus}n_{\htwo} + k_{11} n_{\hthreeplus} n_{\h}, \\
        f(n_{\hthreeplus}) &= \frac{d n_{\hthreeplus}}{dt} = k_8 n_{\htwoplus}n_{\htwo} - k_{11} n_{\hthreeplus} n_{\h}, \\
        f(n_{\h}) &= \frac{d n_{\h}}{dt} = k_8 n_{\htwoplus}n_{\htwo} - k_{11} n_{\hthreeplus} n_{\h}.
    \end{split}
\end{equation}

As applied to our problem, the Jacobian matrix $\partial \vv{f}/\partial \vv{y}$ is the matrix of partial derivatives of the time rates of change of the concentrations of each species with respect to the concentrations of all of the other species in the network. The Jacobian matrix for this system of equations is:
\begin{equation} 
\begin{pmatrix}
    \frac{d f(n_{\htwoplus})}{d n_{\htwoplus}} & \frac{d f(n_{\htwoplus})}{d n_{\htwo}} & \frac{d f(n_{\htwoplus})}{d n_{\hthreeplus}} & \frac{d f(n_{\htwoplus})}{d n_{\h}}\\
    \frac{d f(n_{\htwo})}{d n_{\htwoplus}} & \frac{d f(n_{\htwo})}{d n_{\htwo}} & \frac{d f(n_{\htwo})}{d n_{\hthreeplus}} & \frac{d f(n_{\htwo})}{d n_{\h}}\\
    \frac{d f(n_{\hthreeplus})}{d n_{\htwoplus}} & \frac{d f(n_{\hthreeplus})}{d n_{\htwo}} & \frac{d f(n_{\hthreeplus})}{d n_{\hthreeplus}} & \frac{d f(n_{\hthreeplus})}{d n_{\h}}\\
    \frac{d f(n_{\h})}{d n_{\htwoplus}} & \frac{d f(n_{\h})}{d n_{\htwo}} & \frac{d f(n_{\h})}{d n_{\hthreeplus}} & \frac{d f(n_{\h})}{d n_{\h}}\\
\end{pmatrix} =
\begin{pmatrix}
    -k_8 n_{\htwo} & -k_8 n_{\htwoplus} & k_{11}  n_{\h} & k_{11} n_{\hthreeplus}\\
    -k_8 n_{\htwo} & -k_8 n_{\htwoplus} & k_{11}  n_{\h} & k_{11} n_{\hthreeplus}\\
    k_8 n_{\htwo} & k_8 n_{\htwoplus} & -k_{11}  n_{\h} & k_{11} n_{\hthreeplus}\\
    k_8 n_{\htwo} & k_8 n_{\htwoplus} & -k_{11}  n_{\h} & k_{11} n_{\hthreeplus}\\
\end{pmatrix}
\end{equation}

Specifically in the \citet{press2007numerical} implementation, we use the StepperSie method\footnote{As the grid cells are independent, we are able to run the chemical network in parallel.} with the parameters as defined in \citet{press2007numerical}: $\mathrm{rtol}=1.0\times 10^{-7}$, $\mathrm{atol}=1.0\times 10^{-4} \, \mathrm{rtol}$, $\mathrm{h}_1=1.0\times 10^{-6}$, $\mathrm{hmin}=0.0$.

\section{Tidal gravity term}\label{appendix:tidal}
Consider the substellar ray of length $a$ from the center of the planet (with mass $M_P$, radius $R_p$) to the center of the star (with mass $M_\star$), along with a test particle that is a distance $r\ll a$ from the planet. In the frame centered on the planet and co-rotating with the planet, the forces on the particle are the gravitational force from the planet and the star, as well as the centrifugal force. Though present in the co-rotating frame, we neglect the Coriolis term, as we are working in one dimensions. We make the approximation $M_p \ll M_{\star}$ and $r \ll a$ to obtain the total force term: 

\begin{equation} \label{eq:totalforce}
    F_{\rm{tot}} = -\frac{GM_{P}}{r^2} -\frac{GM_{\star}}{(a-r)^2}+\frac{G(M_{P}+M_{\star})}{a^3} \left(r-a\frac{M_{\star}}{M_{P}+M_{\star}}\right) \simeq -\frac{GM_{P}}{r^2}+\frac{3rGM_{\star}}{a^3}.
\end{equation}

\noindent The last term in Eq. \eqref{eq:totalforce} is the tidal term: it is the sum of the gravitational term due to the star and the centrifugal term.

\section{Escape of Lyman-$\alpha$ photons from the outflow}\label{appendix:lyaesc}

We have calculated the Lyman-$\alpha$ cooling rate (Equation \eqref{eq:lyman_alpha_cooling_rate}) under the approximation that all Lyman-$\alpha$ photons that are produced in collisions ultimately escape from the wind.  This assumption is justified by the idea, demonstrated in \citet{rmc2009} Appendix C using an order-of-magnitude approximation, that these photons scatter into the Lyman-$\alpha$ line wings and escape from the outflow before collisions can thermalize their energy.   Here, we conduct a Monte Carlo radiative transfer simulation to verify this conclusion and determine the fraction of Lyman-$\alpha$ photons that escape from the wind in our fiducial outflow (Figure \ref{fig:rhovpt_fiducial}).  We treat the outflow as spherically-symmetric, with our computed profile applying in all directions.  Because the atmosphere is asymmetrically heated, this is an approximation, but our results nevertheless give a more accurate estimate of the Lyman-$\alpha$ escape fraction than can be obtained via order-of-magnitude estimation.

We start 1000 photons at radius $r_0 = 1.05\times10^{10}$ cm, at the base of the region where Lyman-$\alpha$ cooling is significant (Figure \ref{fig:energy_balance_reactions}).  For each photon, we calculate a random walk trajectory using the following procedure, until the photon escapes from the atmosphere, is thermalized by collisional de-excitation, or reaches the region of the atmosphere where bolometric heating and cooling dominates the energetics (near the base of the simulation).  We consider photons that leave the bottom boundary to be thermalized in the bolometrically-dominated region, where their energy contribution is not significant, and thus ``lost" from the wind.  For our fiducial planet, bolometric heating and cooling dominates below $r = 1.005\times 10^{10}$ cm, and we use this radius as the lower boundary for our Monte Carlo calculation.  However, the energy balance in this lower region of planetary atmospheres merits additional study, so we quote rates of direct loss to space and loss into the bolometric region separately for reference.  

Each time they are absorbed and re-emitted (colloquially ``scattered"), resonance line photons have their wavelengths redistributed according to the probabilities in the line profile \citep[e.g.,][]{harrington1973scattering} Thus, for each step in the random walk, we first draw a photon frequency, $\nu$, from the Voigt probability distribution, $\phi(\nu)$, for hydrogen Lyman-$\alpha$, with thermal broadening determined by the local temperature and natural broadening adjusted by the local collision rate.  This frequency applies in the frame of reference moving with the local outflow velocity.

We then generate a ray through the atmosphere.  The direction of the ray is drawn randomly assuming isotropic scattering by drawing each component of the direction vector in Cartesian coordinates from a normal distribution. A ray along this direction is constructed by choosing a set of points separated by distances along the ray of $\Delta s = 0.01r_0$, running from the photon's initial position to a position with planet-centric radius $r_{\rm max} = 10r_0$, beyond which we consider the photon to have escaped.  We approximate the outflow structure as spherically symmetric with the computed flow properties only a function of planet-centered radius, $r$.  We translate our photon frequency into the intertial frame by adding a Doppler shift due to the component of the outflow velocity along the ray at the point of emission.

For each point along the ray, we compute the absorption cross-section, 
\begin{equation}
\sigma = \frac{1}{8\pi}\frac{g_2}{g_1}\left(\frac{c}{\nu}\right)^2 A_{21} \phi(\nu) \;\;, 
\end{equation}
for our chosen frequency, where the degeneracy factors $g_2 = 6$ and $g_1 = 2$, $c$ is the speed of light, and $A_{21} = 6.265 \times 10^8 \, \mathrm{s^{-1}}$ is the spontaneous decay rate appropriate for hydrogen Lyman-$\alpha$ (dominated by de-excitations from the 2p state).  In calculating $\phi(\nu)$, we take into account the gas properties at the location along our ray and we Doppler shift the result according to the component of the local (at the position on the ray) outflow velocity in the direction of motion of the photon (i.e., along the ray).  Using this cross-section, we calculate the the optical depth $\Delta \tau = n_0 \sigma \Delta s$ and hence the probability of absorption $P_{\rm abs} = 1 - e^{-\Delta \tau}$ for each bin along the ray.  We draw a sent of uniform random numbers between 0 and 1 for each bin along the ray and consider the photon to be absorbed in the first bin where the drawn random number is less than $P_{\rm abs}$.  

Using the properties of the outflow at the absorption radius, $r$, we then probabilistically determine whether the photon is collisionally thermalized or re-emitted as a new Lyman-$\alpha$ photon.  We take the spontaneous decay rate to be $A_{21}$, and we calculate the  the collision rate, $\gamma_{\rm col}$ using the sum of the rate of collisions with electrons (cross-section $2\times 10^{-15}$ cm$^2$; \citealt{brackmann1958collisions}), $\htwo$ (cross-section $1\times 10^{-14}$ cm$^2$; \citealt{terazawa1993excitation} Figure 6 for H(2p) with relative velocity $\approx 4.5$ km s$^{-1}$), protons (cross-section $2\times 10^{-14}$ cm$^2$; \citealt{hunter1977proton}), and other species (cross-section $2\times 10^{-15}$ cm$^2$ similar to elastic collisions between electrons and neutral hydrogen).  Each rate is calculated as $n_i\sigma_i v_{{\rm th},i}$ where for component $i$, $n_i$ is the number density, $\sigma_i$ is the collisional cross-section, and the thermal velocity $v_{{\rm th},i} = (2k_B T/\mu_i)^{1/2}$ is calculated using the mean molecular weight, $\mu_i$, of the component.

We take the collisional de-excitation probability to be $P_{\rm deex} = 1-e^{-\gamma_{\rm col}/A_{21}}$, draw a uniform random number between 0 and 1, and consider the photon to be thermalized if the number is less than $P_{\rm deex}$.  If the photon is not thermalized and remains in the domain of the calculation (i.e., has not been lost out of the top or bottom), we draw a new photon frequency and a new direction of propagation, and repeat this process. 

For our fiducial atmosphere, we find that 78\% of photons beginning at $r_0 = 1.05\times10^{10}$ cm escape to space, 11\% escape out of the bottom of the flow to the bolometric region, and 11\% are thermalized. The total loss rate is 89\%, justifying our use of Equation \ref{eq:lyman_alpha_cooling_rate}. The typical number of steps taken before escape to space is $\approx$$10^5$, consistent with the estimate in \citet{rmc2009}, when calculated using the density where we launch our photons, which is approximately ten times higher than chosen in that work.

We are also interested in whether Lyman-$\alpha$ photons produced following excitations by secondary electrons escape from the outflow, as we have assumed in Section \ref{sec:secondarye}.  These photons are primarily produced in the molecular layer (see Figure \ref{fig:h2rates_selected_fiducial}).  We therefore repeat our calculation with $r_0 = 1.01 \times 10^{10}$ cm and find that 36\% escape to space, 27\% escape out the bottom of the flow, and 37\% are thermalized.  While the majority of the Lyman-$\alpha$ photons escape even from this deep region, about a third do not, meaning that we are modestly underestimating the heating generated by secondary electrons in the molecular layer. Collisional excitation Lyman-$\alpha$ from secondary photoelectrons accounts for $\sim$15\% of the total energy deposited at that radius.  We have verified that errors in the loss of secondary electron energy at this level do not qualitatively change our results.

  \label{fig:rhovpt_fiducial_4case}

\newpage
\bibliography{refs}{}
\bibliographystyle{aasjournal}



\end{document}